\documentclass[12pt,preprint]{aastex}
\usepackage[makeroom]{cancel}
\usepackage{graphicx}
\usepackage{epstopdf}
 \usepackage{amssymb, latexsym, amsmath}
 \usepackage[dvips]{color}
 \usepackage{graphicx}
  \usepackage{color}

 \newcommand{\inc}{{\it i}}

 \newcommand{\be}{\begin{equation}}
 \newcommand{\ee}{\end{equation}}
 \newcommand{\ba}{\begin{eqnarray}}
 \newcommand{\ea}{\end{eqnarray}}
 \newcommand{\bs}{\begin{subequations}}
 \newcommand{\es}{\end{subequations}}

 \newcommand{\erbold}{\mbox{{\boldmath $
 r$}}}
 \newcommand{\rbold}{\mbox{{\boldmath $
 r$}}}

 \newcommand{\xbold}{\mbox{{\boldmath $
 x$}}}
   \newcommand{\Xbold}{\mbox{{\boldmath $
 X$}}}

  \newcommand{\ubold}{{\bf{u}}}
  \newcommand{\Ubold}{{\bf{U}}}

  \newcommand{\fbold}{\mbox{\boldmath ${\boldmath{f}}$}}
  
  \newcommand{\eRbold}{\mbox{{\boldmath $
  {R}$}}}
  \newcommand{\Rbold}{\mbox{{\boldmath $
  {R}$}}}

  \newcommand{\vbold}{{\bf{v}}}
  \newcommand{\Sbold}{{\bf{S}}}
  \shorttitle{Tidal heating}
  \shortauthors{Efroimsky \& Makarov}

\begin{document}
  \title{${{~~~~~~~~~~~~~~~~~~~~~~~}^{
          {\rm
            {
       Extended~version~of~a~paper~
       published~in~
                the~Astrophysical~Journal
        \,~795\;:\;6\,~(2014)
                  }}}}$
                  ~\\
 {\Large{\textbf{{{
 Tidal dissipation in a homogeneous spherical body.\\
 I. Methods}
 \vspace{1.5mm}
  ~\\}
            }}}}
\author{Michael Efroimsky $\,$and$\,$ Valeri V. Makarov ~~~}
\affil{US Naval Observatory, 3450 Massachusetts Avenue NW, Washington DC 20392}
\email{michael.efroimsky@usno.navy.mil$\,$, ~~vvm@usno.navy.mil}
     \date{}

\begin{abstract}

 A formula for the tidal dissipation rate in a spherical body is derived from first principles, to correct some mathematical inaccuracies found in the literature. The
 development is combined with the Darwin-Kaula formalism for tides. Our intermediate results are compared with those by Zschau (1978) and Platzman (1984). When restricted to
 the special case of an incompressible spherical planet spinning synchronously without libration, our final formula can be compared with the commonly used expression from
 Peale \& Cassen (1978, Eqn. 31). The two turn out to differ. While the said expression from {\it{Ibid.}} was intended solely for a synchronously rotating body,
 our formula is for an arbitrary spin state. 
 Examples of the application of our expression for the tidal damping rate are provided in the work by Makarov \& Efroimsky (2014).

 \end{abstract}


 \section{Motivation and plan}

 The tidal heating of planets and moons has long been a key area of planetary science. Accurate investigation into this process requires numerical integration of dissipation
 over layers of the perturbed body. At the same time, it is common to infer qualitative conclusions from approximations based on modeling the body with a homogeneous sphere
 of a certain rheology. However, the simplistic nature of the approach limits the precision of the ensuing conclusions. For example, the presence of a sizable molten core,
 like in Mercury, may increase the damping rate, compared to a homogeneous body. Still, estimates obtained with our simplified, homogeneous-sphere model
 should be accurate within a factor of several --- thus (1) serving as a useful guidance for solar system bodies and (2) being completely legitimate for
 exoplanets, as our knowledge of their structure is speculative at best.

 \vspace{3mm}
 In our paper, we derive from the first principles a formula for the tidal heating rate in a tidally perturbed homogeneous sphere. We compare our result with the formulae used in the literature and point out the differences.

 \vspace{3mm}
 In Sections \ref{prelude2} - \ref{sphere}, we present an accurate re-examination of the standard integral expression for the damping rate in a homogeneous incompressible sphere subject to tides. The check is necessary because in previous studies the expression was derived in an {\it ad hoc} manner, sometimes with demonstrable mathematical inaccuracies. The conventional derivation begins with the general formula for the power
 \ba
 P\;=\;
 \int\,\rho_{_E}\;\vbold\,\cdot\,\nabla {V_{_E}}'\;d^3r
 \nonumber
 \ea
 written in the Eulerian description (i.e., via coordinates associated with a deformed body). Its time average is then cast into the form of
 \ba
 \langle P\rangle~=~\frac{1}{4\pi G R}~\sum_{l=2}^{\infty}\,(2\,l\,+\,1)\,\int~\left\langle~W_l~\stackrel{\bf\centerdot}{U}_l~\right\rangle~dS
 \nonumber
 \ea
 which is in the Lagrangian language (an integral over an undeformed body). In the former equation, $\,\rho_{_E}\,$ is the Eulerian density, $\,\vbold\,$ is the Eulerian velocity, $\,{V_{_E}}'\,$ is the Eulerian perturbation of the potential (perturbation assembled of the tide-raising potential and the resulting additional tidal potential of the deformed body), and $\,r\,$ is a {\it{perturbed}} position in the body frame. In the latter equation, $\,W_l\,$ and $\,U_l\,$ are the degree-$l\,$ components of the tide-raising and additional tidal potentials, $\,G\,$ is the Newton gravity constant, $\,R\,$ is the radius of the planet, and $\,dS\,$ is an element of the {\it{undeformed}} surface of the sphere.

  The transition from the former formula to the latter requires the use of the boundary conditions on the free surface. At that point, integration is already carried out within the Lagrangian description (over an undeformed surface), but the boundary conditions are nonetheless imposed on the Eulerian potential and its gradient. (The boundary conditions are much simpler in the Eulerian form.) This mixed treatment requires attention, and its employment by the early authors (Zschau 1978, Platzman 1984) contained inaccuracies. However none of those turned out to be critical, and the above expression for the average power $\,\langle P\rangle\,$ is correct for small deformations.

 \vspace{3mm}
 In Section \ref{next} we explore the standard way of casting the above integral into a spectral sum over the tidal Fourier modes $\,\omega\,$. It is commonly assumed that the result should read as in Platzman (1984):
 \ba
 \langle P\rangle~=~\frac{1}{4\pi G R}~
 {\sum_{\omega}}
 (2l+1)\,\frac{\omega}{2}\,W^{\,2}_l(\omega)\,k_l(\omega)~\sin\epsilon_l(\omega)\,~.
 \nonumber
 \ea
 Here $\,k_l(\omega)\,$ and $\,\epsilon_l(\omega)\,$ are the Love number and phase lag corresponding to the Fourier mode $\,\omega=\omega_{lmpq}\,$, with $\,lmpq\,$ being the four integers wherewith the Fourier modes are numbered in the Darwin-Kaula theory of tides (see Efroimsky \& Makarov 2013 and references therein). However, an accurate investigation demonstrates that the spectral sum differs from the above. The difference originates for two reasons. One is the degeneracy, i.e., the fact that several different Fourier modes $\,\omega_{lmpq}\,$ share a numerical value $\,\omega\,$, so the
 structure of the above sum is more complex. \footnote{~When calculating $\,W_l\,$, one has first to group together and sum all the terms corresponding to a particular value of $\,\omega\,$. Each such sum should be squared and averaged, and only after that should the final summation over the distinct values of $\,\omega\,$ be carried out. In the original expression for
the average power,
 $\,~(\textstyle 4\pi G R)^{-1}~{\sum_{\omega}} (2l+1)\,\frac{\textstyle \omega}{\textstyle 2}\,W^{\,2}_l(\omega)\,k_l(\omega)~\sin\epsilon_l(\omega)\,~$, the $W^{\,2}_l(\omega)$ term should be replaced with the squared sum of all the harmonics of $\,W\,$ that correspond to a particular value of $\,\omega\,$.} The second reason is that
the modes can be of either sign, not necessarily positive. So the resulting power will contain seemingly strange terms with $\,W_l(\omega)\,W_l(-\omega)\,k_l(\omega)~\sin\epsilon_l(\omega)\,$.

 These difficulties were noticed and analysed by Peale and Cassen (1978) who derived the dissipation rate in a synchronously rotating body.$\,$\footnote{~In the expression for $\,\langle P \rangle\,$, an input from each value of $\,\omega_{lmpq}\,$ must be non-negative. This can be observed from the fact that the mode $\,\omega=\omega_{lmpq}\,$ and the corresponding phase lag $\,\epsilon_l(\omega)\equiv\omega\,\Delta t_l(\omega)\,$ are always of the same sign (the time lag $\,\Delta t_l(\omega)\,$
 being positive definite due to causality). Thus the product $\,\omega\,\epsilon_l(\omega)\,=\,\omega_{lmpq}\,\epsilon_l(\omega_{lmpq})\,$ in the spectral sum can always be
 rewritten as $\,|\omega_{lmpq}|/Q_{lmpq}\,$, with the tidal quality factor being defined via $\,1/Q_{lmpq}\,=\,|\,\sin\epsilon_l(\omega_{lmpq})\,|\,$. In their spectral sum, Peale \& Cassen (1978, eqn 31) have just $\,\omega_{lmpq}/Q_{lmpq}\,$, ~and not $\,|\omega_{lmpq}|/Q_{lmpq}\,$. The reasons for this is that they are employing a nonstandard convention $\,1/Q_{lmpq}\,=\,\sin\epsilon_l(\omega_{lmpq})\,$ wherein the inverse quality factors incorporate the signs of the lags and, thus, are not positive definite.} One of our goals is to generalise their result to an arbitrary spin state.

 The calculation of the power production, developed by Peale \& Cassen (1978), implies averaging not only over the tidal period but also over the apsidal period. This can be observed from the formulae (20 - 21) in their work. In our paper, however, we consider two separate cases: those with and without apsidal precession. In the first case, the period of the apsidal precession is shorter than the typical time of relaxation in the mantle (which may be identified with the Maxwell time). The argument of the pericentre of the perturber, $\,{\omega}^*\,$, cannot be treated as constant, wherefore the formula for the mean power should be averaged not only over the tidal period, but also over the period of the pericentre motion. (We assume this motion steady.) In the second case, the evolution of the line of apsides is slow, with its period being longer than the Maxwell time. The argument of the pericentre should be regarded as a constant. Accordingly, in the latter case the tidal dissipation formula is more complicated, because it includes explicit dependence of Fourier terms on the argument of pericentre.

 In a subsequent work, Makarov \& Efroimsky (2014),
 we apply our results in three case studies: Io, Mercury, and Kepler-10$\,$b. In that paper we, among other things, hypothesise that the tidal heating rate at spin-orbit resonances is greatly influenced by libration and, therefore, by the triaxiality of the tidally perturbed body.

 \section{The Darwin-Kaula formalism in brief}

 Describing of linear bodily tides consists of two steps. First, it is necessary to Fourier-expand both the tide-raising potential and the induced additional potential of
 the tidally perturbed body. Second, it is necessary to link each Fourier component of the additional tidal potential to an appropriate Fourier component of the tide-raising potential. This means: establishing the phase lag and the ratio of magnitudes called the {\it{dynamical Love number}}.

 Due to interplay of rheology and self-gravitation, the phase lags and Love numbers have nontrivial frequency dependencies. Things are complicated even further
 because different mechanisms of friction become leading over different frequency bands, wherefore the tidal response cannot be described by one simple dissipation model (Efroimsky 2012$\,$a,b).

 \subsection{Generalities}

 The development of the mathematical theory of bodily tides was started by Darwin (1879) who derived a partial sum of the Fourier expansion of the additional potential of a tidally perturbed sphere. A decisive contribution into this theory was offered almost a century later by Kaula (1964) who wrote down a complete series. In a previous paper (Efroimsky \& Makarov 2013), we provided a detailed presentation of the Darwin-Kaula expansion and explained how tidal friction and lagging are built
 into it. We compared the Darwin-Kaula theory with the one by MacDonald (1964) and demonstrated that the former theory is superior to the latter, because it can, in principle, be combined with an arbitrary rheology. Referring the reader to the afore-cited literature for details, we present several central formulae that will be necessary.

 An external body of mass $\,M^{\,*}\,$, located in $\,{\erbold}^{\;*} = (r^*,\,\lambda^*,\,\phi^*)\,$, generates the following disturbing
 potential in a point $\,\Rbold = (R,\phi,\lambda)\,$ on the surface of a sphere of radius $\,R\,<\,r^*~$:
 \ba
 \nonumber
 W(\eRbold\,,\,\erbold^{~*})~=~\sum_{{\it{l}}=2}^{\infty}~W_{\it{l}}(\eRbold\,,~\erbold^{~*})~=~-~\frac{G\;M^*}{r^{
 \,*}}~\sum_{{\it{l}}=2}^{\infty}\,\left(\,\frac{R}{r^{~*}}\,\right)^{\textstyle{^{\it{l}}}}\,P_{\it{l}}(\cos \gamma)~=
 ~\quad~\quad~\quad~\quad~\\
 \nonumber\\
 \nonumber\\
 -\,\frac{G~M^*}{r^{\,*}}\sum_{{\it{l}}=2}^{\infty}\left(\frac{R}{r^{~*}}\right)^{\textstyle{^{\it{l}}}}\sum_{m=0}^{\it l}
 \frac{({\it l}-m)!}{({\it l}+m)!}(2-\delta_{0m})P_{lm}(\sin\phi)P_{lm}(\sin\phi^*)~\cos m(\lambda-\lambda^*)~~.~\quad~
 \label{1a}
 \label{101a}
 \ea
 Here $\,G
 \,$ denotes Newton's gravity constant,
 $\,\phi\,$ is the latitude reckoned from the spherical body's equator, $\,\lambda\,$ is the longitude measured from a fixed meridian, and $\gamma$ is the angular
 separation between the vectors $\,{\erbold}^{\;*}\,$ and $\,\Rbold\,$ pointing from the perturbed body's centre. The definitions of the Legendre polynomials $\,P_l(\cos\gamma)\,$ and the associated Legendre polynomials $\,P_{lm}(\sin\phi)\,$ are given in Appendix \ref{appB}.

 While in the above formula the location of the perturber on its trajectory is expressed through the spherical coordinates $\,{\erbold}^{\;*} = (r^*,\,\lambda^*,\,\phi^*)
 \,$, a trigonometric transformation (developed by Kaula 1961) enables one to switch to the perturber's orbital elements  $\,\erbold^{\;*}=(\,a^*,\,e^*,\,\inc^*,\,\Omega^*,\,\omega^*,\,{\cal M}^*\,)\,$. In terms thereof, the disturbing potential is expressed as
 \ba
 \nonumber
  W(\eRbold\,,\;\erbold^{\;*})\;=\;\sum_{lmpq}\,W_{lmpq}\;=\;-\;
  \frac{G\,M^*}{a^*}\;\sum_{{\it
  l}=2}^{\infty}\;\left(\,\frac{R}{a^*}\,\right)^{\textstyle{^{\it
  l}}}\sum_{m=0}^{\it l}\;\frac{({\it l} - m)!}{({\it l} + m)!}\;
  \left(\,2  \right. ~~~~~~~~~~~~~~~~~~~~~~~~~~~~~~~~~~~~~~\\
                                   \nonumber\\
                                   \nonumber\\
       \left.
  ~~~ -\;\delta_{0m}\,\right)\;P_{{\it{l}}m}(\sin\phi)\;\sum_{p=0}^{\it
  l}\;F_{{\it l}mp}(\inc^*)\;\sum_{q=\,-\,\infty}^{\infty}\;G_{{\it l}pq}
  (e^*)
  \left\{
  \begin{array}{c}
   \cos   \\
   \sin
  \end{array}
  \right\}^{{\it l}\,-\,m\;\;
  \mbox{\small even}}_{{\it l}\,-\,m\;\;\mbox{\small odd}} \;\left(
  v_{{\it l}mpq}^*-m(\lambda+\theta^*)  \right)
 ~~~,~~~~~~
 \label{1b}
 \label{101b}
  \label{1}
 \label{101}
 \ea
   where $\,\theta^*\,$ is the rotation angle of the tidally perturbed body,$\,$\footnote{~When the equinoctial precession may be neglected, $\,\theta^*\,$ may be regarded as the sidereal angle.} while $\,F_{lmp}(\inc^*)\,$ and $\,G_{lpq}(e^*)$ are the inclination functions and the eccentricity polynomials,
 respectively. The auxiliary linear combinations $\,v_{lmpq}^*\,$ are defined by
 \ba
 v_{lmpq}^*\;\equiv\;(l-2p)\,\omega^*\,+\,(l-2p+q){\cal M}^*\,+\,m\,\Omega^*~~~.
 \label{2}
 \label{102}
 \ea
 Conventionally,
 the letters denoting the elements of the perturber are accompanied with asterisks:
 $\,a^*,\,e^*,\,\inc^*,\,\Omega^*,\,\omega^*,\,{\cal M}^*\,$. Following Kaula (1964), the sidereal angle also acquires an asterisk, when it appears in a
 combination $~v_{lmpq}^*-\,m\,\theta^*~$ with the perturber's elements.

 The angle $\,\theta\,$, however, does not acquire an asterisk, when it appears in a linear combination $~v_{lmpq}-\,m\,\theta~$ with the orbital elements of a test body
 subject to the additional tidal potential of the perturbed body. This strange nomenclature introduced by Kaula (1964) --- two different notations for one angle --- turns out to
 be helpful and convenient in the calculation of the back-reaction experienced by the perturber. For comprehensive explanation of this obscure point, see Section 5 in Efroimsky \& Makarov (2013).

 Over timescales shorter than the apsidal-motion period, the expression in round brackets in the formula (\ref{1}) can be linearised as
 \ba
 v_{lmpq}^*-m(\lambda+\theta^*)\,=\,\omega_{lmpq}\,(t\,-\,t_0)~-~m~\lambda~+~v_{lmpq}^*(t_0)~-~m~\theta^*(t_0)~~,
 \label{3}
 \label{103}
 \ea
 where the following quantities act as the Fourier tidal modes:
 \ba
 \omega_{\textstyle{_{lmpq}}}\;\equiv~\stackrel{\bf\centerdot~~~~}{v^*_{lmpq}}\,-~m\,\stackrel{\bf\centerdot\,}{\theta^*}
 ~=\;(l-2p)\;\dot{\omega}^*\,+\,(l-2p+q)\;{\bf{\dot{\cal{M}}}}^{\,*}\,+\,m\;(\dot{\Omega}^*\,-\,\dot{\theta}^*)~~,
 \label{4a}
 \label{104a}
 \label{504a}
 \label{504}
 \ea
 ${\bf{\dot{\cal{M}}}}^{\,*}\,$ being the perturber's ``anomalistic" mean motion (see Section \ref{diff} below), and $\,t_0\,$ being the time of pericentre passage. (As ever, we set $\,{\cal{M}}^{\,*}=\,0\,$ in the pericentre.) The modes $\,\omega_{\textstyle{_{lmpq}}}\,$ can assume either sign, but the physical forcing frequencies are positive definite:
 \ba
 \chi_{\textstyle{_{lmpq}}}\,=\,|\,\omega_{\textstyle{_{lmpq}}}\,|\,~.
 \label{chi}
 \ea

 \subsection{Simplifying the notation: $\,$less asterisks}

 In the preceding subsection, we obeyed the convention by Kaula (1964) and marked with asterisk the orbital elements of the tide-raising body. Kaula introduced this notation, because within his model he also considered another exterior body that was disturbed by the tides generated on the planet by the tide-raising body. This exterior body's elements were denoted by the same letters, but without an asterisk.

 When the two outer bodies coincide, the asterisks may be dropped, except on two occasions. The first is writing the masses -- while the mass of the planet is denoted with $\,M\,$, the mass of the perturber (the star) will be written as $\,M^{\,*}\,$. The other occasion requires writing the additional tidal potential of the perturbed body -- the potential will have a value $\,U(\erbold,\,\erbold^{\,*})\,$ in a point $\,\erbold\,$,
 provided the perturber (the star) is located in an exterior point $\,\erbold^{\,*}\,$ (both vectors being planetocentric). The planet's rotation rate $\,\theta\,$, as well
 as the orbital elements of the star as seen from the planet, will hereafter be written without asterisks.

The most important notations employed in this paper are connected in Table \ref{nota.tab}.

 \subsection{Difficulties}\label{diff}

 At this point, a word of warning is necessary. Deriving the equation (\ref{504a}), we differentiated the expression (\ref{2}), which gave us the terms with $\,\dot{\omega}\,$
 and $\,\dot{\Omega}\,$. Including these terms in the equation (\ref{504a}) acknowledges the fact that the perturber's trajectory is disturbed, not Keplerian.
 The disturbance may come solely from tides, as in Kaula (1964, eqn 38), or from both tides and other sources. One way or another, the perturber's mean anomaly
 $\,{\cal{M}}\,$ is no longer equal to $\,n\,t\,$, $\,$but is now given by
 $~\,
 {\cal{M}}\,=~{\cal{M}}_0\,+~\int^{\,t}_{t_0} n(t)~dt
 ~\,$,
 ~where
 $
 n(t)\,\equiv\,\sqrt{G\,(M\,+\,M^*)~a^{-3}(t)\,}~.
 \,$
 Accordingly, the expression for the modes becomes:
 \ba
 \omega_{\textstyle{_{lmpq}}}\;\equiv~\stackrel{\bf\centerdot~~~~}{v_{lmpq}}\,-~m\,\stackrel{\bf\centerdot\,}{\theta}
 ~=\;(l-2p)\;(\dot{\omega}\,+\,\dot{\cal{M}}_0)\,+\,q\,\dot{\cal{M}}_0
 \,+\,(l-2p+q)\;n\,+\,m\;(\dot{\Omega}\,-\,\dot{\theta})~~.\quad
 \label{4b}
 \label{104b}
 \label{504b}
 \ea
  It is, of course, tempting to assume that $\,\stackrel{\bf\centerdot}{{\cal{M}}_0}\,\ll\,n\,$, thus accepting the approximation
 \ba
 \stackrel{\bf\centerdot}{\cal{M}\,}\,=~\stackrel{\bf\centerdot}{{\cal{M}}_0}\,+~n~\approx~n~~,
 \label{105}
 \ea
 as Kaula (1964) did in his equations (46 - 47). Within his theory, however, this approximation could not be used.$\,$\footnote{~In his books, Kaula (1966, 1968)
 corrected this oversight. There, he kept the notation $\,n\,$ for the mean motion defined as in the Kepler law, and never confused it with $\,\stackrel{\bf\centerdot}{\cal{M}\,}\,$.} This is explained in Appendix \ref{difficulty} where we consider two examples. One is the case where perturbation of an orbit of a moon is mainly due to the tides the moon creates in the planet. In that situation, $\,\dot{\omega}\,$ and $\,\dot{\cal{M}}_0\,$ are of the same order but of opposite signs, so they largely compensate one another. This suggests a simultaneous neglect of $\,${\it{both}}$\,$ rates. The second example is when the dominant perturbation of the orbit comes from the oblateness of the primary. In this case, $\,\dot{\omega}\,$ and $\,\dot{\cal{M}}_0\;$ turn out to be of the same order and of the same sign -- so keeping one of these terms requires keeping the other.

 Whether one or both of these rates should be included depends on a particular setting, and each practical case must be examined separately. In general, both rates should be kept.

 While keeping $\,\stackrel{\bf\centerdot\,}{\omega}\,$ complicates the formalism, the emergence of $\,\stackrel{\bf\centerdot}{{\cal{M}}_0}\,$ complicates the treatment even further. To sidestep this issue, we shall $\,${\it{define}}$\,$ the mean motion via
 \ba
 n\,\equiv\,{\bf{\dot{\cal{M}}}}~~.
 \label{}
 \ea
 This, the so-called $\,${\it{anomalistic}}$\,$  mean motion differs from  $\,\sqrt{G(M+M^*)\,a^{-3}(t)\,}~$.

 We shall derive the heat-production formulae for two different settings -- with a fixed pericentre $\,\omega\,$ and with $\,\omega\,$ moving uniformly.

 \subsection{Lagging}

 For a static tide, the incremental tidal potential of the perturbed body mimics the perturbation (\ref{1}), except that each term $\,W_l\,$ is now equipped with a
 mitigating multiplier $\,k_{\textstyle{_l}}\left({R}/{r}\right)^{\,l+1}\,$, where $\,k_{\textstyle{_l}}\,$ is an $\,l-$degree Love number. With the star located in $\,\erbold^{\,*}\,$, the additional potential in a point $\,\erbold\,$ will read as
 \ba
 U(\erbold\,,\;\erbold^{\;*})&=&\sum_{{\it l}=2}^{\infty}~U_{\it{l}}(\erbold)~=~\sum_{{\it l}=2}^{\infty}~k_{\it
 l}\;\left(\,\frac{R}{r}\,\right)^{{\it l}+1}\;W_{\it{l}}(\eRbold\,,\;\erbold^{\;*})~~.~~~~~~~~
 ~~~~~~~~~~~~~~~~
  \label{dr}
 \ea
 For time-dependent tides, this expression acquires an extra amendment: the reaction must lag, compared to the action. Naively, this would imply taking each $\,W_l\,$ at an earlier instant of time. However, in reality lagging depends on frequency; so each $\,W_l\,$ must be first expanded into a Fourier series over tidal modes, whereafter each term of the series should be delayed separately. The magnitude of the tidal reaction is frequency dependent too; so each term of the Fourier series will be multiplied by a
 dynamical Love number of its own. Symbolically, this may be written in a manner similar to the static expression:
 \ba
 U(\erbold\,,\;\erbold^{\;*})&=&\sum_{{\it l}=2}^{\infty}~U_{\it{l}}(\erbold\,,\;\erbold^{\;*})~=~\sum_{l=2}^{\infty}~\left(\,\frac{R}{r}\,\right)^{{\it l}+1}\;\hat{k}_{l}\;W_{\it{l}}(\eRbold\,,\;\erbold^{\;*})~~.~~~~~~~~
 ~~~~~~~~~~~~~~~~
 \label{505a}
 \label{107a}
 \ea
 The hat above $\,\hat{k}_{l}\,$ means that this is not a multiplier but a linear operator that mitigates and delays each Fourier mode of $\,W_l\,$ differently:
  \ba
 \nonumber
  U(\erbold\,,\;\erbold^{\;*})&=&
     -\;
  \frac{G\,M^*}{a}\;\sum_{{\it
  l}=2}^{\infty}\;\left(\,\frac{R}{r}\,\right)^{\textstyle{^{l+1}}}
  \left(\,\frac{R}{a}\,\right)^{\textstyle{^{\it
  l}}}\sum_{m=0}^{\it l}\;\frac{(l - m)!}{(l + m)!}\;
  \left(\,2~-\;\delta_{0m}\,\right)\;P_{lm}(\sin\phi)~\sum_{p=0}^{\it
  l}\;F_{lmp}(i)~~~~\\
                                   \nonumber\\
                                   \nonumber\\
  &~&\left.~~~\right.\sum_{q=\,-\,\infty}^{\infty}\;G_{lpq}(e)~k_l(\omega_{lmpq})~ \left\{
  \begin{array}{c}
   \cos   \\
   \sin
  \end{array}
  \right\}^{{\it l}\,-\,m\;\;
  \mbox{\small even}}_{l\,-\,m\;\;\mbox{\small odd}} \;\left(\,v_{lmpq}~-\,m\,(\lambda~+~\theta)~-~\epsilon_{l}
  \, \right)~~,~~\qquad~\,
  \label{505b}
  \label{107b}
 \ea
 where the Love numbers $\,k_l(\omega_{lmpq})\,$ and the phase lags $\,\epsilon_{\textstyle{_{l}}}(\omega_{lmpq})\,$ are functions of the Fourier modes. The lags emerge as the products
 \bs
 \ba
 \epsilon_l(\omega_{\textstyle{_{lmpq}}})~=~\omega_{\textstyle{_{lmpq}}}~\Delta t_l(\omega_{\textstyle{_{lmpq}}})~~,
 \label{506a}
 \label{108a}
 \ea
 where $\Delta t_l(\omega_{\textstyle{_{lmpq}}})\,$ is the time delay at the mode $\,\omega_{\textstyle{_{lmpq}}}\,$. In reality, the time delays are functions not
 of the Fourier modes (which can assume either sign), but of the actual physical forcing frequencies $~\chi_{\textstyle{_{lmpq}}}\,=\,|\,\omega_{\textstyle{_{lmpq}}}\,|~$
 which are positive definite. Thus it is more accurate to write the delays not as $\Delta t_l(\omega_{\textstyle{_{lmpq}}})\,$ but as $\Delta t_l(\chi_{\textstyle{_{lmpq}}})\,$. Accordingly, the phase lags become
 \ba
 \epsilon_l(\omega_{\textstyle{_{lmpq}}})~=~\omega_{\textstyle{_{lmpq}}}~\Delta t_l(\chi_{\textstyle{_{lmpq}}})~~.
 \label{506b}
 \label{108b}
 \ea
 The time delays are positive definite due to causality, so the sign of the phase lag always coincides with that of the corresponding Fourier mode. Thus we finally have:
 \ba
 \epsilon_l(\omega_{\textstyle{_{lmpq}}})~=~|\,\omega_{\textstyle{_{lmpq}}}\,|~\,\mbox{Sgn}\,(\omega_{\textstyle{_{lmpq}}})~\,\Delta t_l(\chi_{\textstyle{_{lmpq}}})~=~
 \chi_{\textstyle{_{lmpq}}}~\,\mbox{Sgn}\,(\omega_{\textstyle{_{lmpq}}})~\,\Delta t_l(\chi_{\textstyle{_{lmpq}}})~~,
 \label{506c}
 \label{108c}
 \ea
 \label{506}
 \label{108}
 \es
 where $\,\chi_{\textstyle{_{lmpq}}}\equiv\,|\,\omega_{\textstyle{_{lmpq}}}\,|\,$ are the positive definite forcing frequencies.

  The dynamical Love number $\,k_l(\omega_{\textstyle{_{lmpq}}})\,$ and the phase lag  $\,\epsilon_l(\omega_{\textstyle{_{lmpq}}})\,$ are the absolute value and the negative
 phase of the complex Love number $\,\bar{k}_l(\omega_{\textstyle{_{lmpq}}})\,$ whose functional dependence upon the Fourier mode is solely determined by $\,l\,$, provided the body is spherical.$\,$\footnote{~For oblate celestial bodies, the functional form of the complex $\,\bar{k}_{\textstyle{_{l}}}(\omega_{\textstyle{_{lmpq}}})\,$ is also determined by the order $\,m\,$. In that situation, the right notation for the complex Love number is: $\,\bar{k}_{\textstyle{_{lm}}}(\omega_{\textstyle{_{lmpq}}})\,$. Its absolute value and negative phase will then be denoted with $\,{k}_{\textstyle{_{lm}}}(\omega_{\textstyle{_{lmpq}}})\,$ and $\,\epsilon_{\textstyle{_{lm}}}(\omega_{\textstyle{_{lmpq}}})\,$.}

 \subsection{Physics behind the Love numbers and phase lags}

 As we saw above, to obtain the decomposition (\ref{107b}) from the Fourier series (\ref{101b}), each $\,lmpq\,$ term of the latter had to be endowed with its own mitigating factor $\,k_{\textstyle{_l}}=k_{\textstyle{_l}}(\omega_{\textstyle{_{lmpq}}})\,$ and phase lag $\,\epsilon_{\textstyle{_l}}=\epsilon_{\textstyle{_l}}(\omega_{\textstyle{_{lmpq}}})\,$. In the past, some authors enquired whether this mitigate-and-lag method was general enough to describe tides. It is, as long as the tides are linear. This is explained in Appendix \ref{universality} below.

 The expression (\ref{107b}) for the additional tidal potential contains both sines and cosines of the phase lags, and so does the ensuing expression for the surface elevation. However, the resulting expression for the tidal dissipation rate turns out to contain only the combination $\,{k}_{\textstyle{_l}}(\omega)\;\sin\epsilon_{\textstyle{_l}}(\omega)\,$ which is the negative imaginary part of the complex Love number:
 \ba
 {k}_{\textstyle{_l}}(\omega)\;\sin\epsilon_{\textstyle{_l}}(\omega)\;=\;|\bar{k}_{\textstyle{_l}}(\omega)|\;\sin\epsilon_{\textstyle{_l}}(\omega)\;=
 \;-\;{\cal{I}}{\it{m}}\left[\bar{k}_{\textstyle{_l}}(\omega)\right]~~,~~~\mbox{where}\quad\omega=\omega_{\textstyle{_{lmpq}}}~~.
 \label{gfr}
 \ea
 This quantity is often denoted as $\,k_l/Q\,$, although it would be more reasonable to employ the notation $\,k_l/Q_l\,$, with the tidal quality factors defined through $\,1/Q_l\,\equiv\,\sin|\epsilon_l|\,$.

 A dynamical Love number $\,{k}_{\textstyle{_l}}(\omega_{\textstyle{_{lmpq}}})\,$ is an even function of the tidal mode $\,\omega_{\textstyle{_{lmpq}}}\,$, while a phase lag $\,\epsilon_{\textstyle{_l}}(\omega_{\textstyle{_{lmpq}}})\,$ is odd, as can be observed from the equation (\ref{108b}). Thus the expression for the product $\,{k}_{\textstyle{_l}}\,\sin\epsilon_{\textstyle{_l}}\,$ as a function of the physical frequency $\,\chi=\chi_{\textstyle{_{lmpq}}}\,\equiv\,|\omega_{\textstyle{_{lmpq}}}|\,$ is:
 \ba
 {k}_{\textstyle{_l}}(\omega)\;\sin\epsilon_{\textstyle{_l}}(\omega)\;=\;{k}_{\textstyle{_l}}(\chi)\;\sin\epsilon_{\textstyle{_l}}(\chi)\;\,\mbox{Sgn}\,(\omega)~~,
 \label{ggffrr}
 \ea
 where $\,\epsilon_l(\chi)\,$ is non-negative, because non-negative is the physical frequency $\,\chi\,$.

 The frequency dependence of $\,{k}_{\textstyle{_l}}/Q_l={k}_{\textstyle{_l}}(\chi)\,\sin\epsilon_{\textstyle{_l}}(\chi)\,$ is defined by two major physical circumstances: self-gravitation of the planet and the rheology of its mantle. A rheological law is expressed by a constitutive equation, i.e., by an equation interconnecting the strain and the stress. A particular form of this equation is determined by the friction mechanisms present in the considered medium. A realistic rheological law should contain contributions from elasticity, viscosity, and inelastic processes (mainly, dislocation unjamming). Self-gravitation suppresses the tidal bulge. At low frequencies this effectively adds to the mantle's rigidity, whereas at higher frequencies the interplay of rheology and gravity is more complex (Efroimsky 2012$\,$b, Figure 2).

 The calculation of the frequency dependence $\,{k}_{\textstyle{_l}}(\chi)\,\sin\epsilon_{\textstyle{_l}}(\chi)\,$ for a homogeneous body of a known size, mass and rheology is presented in detail in Efroimsky (2012a,b). See also the Appendix to Makarov \& Efroimsky (2014).

 While quadrupole ($\,l=2\,$) terms are sufficient in most problems, exceptions are known. For the orbital evolution of Phobos, the $\,l=3\,$ and, possibly, even $\,l=4\,$ terms of the Martian tidal potential may be of relevance (Bills et al. 2005). Studying close binary asteroids, Taylor \& Margot (2010) took into account the Love numbers up to $\,l=6\,$.

 The question of how rapidly $\,l>2\,$ terms fall off with the increase of the degree $\,l\,$ is also interesting. Most authors only rely on the geometric factor $\,(R/a)^{2l+1}\,$ to answer this question. As was explained in Efroimsky (2012$\,$b), the $\,l$-dependence of $\,k_l(\omega_{lmpq})\,\sin\epsilon(\omega_{lmpq})\,$, too, comes into play and changes the result considerably.

 \section{The Eulerian and Lagrangian descriptions
  }\label{prelude2}
  \vspace{2.4mm}
  ${\left.~~~~~~\,~~~~~~~~~~~~~~~~~~~~~~~~~~~~~~~~~~~~~~~~~~~~~~~~~~~~~~~~~~~~~~~~
 \,\right.}^{\mbox{\small \it What~we~hope~ever~to~do~with~ease,}}$~~\\
 ${\left.~~~~~~~~~~~~~~~~~~~~~~~~~~~~~~~~~~~~~~~~~~~~~~~~~~~~~~~~~~~~~~~~~~
 ~~~~~\,\right.}^{\mbox{\small \it we~must~learn~first~to~do~with~diligence.}}$
 ~\\
 ${\left.~~~~~~~~~~~~~~~~~~~~~~~~~~~~~~~~~~~~~~~~~~~~~~~~~~~~~~~~~~~~~~~~~~~~~~~
 ~~~~~~~~~~~~~~~~~~~~~~~~~~~\right.}^{
 \mbox{\small\it ~~~Samuel~Johnson
 }}$

 \subsection{Notations and definitions}

 To compare the varying shape of a deformable body against some benchmark configuration, we use $\,\Xbold\,$ to denote the initial position occupied by a particle at $\,t=0\,$. At  another time $\,t\,$, the particle finds itself in a new place
 \ba
 \xbold~=~\fbold(\Xbold,\,t)\,~,
 \label{change}
 \nonumber
 \ea
 where the function $\,\fbold(\Xbold,\,t)\,$ is a trajectory, i.e., a solution to the equation of motion, with the initial condition $\,\xbold=\Xbold~$ set at $\,t=0\,$.

 The current values of all physical and kinematic properties of the medium can be expressed as functions of the instantaneous coordinates $\,\xbold\,$ of a point where these properties are being measured at the present moment $\,t\,$. When referred to the present time and position, such properties are named {\it{Eulerian}} and are equipped with a subscript {\it{E}}$\,$; for example: $\,q_{_E}(\xbold,\,t)\,$. The Eulerian description is fit to answer the question ``where" and therefore is convenient in fluid dynamics where the displacement $\,\xbold-\Xbold\,$ can become arbitrarily large and the initial position $\,\Xbold\,$ is soon forgotten.

 While $\,\xbold\,$ denotes a place in space, the initial condition $\,\Xbold\,$ acts as the ``number" of a particle presently residing at the place $\,\xbold\,$. Although located now at $\,\xbold\,$, the particle originally came from $\,\Xbold\,$ and will carry the label $\,\Xbold\,$ forever.

 Knowing the trajectories of all particles, we can express the properties as functions of the time $\,t\,$ and the initial conditions $\,\Xbold~$. To that end, we employ the change of variables $\,\xbold~=~\fbold(\Xbold,\,t)\,$. Expressed through the initial conditions, a property $\,q\,$ will be termed as {\it{Lagrangian}} and equipped with the subscript $\,L~$:
 \bs
 \ba
 q_{_L}(\Xbold,\,t)~\equiv~q_{_E}(\xbold\,,~t)\,~~~~~~~~~
 \label{109a}
 \label{416a}
 \ea
 or, in more detail:
 \ba
 q_{_L}(\Xbold,\,t)~\equiv~q_{_E}(\,\fbold(\Xbold\,,\,t)\,,~t\,)\,~.
 \label{109b}
 \label{416b}
 \ea
 \label{109}
 \label{416}
 \es
 So $\,q_{_L}\,$ has the same value as $\,q_{_E}\,$, but has a different functional form, as it is now understood as a function of the initial conditions
  (the particles' $\,$`numbers') $\,\Xbold\,$, and not of the present-time coordinates $\,\xbold\,$. Relating the quantities to the initial positions $\,\Xbold\,$, the Lagrangian description tells us ``which particle" and is thus practical in description of deformable solids.

 In anticipation of perturbative treatment, we regard the trajectory $\,\xbold~=~\fbold(\Xbold,\,t)\,$  as fiducial and equip the appropriate functional dependencies with a superscript $\,0\,$:
  \bs
 \ba
 q^0_{_L}(\Xbold,\,t)~\equiv~q^0_{_E}(\xbold\,,~t)~~~~~~~~~~~
 \label{equality_a}
 \label{110a}
 \label{417a}
 \ea
 which is:
 \ba
 q^0_{_L}(\Xbold,\,t)~\equiv~q^0_{_E}(\,\fbold(\Xbold,\,t)\,,~t\,)\,~.
 \label{equality_b}
 \label{110b}
 \label{417b}
 \ea
 \label{110}
 \label{417}
 \label{equality}
 \es

 \subsection{Perturbative approach}\label{approach}

\noindent
 Under disturbance, two changes will take place in a point $\,\erbold\,$ at a time $\,t~$:
 \begin{itemize}
 \item[\bf 1.~] Properties will now assume different values in this point at this time.
 So we substitute the unperturbed Eulerian dependencies $\,q^0_{_E}(\rbold\,,~t)\,$ with
 \ba
 q_{_E}(\rbold\,,~t)~=~q^0_{_E}(\rbold\,,~t)~+~q\,'_{_E}(\rbold\,,~t)\,~.
 \label{111}
 \ea
 This equality, in fact, serves as a definition of the variation $\,q\,'_{_E}(\rbold,~t)~$: the variation is a change in the functional dependence of a physical property upon the present position $\erbold$
 ~\\

 \item[\bf 2.~] A different particle will now appear in the point $\,\erbold\,$ at the time $\,t\,$. It will not be the same particle as the one expected there at the time $\,t\,$ in the absence of perturbation.\\
      ~\\
      Accordingly, a particle, which starts in $\,\Xbold~$ at $~t=0\,$, will show up, at the time $\,t\,$, not in the point $\,\xbold=\fbold(\Xbold,\,t)\,$ but in some other location displaced by $\,\ubold~$:
 \ba
 \erbold~=~\xbold~+~\ubold
 ~=~\fbold(\Xbold,\,t)~+~\ubold(\Xbold,\,t)\,~.
 \label{112}
 \ea
 \end{itemize}
 Both OF these changes, {\bf{1}} and {\bf{2}}, will affect the Lagrangian dependencies of the properties upon the initial conditions, so the dependency of each property will acquire a variation $~q\,'_{_L}(\Xbold,\,t)~$:
 \ba
 q_{_L}(\Xbold,\,t)~=~q^0_{_L}(\Xbold,\,t)~+~q\,'_{_L}(\Xbold,\,t)\,~.
 \label{113}
 \ea
 In Appendix \ref{EL}, we provide a self-sufficient introduction into the perturbative treatment of a deformable body, both in the Eulerian and Lagrangian languages. There we derive a relation between the perturbations of the Lagrangian and Eulerian quantities:
 \ba
 q\,'_{_L}(\Xbold,\,t)~=~q\,'_{_E}(\,\fbold(\Xbold,\,t)\,,~t\,)~+~\ubold(\Xbold,\,t)~\nabla_{\textstyle{_x\,}} q_{_E}~+~O(\ubold^2)~~,
 \label{duga}
 \ea
 with the gradient in the second term acting on the unperturbed history: \footnote{~To derive (\ref{duga}), we expanded $\,q_{_E}(\erbold,\,t)=q_{_E}(\xbold+\ubold,\,t)$ into the Taylor series near the unperturbed $q_{_E}(\xbold,\,t)$.}
  \ba
 \nabla_{\textstyle{_{x\,}}} q_{_E}\,\equiv\,\nabla_{\textstyle{_x\,}}q_{_E}(\xbold,\,t)~\,~,\,\quad\mbox{where}\qquad \xbold\,=\,\fbold(\Xbold,\,t)~~.
 \label{116}
 \ea
  In the formula (\ref{duga}), the first term on the right-hand side, $~q\,'_{_E}(\xbold,\,t)~$, accounts for the change of the final spatial distribution of properties. The other two terms show up because
 perturbation alters the mapping from $\,\Xbold\,$ to the present position.

 \subsection{Summary of linearised formulae for the density\\
 of a periodically deformed solid}

 We need several formulae for density perturbations, which are obtained in Appendix \ref{EL}.

 \vspace{3mm}
 ${\bf{
 \divideontimes
 }}~~~$ {\underline{In the Eulerian description:}}
 \ba
 \rho_{_E}(\rbold,\,t)\,=~\rho^{\,0}_{_E}(\rbold)~+~{\rho_{_E}}'(\rbold,\,t)\,~,~\qquad~\quad~
 \label{141}
 \ea
 \ba
 \rho_{_E}\,'\,+\,\nabla_{\textstyle{_r\,}}\cdot(\rho^{\,0}_{_E}\,\ubold)\,=\,0\,~,~\qquad~\quad~
 \label{142}
 \ea

 Formula (\ref{141}) renders the interrelation between the functions of the same variable. The unperturbed density $\,\rho^{\,0}_{_E}\,$ appears here as a function of the perturbed present positions $\,\erbold\,$, not of the unperturbed reference positions $\,\xbold\,$. This can be traced through the derivation (\ref{132} - \ref{135}). There, the unperturbed density initially shows up as a function of $\,\xbold=\erbold-\ubold\,$. It then ends up as a function of $\,\erbold\,$, after the Taylor expansion around $\,\rbold\,$ over powers of $\,\ubold\,$ is performed.

 Accordingly, the symbol $\,\nabla_{\textstyle{_r\,}}\,$ denotes differentiation with respect to the perturbed position $\,\erbold\,$ upon which $\,\rho^{\,0}_{_E}\,$ is
 set to depend in the above equations. Also remember that in $\,\ubold(\xbold,\,t)=\ubold(\erbold,\,t)\,+\,O(\ubold^2)\,$ we can neglect $\,O(\ubold^2)\,$, in the linear approximation. Thus the Lagrangian and Eulerian values of the displacement coincide in the first order. Specifically, in the equation (\ref{142}), our $\,\ubold\,$ can be treated as a function of $\,\erbold\,$. So all entities in that equation are functions of the same variable, the perturbed location.

 \vspace{3mm}
 ${\bf{
 \divideontimes
 }}~~~$ {\underline{In the Lagrangian description:}}\vspace{3mm}
 \ba
 \rho_{_L}(\Xbold,\,t)\,=~\rho^{\,0}_{_E}(\Xbold)~+~{\rho_{_L}}'(\Xbold,\,t)\,~,~\quad~\quad
 \label{143}
 \ea
 \ba
 \rho_{_L}\,'\,+\,\rho^{\,0}_{_E}\,\nabla_{\textstyle{_X\,}}\cdot\ubold\,=\,0\,~.~\qquad~~\quad~
 \label{144}
 \ea

 Recall that this is an interrelation between functions of the same variable. This time, it is the initial position
 $\,\Xbold\,$. Had we altered the notation from $\,\Xbold\,$ to $\,\erbold\,$, nothing would have changed (except that we would write $\,\nabla_{\textstyle{_r\,}}\,$
 instead of $\,\nabla_{\textstyle{_X\,}}\,$) --- it would still be the same relation between three functions of the same argument.

 \vspace{3mm}
 ${\bf{
 \divideontimes
 }}~~~$ {\underline{Relation between the increments $\,\rho_{_L}\,'\,$ and $\,\rho_{_E}\,'\,\;$:}}
  \vspace{4mm}~\\
 This relation originates from the general formula (\ref{duga}). In our case the reference trajectory $\,\xbold=\fbold(\Xbold,\,t)\,$ stays identical to the initial position
 $\,\Xbold\,$, so we obtain:
 \ba
 \rho_{_L}\,'(\xbold,\,t)~=~\rho_{_E}\,'(\xbold,\,t)~+~\ubold(\xbold,\,t)\cdot\nabla_{\textstyle{_x\,}}\rho^{\,0}_{_E}(\xbold,\,t)\,~.
 \label{145}
 \ea
 Once again, we are dealing with a relation between several functions taken all at one and the same point. Here the point is denoted with $\,\xbold\,$.  Had we denoted it with $\,\erbold\,$, the only change would be a switch from $\,\nabla_{\textstyle{_x}}\,$ to $\,\nabla_{\textstyle{_r}}\,$,
 no matter what meaning we instill into these $\,\xbold\,$ and $\,\erbold\,$.

 For an initially homogeneous body, $\,\nabla\rho^{\,0}_{_E}\,=\,0\,$; so the forms (\ref{142}) and (\ref{144}) of the linearised conservation law coincide and can both
 be conveniently written as
 \ba
 \rho^{\,0}~\nabla_{\textstyle{_r}}\cdot{\vbold}\,+\,\frac{\textstyle \partial \rho}{\textstyle\partial t\,}\,=\,0\,~,
 \label{146}
 \ea
 where $~\rho^{\,0}\equiv\rho^{\,0}_{_E}~$ and the velocity is
 \ba
 \vbold~=~\frac{\partial \ubold}{\partial t}~~.
 \label{147}
 \ea

 \subsection{Potentials and their increments}

 In each point, the density $\,\rho\,$
 and potential $\,V\,$ comprise a mean value and a perturbation:
 \begin{subequations}
 \ba
 \mbox{density:}\qquad\quad\rho&=&\rho^{\,0}~+~\rho\,'\,~,\,\qquad\qquad
 \label{8a}
 \label{148a}\\
 \nonumber\\
 \mbox{potential:}\qquad\quad V&=&V^{\,0}\,+~V\,'\,=~V^{\,0}\,+~(\,W~+~U\,)\,~,
 \label{8d}
 \label{148b}
 \ea
 \label{8}
 \label{148}
 \end{subequations}
 where $\,V^{\,0}\,$ is the constant-in-time spherically symmetrical potential of an undeformed body, while $\,V\,'\,$ denotes the potential's perturbation. The
 perturbation consists of the external tide-raising potential $\,W\,$ and the resulting additional potential $\,U\,$ of the perturbed body:
 \ba
 V\,'~=~W~+~U\,~.
 \label{9}
 \label{149}
 \ea
 The potentials and densities will be endowed with a subscript {\it{$\,$``$\,L\,$"$\,$}} or {\it{$\,$``$\,E\,$"$\,$}} pointing at the Lagrangian or Eulerian descriptions, accordingly.
 Owing to the general expression (\ref{duga}), we have:
 \ba
 {V_{_E}}'(\erbold,\,t)~=~{V_{_L}}'(\xbold,\,t)~-~\ubold\cdot\nabla_{\textstyle{_x\,}} V^{\,0}_{_E}\,~,
 \label{150}
 \ea
 the same being valid for
 $\,\rho\,$, see the equation (\ref{145}).
 For unperturbed properties, however, subscripts may be dropped without causing any confusion:
 \ba
 V^{\,0}~\equiv~V^{\,0}_{_E}\,\quad,\qquad\rho^{\,0}~\equiv~\rho^{\,0}_{_E}\,~.
 \label{151}
 \ea

 \subsection{The Poisson equation in the Eulerian description}

 In both the perturbed and unperturbed settings, the density and potential are always linked through the Poisson equation:
 \begin{subequations}
 \ba
 \nabla_{\textstyle{_r\,}}^{\,2}\,V_{_E}&=&-~4\,\pi\,G\,\rho_{_E}~~,
 \label{152a}\\
 \nonumber\\
 \nabla_{\textstyle{_r\,}}^{\,2}\,V^{\,0}_{_E}&=&-~4\,\pi\,G\,\rho^{\,0}_{_E}~~,
 \label{152b}
 \ea
 while the perturbing potential $\,W\,$ obeys the Laplace equation outside the perturber:
 \ba
 \nabla_{\textstyle{_r\,}}^{\,2}\,W_{_E}&=&0~~.~\qquad\qquad
 \label{152c}
 \ea
 \label{152}
 \end{subequations}
 Subtraction of (\ref{152b}) from (\ref{152a}) results in a Poisson equation for the density perturbation:
 \ba
 \left.~\quad~\right. \nabla_{\textstyle{_r\,}}^{\,2}\,{V_{_E}}'~=~-~4~\pi~G~{\rho_{_E}}'~~~
 \label{153}
 \ea
 The Poisson equation in the Lagrangian description is presented in Appendix \ref{EL}.

 \section{The power produced by the tidal force}

 \subsection{In the Eulerian description}

 The power $\,P\,$ exerted on the perturbed body is an integral, over its volume, of the rate of working by tidal forces on displacements.
 In the Eulerian language, the power reads as
 \ba
 P\;=\;
 \int\,\rho_{_E}\;{\vbold}\,\cdot\,\nabla_{\textstyle{_r\,}} {V_{_E}}'\;d^3r\,~,
 \label{157}
 \ea
 the integration being performed over an instantaneous, deformed volume. Together with
 \ba
 \rho_{_E}~{\vbold}\cdot\nabla_{\textstyle{_r\,}} {V_{_E}}'~=~
 \nabla_{\textstyle{_r\,}}\cdot(\,\rho_{_E}\,{\vbold}~{V_{_E}}'\,)~-~{V_{_E}}'~\nabla_{\textstyle{_r\,}}\cdot(\,\rho_{_E}\,{\vbold}\,)\,~,~~
 \label{13}
 \label{158}
 \ea
 the mass-conservation law
 \ba
 \nabla_{\textstyle{_r\,}}\cdot(\,\rho_{_E}\,\vbold\,)\,+\,\frac{\textstyle \partial \rho_{_E}}{\textstyle\partial t\,}\,=\,0\,~
 \label{14}
 \label{159}
 \ea
 simplifies the expression under the integral to the following form:
 \bs
 \ba
 \rho_{_E}~{\vbold}\cdot\nabla_{\textstyle{_r\,}} {V_{_E}}'
 ~=~\nabla_{\textstyle{_r\,}}\cdot(\,\rho_{_E}\,{\vbold}\,{V_{_E}}'\,)\,+\,{V_{_E}}'~\frac{\,\partial {\rho_{_E}}'}{\partial t\,}
 \;\;~.
 \label{160a}
 \ea
 Further employment of the Poisson equation in the Eulerian form, (\ref{153}), gives us
 \ba
 \rho_{_E}~{\vbold}\cdot\nabla_{\textstyle{_r\,}} {V_{_E}}'
 ~=~\nabla_{\textstyle{_r\,}}\cdot(\,\rho_{_E}\,{\vbold}\,{V_{_E}}'\,)~-~\frac{1}{4\,\pi\,G}~{V_{_E}}'~\frac{\partial}{\partial t}\,\nabla_{\textstyle{_r\,}}^2 {V_{_E}}'\;\;~.
 \label{160b}
 \ea
 \label{15}
 \label{160}
 \es
 So the power becomes
 \bs
 \ba
 P&=&
 \int\,\nabla_{\textstyle{_r\,}}\cdot(\,\rho_{_E}\,{\vbold}\,{V_{_E}}'\,)\;d^3r
 ~+~
  \int\,{V_{_E}}'~\frac{\,\partial {\rho_{_E}}'}{\partial t\,}~d^3r
 \label{161a}\\
 \nonumber\\
 \nonumber\\
 &=&
 \int_{\Sigma^{t}}\,\rho_{_E}~{V_{_E}}'\;{\vbold}\cdot d\Sbold^t
 ~-~\frac{1}{4\,\pi\,G}~
  \int\,{V_{_E}}'~\frac{\partial}{\partial t\,}\,\nabla_{\textstyle{_r\,}}^2 {V_{_E}}'~d^3r
 \,~,
 \label{161b}
 \ea
 \label{161}
 \es
 where $\,d\Sbold^t\,\equiv~ {\bf{\hat{n}}}^{t}\;d\Sigma^{t}\,$, ~with
 $\,{\bf{\hat{n}}}^{t}\,$ and $\,d\Sigma^{t}\,$  being a unit normal to the deformed surface and an element of area on that surface, both taken at the time $\,t\,$.
 Correct to the first order in the displacement $\,\ubold\,$, these are related to their unperturbed analogues via
 \ba
 {\bf{\hat{n}}}^{t}\,=\;(\,1~-~\nabla^{\Sigma}\otimes\ubold\,)\,{\bf{\hat{n}}}\qquad~\mbox{and}~\qquad d\Sigma^{t}\,=\;\left(\,1~+~\nabla^{\Sigma}\cdot\ubold\,\right)\,d\Sigma~~,
 \label{162}
 \ea
 where the surface gradient is defined as
 \ba
 \nabla^{\Sigma}\,\equiv\,\nabla_{\textstyle{_x\,}}\,-~{\bf{\hat{n}}}~\partial_{{\bf{\hat{n}}}}\,~,
 \label{163}
 \ea
 so $\,\nabla^{\Sigma}\otimes\ubold\,$ is a three-dimensional second-rank tensor (Dahlen \& Tromp 1998). Altogether,
 \bs
 \ba
 d\Sbold^t\,\equiv~ {\bf{\hat{n}}}^{t}\;d\Sigma^{t}\,=\,\left(\,1\,+\,\nabla^{\Sigma}\cdot\ubold\,\right)\,{\bf{\hat{n}}}~d\Sigma~-~(\nabla^\Sigma\otimes\ubold)\,{\bf{\hat{n}}}~d\Sigma~=~
 \left(\,1\,+\,\nabla^{\Sigma}\cdot\ubold\,\right)\,d{\bf S}~-~(\nabla^\Sigma\otimes\ubold)\,d{\bf S}
 \,~,~~\,~
 \label{164a}
 \ea
 with $d{\bf S}\equiv{\bf{\hat{n}}}d\Sigma$ pertaining to the unperturbed surface. In a shorter form, the above reads as
 \ba
 d\Sbold^t\,=~{\mathbb{J}}~d{\bf S}~~,
 \label{164b}
 \ea
 \label{164}
 \es
 where the three-dimensional second-rank tensor
 \ba
 {\mathbb{J}}~\equiv~(\,1\,+\,\nabla^{\Sigma}\cdot\ubold\,)\,{\mathbb{I}}\,-\,\nabla^\Sigma\otimes\ubold
 \label{165}
 \ea
 is, loosely speaking, playing the role of a Jacobian for elements of area. This is fully analogous to the formula
 \ba
 d^3r\,=\,J\,d^3x\,=\,(1\,+\,\nabla_{\textstyle{_x}}\cdot\ubold)\,d^3x\,=\,[1\,+\,\nabla_{\textstyle{_r}}\cdot\ubold\,+\,O(u^2)\,]\,d^3x
 \label{uu}
 \ea
 linking the deformed volume $\,d^3r\,$ to the undeformed volume $\,d^3x\,$. (See Appendix \ref{continuity}.)

 \subsection{In the Lagrangian description}

 Applied to the density, the general formula (\ref{416}) renders:
 \ba
 \rho_{_L}(\Xbold,\,t)\,\equiv\,\rho_{_E}(\erbold,\,t)\,~~.
 \label{445}
 \ea
 This, together with the formula (\ref{150}) for the potential perturbation, enables us to express the power in the Lagrangian description:
 \ba
 P\;=\;
 \int\,\rho_{_L}\;{\vbold}\,\cdot\,\nabla_{\textstyle{_x\,}}{V_{_L}}'\;d^3x~-~\int\,\rho_{_L}\;{\vbold}\,\cdot\,\nabla_{\textstyle{_x\,}}(\,\ubold\cdot
 \nabla_{\textstyle{_x\,}} V^0\,)\;d^3x\,~,
 \label{166}
 \label{446}
 \ea
 the integral now being taken over the undeformed body. Be mindful that $\,d^3 r\,\nabla_{\textstyle{_r\,}}\,=\,d^3x\,\nabla_{\textstyle{_x\,}}\,$, so no Jacobian shows up
 on the right-hand side.

 The velocity and displacement being in quadrature, the second term should be dropped after time averaging (denoted with angular brackets):
 \bs
 \ba
 \langle P\rangle\;=\;
 \int\,\rho_{_L}\;{\vbold}\,\cdot\,\nabla_{\textstyle{_x\,}}{V_{_L}}'\;d^3x\,~,
 \label{167a}
 \label{447a}
 \ea
 For a periodically deformed solid, we set the equilibrium state to play the role of the unperturbed configuration, for which reason
 \footnote{~The mass is conserved along both trajectories, perturbed and unperturbed. So both $\,\rho_{_E}(\erbold,\,t)\,d^3\erbold\,$ and $\,\rho^{\,0}_{_E}(\xbold,\,t)\,d^3\xbold\,$ must be equal to the initial mass $\,\rho^{\,0}_{_E}(\Xbold)\,d^3\Xbold\,$, and therefore to one another:
  $\,\rho_{_E}(\erbold,\,t)\,d^3\erbold\,=\,\rho^{\,0}_{_E}(\xbold,\,t)\,d^3\xbold\,$. Thence,
  $\,\rho_{_E}(\rbold\,,~t)~J~=~\rho^{\,0}_{_E}(\Xbold)\,$, where $\,J\equiv d^3\erbold/d^3\xbold\,$. In combination with (\ref{445}), this yields:
 \ba
 \rho_{_L}(\Xbold\,,~t)~J~=~\rho^{\,0}_{_E}(\xbold\,,~t)~~.
 \nonumber
 \ea
 When the unperturbed configuration is the equilibrium state, $\,\xbold=\fbold(\Xbold,\,t)\,$ coincides with $\,\Xbold\,$ at all times. So $\,\rho^{\,0}_{_E}(\xbold,\,t)\,$ bears no dependence on time, and the above equality becomes simply $\,\rho_{_L}(\Xbold\,,~t)~J~=~\rho^{\,0}_{_E}(\xbold)\,$.
  See Appendix \ref{continuity} for a detailed discussion.
 \label{7}}
 $\,~\rho_{_L}(\Xbold\,,~t)~J~=~\rho^{\,0}_{_E}(\xbold)\,$. Insertion of this equality into the expression (\ref{167a}) gives us:
 \ba
 \langle P\rangle\;=\;
 \int\,\rho^{\,0}\;\,{\vbold}\,\cdot\,\nabla_{\textstyle{_x\,}}{V_{_L}}'\;J^{-1}\,d^3x\,~.
 \label{167b}
 \label{447b}
 \ea
 \label{167}
 \label{447}
 \es
 The dot-product can be easily rearranged via the formulae analogous to (\ref{158} - \ref{160}). Due to
 \ba
 \rho^{\,0}~{\vbold}\cdot\nabla_{\textstyle{_x\,}} {V_{_L}}' ~=~
 \nabla_{\textstyle{_x\,}}\cdot(\,\rho^{\,0}\,{\vbold}\,{V_{_L}}'\,)\,-\,{V_{_L}}'\,\nabla_{\textstyle{_x\,}}\cdot(\,{\vbold}\,\rho^{\,0}\,)
 \label{168}
 \label{448}
 \ea
 and
 \ba
 \rho^{\,0}~\nabla_{\textstyle{_x\,}}\cdot{\vbold}\,+\,\frac{\textstyle \partial \rho}{\textstyle\partial t\,}\,=\,0\,~,
 \label{169}
 \label{449}
 \ea
 the expression under the integral becomes
 \ba
 \rho^{\,0}~{\vbold}\cdot\nabla_{\textstyle{_x\,}} {V_{_L}}'
 ~=~\nabla_{\textstyle{_x\,}}\cdot(\,\rho^{\,0}\,{\vbold}\,{V_{_L}}'\,)\,+\,{V_{_L}}'~\frac{\,\partial \rho_{_L}\,'}{\partial t\,}
 \;\;~,
 \label{170}
 \label{450}
 \ea
 provided we set $\,\nabla_{\textstyle{_x\,}}\rho^{\,0}\,=0\,$, i.e., provided we assume that the unperturbed body is homogeneous.$\,$\footnote{~No such assumption was required to obtain the Eulerian analogue (\ref{159}) of the Lagrangian formula (\ref{169}).} Then the time-averaged power, for an initially homogeneous body, acquires the form of
 \ba
 \langle P\rangle\;=\;
 \int\,\nabla_{\textstyle{_x\,}}\cdot(\,\rho^{\,0}\,{\vbold}\,{V_{_L}}'\,)\;d^3x
 ~+~
 \int\,{V_{_L}}'~\frac{\,\partial \rho_{_L}\,'}{\partial t\,}\;d^3x
 \,~,
 \label{171}
 \ea
 where we approximated the Jacobian with unity, thus neglecting higher-order terms.

 \section{Tidal dissipation rate in a homogeneous sphere}\label{sphere}

 Although the Eulerian and Lagrangian descriptions are equivalent,
 the boundary conditions look simpler in the Eulerian picture. On the other hand, for periodic deformations, practical calculations are easier carried out in the
 Lagrangian description, as it implies integrations over the unperturbed volume and surface corresponding to the equilibrium shape. It is, unfortunately, not unusual for the authors to refrain from pointing out which description is employed, leaving this to the discernment of the readers. The easiest way to trace an author's choice is to look at the way they write the expression for the power and the Poisson equation.

 The often-cited authors Zschau (1978) and Platzman (1984) started in the Eulerian language and then switched to the Lagrangian description. This can be seen from the fact that the time-average power was eventually written by both of them as an integral over the $\,${\it{undeformed}}$\,$ body. Both works contained some mathematical omissions which, fortunately, did not influence the final form of the integral.

 Below we present these authors' method in a more mathematically complete manner. While our expression for the power, written as an integral over the unperturbed surface, will coincide with the integrals derived by the said authors, our final result (the power written as a spectral sum over the Fourier modes) will differ. In one important detail, our result also differs from that by Peale \& Cassen (1978).

 \subsection{A mixed, Eulerian-Lagrangian treatment}\label{zp}\label{p}\label{z}

 Similar to Zschau (1978, eqn. 2), we begin with the formula (\ref{157}) for the power in the Eulerian variables. The next natural step is (\ref{161}), whereafter integration by parts renders:
 \begin{subequations}
 \ba
 P~=\int\rho_{_E}\,{V_{_E}}'\;{\vbold}\cdot d{\bf{S}}^t\,-~\frac{1}{4\pi G}\,\int d^3r~ \nabla_{\textstyle{_r}}\cdot\left({V_{_E}}'~\,\frac{\partial \nabla_{\textstyle{_r\,}} {V_{_E}}'}{\partial t}\,\right)
 ~+~\frac{1}{4\pi G}\,\int d^3r~\frac{\partial \nabla_{\textstyle{_r\,}} {V_{_E}}'}{\partial t}\,\cdot\,\nabla_{\textstyle{_r\,}} {V_{_E}}'~~~~~~~
 \label{172a}\\
 \nonumber\\
 \nonumber\\
 =\int\rho_{_L}\,{V_{_E}}'\,{\vbold}\cdot\left(\,{\mathbb{J}}~d{\bf{S}}\,\right)\,-~\frac{1}{4\pi G}\,\int {V_{_E}}'~\frac{\partial \nabla_{\textstyle{_r\,}} {V_{_E}}'}{\partial t}\,\cdot\left(\,{\mathbb{J}}\,d{\bf{S}}\,\right)\,+~\frac{1}{8\pi G}~\frac{\partial}{\partial t}\int (J\,d^3x)~\nabla_{\textstyle{_r\,}} {V_{_E}}'\cdot\nabla_{\textstyle{_r\,}} {V_{_E}}'\,~.~~~~~
 \label{172b}
 \ea
 \label{172}
 \end{subequations}
 {\it{En route}} from the former expression to the latter, we switch from $\,d{\bf{S}}^t\,$ and $\,d^3r\,$ to $\,{\mathbb{J}}~d{\bf{S}}\,$ and $\,J\,d^3x\,$, respectively. Thereby we switch from integration over a deformed body to that over the undeformed one. So $\,\rho_{_E}\,$ becomes $\,\rho_{_L}\,$,
 see the equation (\ref{445}). A similar switch from $\,{V_{_E}}'\,$ to $\,{V_{_L}}'\,$ can be performed using the equation (\ref{150}), but we prefer to stick to $\,{V_{_E}}'\,$ for some time, for it will be easier to impose the boundary conditions on the Eulerian potential.

 In a leading-order calculation, both the Jacobian and its tensorial analogue may be set unity: $\;{\mathbb{J}}\approx{\mathbb{I}}\;$ and $\;J\approx 1~$,
 $\,$as evident from the formulae (\ref{165}) and (\ref{uu}). In the same order, we can substitute $\,\nabla_{\textstyle{_r\,}}\,$ with $\,\nabla_{\textstyle{_x\,}}\,$. In addition, as was explained in Footnote \ref{7}, we can substitute $\,\rho_{_L}\,=\,\rho^{\,0}/J\,$ with $\,\rho^{\,0}\,$, and can treat the latter as time-independent. Thus the time average of the power becomes:
 \begin{subequations}
 \ba
 \langle P\rangle &=&\int\,\left\langle\,\rho^{\,0}~{V_{_E}}'\;{\vbold}\,\right\rangle\,\cdot\,d{\bf{S}}~-~\frac{1}{4\pi G}~\int ~\left\langle\,{V_{_E}}'~\,\frac{\partial \,\nabla_{\textstyle{_x}} {V_{_E}}'}{\partial t}\,\right\rangle\,\cdot\,d{\bf{S}}
 \label{19a}
 \label{173a}
 \ea
 \ba
 &=&-~\frac{1}{4\pi G}~\int ~\left\langle~{V_{_E}}'~\,\frac{\partial}{\partial t}\,\left(\,\nabla_{\textstyle{_x\,}} {V_{_E}}'\,-\,4\,\pi\,G\,\rho^{\,0}\,\ubold
 \,\right)\;\right\rangle\,\cdot\,d{\bf{S}}\,~.
 \label{19b}
 \label{173b}
 \ea
 with the volume integral dropped.$\,$\footnote{~As previously agreed, in our approximation the Jacobian is set unity. The potential variation $\,{V_{_E}}'\,$ is a sum of sinusoidal harmonics, and so is its gradient $\,\nabla_{\textstyle{_x\,}} {V_{_E}}'\,$. After time averaging of (\ref{172b}), the cross terms in the product $\,\nabla_{\textstyle{_x\,}} {V_{_E}}'\,\cdot\,\nabla_{\textstyle{_x\,}} {V_{_E}}'\,$ will vanish, while the products of harmonics of the same frequency will render constants.} The potential $\,{V_{_E}}'\,$ in the above developments was the $\,${\it{interior}}$\,$ potential,
 so the above formula should, rigorously speaking, have been written as
  \ba
 \langle P\rangle~=~-~\frac{1}{4\pi G}~\int ~\left\langle~{{V_{_E}}'}^{\textstyle{^{\,(interior)}}}~\frac{\partial}{\partial t}\,\left(\,\nabla_{\textstyle{_x\,}} {{V_{_E}}'}^{\textstyle{^{\,(interior)}}}\,-\,4\,\pi\,G\,\rho^{\,0}\,\ubold
 \,\right)\;\right\rangle\,\cdot\,d{\bf{S}}\,~.
 \label{19c}
 \label{173c}
 \ea
 \label{19}
 \label{173}
 \end{subequations}

   The expression (\ref{19c}) is somewhat formal. On the one hand, it contains integration over an undeformed surface, an operation appropriate to the Lagrangian description. On the other hand, the quantity under the integral is Eulerian, i.e., is a function of the perturbed positions. Thus, to employ the expression (\ref{173c}) in practical calculations, one would first have to express the integrated average product $~\left\langle~{{V_{_E}}'}^{\textstyle{^{\,(interior)}}}~\frac{\textstyle\partial}{\textstyle\partial t}\,\left(\,\nabla_{\textstyle{_x\,}} {{V_{_E}}'}^{\textstyle{^{\,(interior)}}}\,-\,4\,\pi\,G\,\rho^{\,0}\,\ubold
 \,\right)\;\right\rangle~$ as a function of the unperturbed positions, i.e., of the coordinates on the undeformed surface. Simply speaking, one would have to switch from a Eulerian function under the integral to a Lagrangian function, using the formula (\ref{150}). The reason for our procrastination with this step is the convenience of the Eulerian description for imposing boundary conditions.

 \subsection{Comparing the intermediate result (\ref{173c}) with analogous\\ formulae from Zschau (1978) and Platzman (1984)}

 Our expression (\ref{173c}) is equivalent to the formula (12) in Zschau (1978). The sole difference is how we justify the substitution of the Lagrangian density $\,\rho_{_L}\,$ with the unperturbed $\,\rho^{\,0}\,$. Whereas we approximated the Jacobian with $\,1+O(|\ubold |)\,$, Zschau (1978, eqn. 10) employed a clever trick that did not rely on the smallness of disturbance. In our notation, the trick looks like this: if in the first term of our expression (\ref{173a}) we also keep the first-order perturbation $\,{\rho_{_L}}'\,$ of the density, the time average of the product $\,{\rho_{_L}}'\,\vbold\,{V_{_E}}'\,\,$ will always be zero, provided all three oscillate at the same frequency. While elegant, Zschau's argument works only for a perturbation at one frequency, not for a spectrum of frequencies.

 The treatment by Platzman (1984) contains more inaccuracies. The author's formula (2) looks like our equation (\ref{167b}), with the actual density substituted from the beginning by its unperturbed value $\,\rho^{\,0}\,$. Such a start indicates the use of the Lagrangian description. This however comes into contradiction with the way the author writes down the conservation law. Platzman's form of that law is equivalent to our equation (\ref{142}), i.e., is  written in the Eulerian language. The following Poisson equation, too, is Eulerian. That the author eventually arrives at the right integral expression (equation 5 in {\it{Ibid.}}) is more due to luck than to accuracy. In the subsequent derivation, the author's formulae (7) and (10) are incorrect, because the fact that the Fourier modes in the Darwin-Kaula theory can be of either sign is neglected. We shall address this point at the end of Section \ref{foll}.

 \subsection{Employment of the boundary conditions}

 The Eulerian boundary conditions mimic those from electrostatics (see Appendix \ref{appA}):
 \ba
 \label{174}
 \label{20}
 {{V_{_E}}'}^{\,\textstyle{^{(interior)}}}~=~{{V_{_E}}'}^{\,\textstyle{^{(exterior)}}}~~
 \ea
 and
 \ba
 \left[~\frac{\partial~}{\partial\hat{\bf{n}}}\, {V_{_E}}'~-~4~\pi~G~\rho^{\,0}~{\bf u}~\right]^{\,\textstyle{^{(exterior)}}}\,=~
 \left[~\frac{\partial~}{\partial\hat{\bf{n}}}\, {V_{_E}}'~-~4~\pi~G~\rho^{\,0}~{\bf u}~\right]^{\,\textstyle{^{(interior)}}}\,~.
 \label{21}
 \label{175}
 \ea
 Insertion thereof into the equation (\ref{173c}) makes the power look
 \ba
 \langle P\rangle~=~-~\frac{1}{4\pi G}~\int ~\left\langle~{{V_{_E}}'}^{\textstyle{^{\,(exterior)}}}~\frac{\partial}{\partial t}~\nabla_{\textstyle{_x\,}} {{V_{_E}}'}^{\textstyle{^{\,(exterior)}}}\;\right\rangle\,\cdot\,d{\bf{S}}\,~.
 \label{176}
 \ea
 It is now high time to write the expression under the integral (\ref{176}) as a function of the coordinates on the unperturbed surface, the one over which we integrate. The formula (\ref{150}) prescribes us to substitute $\,{{V_{_E}}'}\,$ with $\,{V_{_L}}'-\,\ubold\cdot\nabla_{\textstyle{_x\,}} V_0\,$. As $\,\ubold\,$ is zero outside the body, we get~\footnote{~For the first multiplier under the integral (\ref{176}), we simply substitute $\,{{V_{_E}}'}^{\textstyle{^{\,(exterior)}}}\,$ with $\,{{V_{_L}}'}^{\textstyle{^{\,(exterior)}}}\,$, omitting the term $\,\left[\,-\,\ubold\cdot\nabla_{\textstyle{_x\,}} V_0\,\right]^{\textstyle{^{\,(exterior)}}}\,$ because $\,\ubold\,$ is zero outside the body.\\
 $\left.~~~~\right.$ The case of the second multiplier, $\,\frac{\textstyle\partial}{\textstyle\partial t}~\nabla_{\textstyle{_x\,}} {{V_{_E}}'}^{\textstyle{^{\,(exterior)}}}\,$, is less obvious. Employment of the formula (\ref{150}) furnishes us $\,\frac{\textstyle\partial}{\textstyle\partial t}\,\left[\,\nabla_{\textstyle{_x\,}}{V_{_L}}'\,-\,\nabla_{\textstyle{_x}}\cdot(\ubold\, V^{\,0})\,\right]^{\textstyle{^{\,(exterior)}}}$.
 The vanishing of $\,\ubold\,$ on the exterior side of the boundary does not imply the vanishing of its gradient there. On the contrary, $\,\nabla_{\textstyle{_x}}\cdot(\ubold\, V^{\,0})\,$ performs a finite step -- but so also does the gradient of $\,{V_{_L}}'\,$, so that altogether the gradient $\,{V_{_E}}'\,$ remains continuous. To sidestep these intricacies, we can expand the volume of integration slightly outward from the actual volume of the planet (Platzman 1984, p. 74).}
 \ba
 \langle P\rangle~=~-~\frac{1}{4\pi G}~\int ~\left\langle~{{V_{_L}}'}^{\textstyle{^{\,(exterior)}}}~\frac{\partial}{\partial t}~\nabla_{\textstyle{_x\,}} {{V_{_L}}'}^{\textstyle{^{\,(exterior)}}}\;\right\rangle\,\cdot\,d{\bf{S}}\,~,
 \label{177}
 \label{22}
 \ea
 To analyse the behaviour of $\,V\,'\,$ outside the perturbed body, recall that its two components, $\,U\,$ and $\,W\,$, scale differently with the planetocentric radius. As can be seen from (\ref{1a}), the degree-$l\,$ Legendre component of the perturbing potential changes as \footnote{~Do not be misled by the planetocentric distance in (\ref{1a}) being denoted with $\,R\,$. There we needed the value of $\,W\,$ on the surface, whereas here we need to know $\,W\,$ at an arbitrary planetocentric distance.} ~$~W_l\,\propto\,\,r^{\,l}\,$.  According to (\ref{505a}), the degree-$l\,$ component of the tidal potential obeys $~U_l\,\propto\,r^{-(l+1)}\,$. All in all, the $\,l-$degree part of the exterior $\,V\,'\,$ assumes the form of
 \ba
 {{V_{_L}}'}^{\,\textstyle{^{(exterior)}}}=~\sum_{l=2}^{\infty}\,\left[\,\left(\,\frac{r}{R}\,\right)^{\textstyle{^{l}}}\,W_l(R)\,+\,\left(\,\frac{r}{R}\,
 \right)^{\textstyle{^{-(l+1)}}}\,U_l(R)\,\right]\,~,
 \label{23}
 \label{178}
 \ea
 while the normal part of its gradient on the free surface is
 \ba
 \frac{\partial\,}{\partial r}~{{V_{_L}}'}_{\textstyle{_l}}^{\,\textstyle{^{(exterior)}}}~=~R^{-1}\,\sum_{l=2}^{\infty}\,\left[\,l~W_l\,-\,(l+1)~U_l\,\right]\,~.
 \label{24}
 \label{179}
 \ea
 Plugging it into (\ref{22}), and benefitting from the orthogonality of surface harmonics, we obtain:~\footnote{~On the boundary, we have: $\,{{V_{_L}}'}(R)\,=\,\sum_{l=2}^{\infty}\,\left[\,W_l(R)\,+\,U_l(R)\,\right]~$, as evident from (\ref{23}). Together with (\ref{24}), this expression was inserted in (\ref{22}). By doing so, we omitted the diagonal products $\,W_l\,\dot{W}_l\,$ and $\,U_l\,\dot{U}_l\,$ that vanish after time averaging. (Indeed, $\,W_l\,$ is in quadrature with $\,\dot{W}_l\,$, while $\,U_l\,$ is in quadrature with $\,\dot{U}_l\,$.) {\it{En route}} from
 (\ref{25a}) to (\ref{25b}), we took into account that the time averages of $\,\partial (U_l\,W_l)/\partial t\,$ also vanish.}
 \begin{subequations}
 \ba
 \langle P\rangle &=&-~\frac{1}{4\pi G R}~\sum_{l=2}^{\infty}\,\int~\left\langle~l~U_l\,\stackrel{\bf\centerdot}{W}_l\,-\,(l+1)~W_l\,\stackrel{\bf\centerdot}{U}_l~
 \right\rangle~dS
 %
 %
 %
 %
 %
 %
 \label{25a}
 \label{180a}
  \ea \ba
 &=&\frac{1}{4\pi G R}~\sum_{l=2}^{\infty}\,(2\,l\,+\,1)\,\int~\left\langle~W_l~\stackrel{\bf\centerdot}{U}_l~\right\rangle~dS
  %
  %
  %
  %
  %
  %
  %
  %
  %
  \,~,
 \label{25b}
 \label{180b}
 \ea
  %
  %
  %
  %
  %
  %
  %
  %
  %
  %
  %
  %
  %
  %
 \label{25}
 \label{180}
 \end{subequations}
 which is equivalent to the formulae (18) in Zschau (1978) and (5) in Platzman (1984). This, however, is the last point on which we are still in agreement with our predecessors.

 \subsection{Writing the integral as a spectral sum}\label{foll}

 Bringing in the dynamical Love numbers $\,k_l\,$
 and
 the phase lags defined in (\ref{505b}), one can express the products $\,W_l(t)\,\dot{U}_l(t)\,$
  via the spectral components of the disturbance $\,W(t)~$.~\footnote{~In this subsection, $\,\omega\,$ is a shortened notation for the mode $\,\omega_{lmpq}\,$, not the argument of the pericentre.}

 Although the formula
 \ba
 \bcancel{
 \sum_{l=2}^{\infty}(2l+1)\,\left\langle W_l(t)\,\dot{U}_l(t)\right\rangle ~=~
 \sum_{\omega}\,(2l+1)\,\frac{\omega}{2}\,W^{\,2}_l(\omega)\,k_l(\omega)~\sin\epsilon_l(\omega)
 }
 ~
 \label{26}
 \ea
 is often used in the literature (Zchau 1978, Platzman 1984, Segatz et al. 1988),
 $\,$\footnote{~Our expression (\ref{26}) is identical to the upper line of the equation (10) in Platzman (1984).
   (Note a misprint on that line of Platzman's equation: a missing factor of $\,\omega\,$.)\\
   $~\quad~~$ Our formula (\ref{26}), when truncated to $\,l=2\,$, also becomes equivalent to the equation (22) in Zschau (1978) and to the equation (12) in Segatz et al. (1988). ~(Both authors kept only the degree-2 terms.)}$\,$
 accurate examination demonstrates that it is incorrect. To appreciate this, one simply has to insert the expansions (\ref{1b}) and (\ref{505b}) into the formula (\ref{25b}) and see what happens.

 That the answer differs from (\ref{26}) was noticed by Peale \& Cassen (1978). However, their development also needs correction. Below, we dwell upon this matter in great detail and provide a full inventory of the terms emerging in the spectral expansion for damping rate. At this point, we only mention the two key circumstances:

 \begin{itemize}

 \item[\bf(a)~] The conventional expression (\ref{26}) ignores the degeneracy of modes, i.e., a situation where several modes $\,\omega_{\textstyle{_{lmpq}}}\,$
 with different sets $\,lmpq\,$ take the same numerical value $\,\omega\,$. As will be demonstrated in the Section \ref{accurate}, the sum over modes $\,\omega\,$ in (\ref{26}) should be substituted with a sum over $\,${\it{distinct$\,$ values}}$\,$ of the modes:
 \ba
 \nonumber
 \mbox{instead~of}~~ \sum_\omega W_l^2(\omega)~~\mbox{in~(\ref{26})$\,,\,$~use~this}~:~~ \sum_\omega\left[\,\sum_{{\omega_{\textstyle{_{_{lmpq}}}}=~\omega}} W_{l}(\omega_{lmpq})\,\right]^2 \ea
 \ba
 \nonumber\\
 \mbox{where}~\sum_{{\omega_{\textstyle{_{_{lmpq}}}}=~\omega}} W_{l}(\omega_{lmpq})~~\mbox{denotes~a~sum~of~all~terms~for~which~}\,\omega_{lmpq}\,~\mbox{takes~a~value~}\,\omega~.\qquad\qquad\qquad\qquad\qquad
 \nonumber
 \ea
 In short: first sum all the terms corresponding to one value of $\,\omega~$, $\,$then square the sum, and only thereafter sum over all the values of $\,\omega\,$.\\

 \item[\bf(b)~] Much less intuitive is the fact that the spectral sum will contain extra terms missing completely in the expression (\ref{26}). As we shall see in Appendix \ref{sketch}, these terms look (up to some caveat) as $\,W_l(\omega)\,W_l(-\omega)\,$. They show up because two modes of  opposite values, $\,\omega\,$ and $~-\omega\,$, correspond to the same physical frequency $\,|\,\omega\,|\,$.\\
 \end{itemize}
 For the time being, we use the notation $\,\sum^{\,\textstyle\sharp}~$:
  \ba
 \sum_{l=2}^{\infty}(2l+1)\,\left\langle W_l(t)\,\dot{U}_l(t)\right\rangle ~=~
 {\sum_{\omega}}^{\textstyle{\sharp}~}(2l+1)\,\frac{\omega}{2}\,W^{\,2}_l(\omega)\,k_l(\omega)~\sin\epsilon_l(\omega)\,~,
 \label{27}
 \ea
 where the superscript $\,^{\textstyle{\sharp}~}\,$ reminds the reader that the spectral sum needs to be amended down the road.

 %
 %
 %
 Insertion of (\ref{27})
 %
 %
 %
 %
 %
 %
 into (\ref{25b}) results in:$\,$\footnote{~Were we using complex potentials, we would
 have $\,W_l\,\dot{U}_l^*\,$ instead of $\,W_l\,\dot{U}_l\,$ in (\ref{25b}), and would have $\,W_l\,W_l^*\,$ instead of $\,W_l\,W_l\,$ in (\ref{27}).}
 \ba
  \langle\,P\,\rangle~=~\frac{1}{8\pi G R}~{\sum_{\omega}}^{\textstyle{\sharp}~}\,(2\,l\,+\,1)~\omega\,k_l(\omega)~\sin\epsilon_l(\omega)\int~W^{{\,2}}_l(\omega)~dS
 %
 %
 %
 %
 %
 %
 %
 %
 %
  \,~ ~.~~
 \label{30a}
 \label{30b}
 \label{30}
 \label{183}
 \ea
 If not for the superscript $\,^{\textstyle{\sharp}~}\,$, this expression would coincide with the results by Zschau (1978) and Platzman (1984).$\,$\footnote{~Our expression (\ref{30}) should be compared to the equation (22) from Zschau (1978), in understanding that our expression furnishes the mean damping rate summed over the entire spectrum, whereas Zschau's formula renders the energy loss over a period, at a certain frequency. With these details taken into account, the formulae are equivalent. They are also equivalent to the formulae (10) and ({12}) in Segatz et al. (1988) and  (10) in Platzman (1984). Note, however, that in the first line of Platzman's formula a factor of $\,\omega\,$ is missing. \label{foot}}
 The superscript reminds us of the important caveat in the evaluation of the sum: the factors $\,W^{\,2}_l(\omega)\,$ should be substituted with more complicated expressions, whereas the sum should be carried not over all modes $\,\omega=\omega_{\textstyle{_{lmpq}}}\,$, but over all $\,${\it{distinct$\,$ values$\,$}} of $\,\omega\,$, see Section \ref{next}.


 \section{Heat production over tidal modes}\label{next}\label{accurate}

 We must insert the expansions (\ref{1b}) and (\ref{505b}) into the formula (\ref{25b}) for the heating rate, in order to obtain a comprehensive version of the
 somewhat symbolic sum (\ref{27}) and to see what the modified sum $\,\sum^{\textstyle{^{\,\sharp}}}\,$ actually means. A sketchy version of this calculation (which takes into account that the modes may have either sign, but neglects the degeneracy of modes) is given in Appendix \ref{sketch}. Extraordinarily laborious, the full calculation is presented in Appendix \ref{appD}. Here we provide the final results.

 In the case of a $\,${\it{uniformly}}$\,$ moving pericentre, the average dissipation rate is:
  \ba
  \nonumber
  \langle\,P\,\rangle~=~
  \ea
  \ba
    \frac{G\,{M^*}^{\,2}}{a  }\sum_{l=2}^{\infty}\left(\frac{R\,}{a}\right)^{\textstyle{^{2l+1}}}\sum_{m=0}^{l}
    \frac{(l - m)!}{({\it l} + m)!}
  \left(2-\delta_{0m}\right)
  \sum_{p=0}^{l}F^{\,2}_{lmp}(i)
  \sum_{q\,=-\infty}^{\infty}G^{\,2}_{lpq}(e)
  \,\chi_{\textstyle{_{lmpq}}}\,
  k_l(\chi_{\textstyle{_{lmpq}}})~\sin\epsilon_l(\chi_{\textstyle{_{lmpq}}})\,~,~~~
  \label{196}
  \ea
 where the physical frequencies are the absolute values of the Fourier modes:
 \ba
 \chi_{\textstyle{_{lmpq}}}\,=~|\,\omega_{\textstyle{_{lmpq}}}\,|~=~|\,
 (l-2p)\;\dot{\omega}\,+\,(l-2p+q)\;{\bf{\dot{\cal{M}}}}\,+\,m\;(\dot{\Omega}\,-\,\dot{\theta})
   \,|~\approx~|\,(l-2p+q)\;n~-~m~\dot{\theta}\,|\,~,~~~
 \label{freq}
 \ea
 and $\,\sin\epsilon_l(\chi_{\textstyle{_{lmpq}}})\,$ is what they often call $\,1/Q_{\textstyle{_{l}}}\,$ in the literature.$\,$\footnote{~It would not hurt to reiterate that the Fourier modes $\,\omega_{\textstyle{_{lmpq}}}\,$ can be of either sign, while the physical forcing frequencies (\ref{freq}) are positive definite. Obviously,
 $~\chi_{\textstyle{_{lmpq}}}\,k_l(\chi_{\textstyle{_{lmpq}}})~\sin\epsilon_l(\chi_{\textstyle{_{lmpq}}})\,=\,\omega_{\textstyle{_{lmpq}}}\,
  k_l(\omega_{\textstyle{_{lmpq}}})~\sin\epsilon_l(\omega_{\textstyle{_{lmpq}}})\,$, because the dynamical Love numbers are even functions, whereas the phase lags are odd and  of the same sign as their argument. This is why the tidal quality factors may be expressed as $\,1/Q_{\textstyle{_{l}}}\,=\,\sin\epsilon_l(\chi_{\textstyle{_{lmpq}}})\,$ and also as $\,1/Q_{\textstyle{_{l}}}\,=\,|\,\sin\epsilon_l(\omega_{\textstyle{_{lmpq}}})\,|\,$, with the absolute value symbols being redundant in the former formula and needed in the latter.}

 In the Appendix \ref{appD}, we also derive a formula for an idle pericentre; but the applicability realm of that formula is limited.$\,$\footnote{~For an idle pericentre, the time-averaged tidal-heating power reads as:
  \ba
  \nonumber
  \langle\,P\,\rangle~=~
  \frac{G\,{M^*}^{\,2}}{a}\,
  \sum_{l=2}^{\infty}\,\left(\frac{R}{a}\right)^{\textstyle{^{2\,l\,+\,1}}}
  \sum_{m=0}^{l}~
  \frac{(l - m)!}{({\it l} + m)!}\;
  \left(\,2 -\delta_{0m}\,\right)
  \,\sum_{p=0}^{l}F_{lmp}(i)\;\sum_{p\,'=0}^{l}F_{lmp\,'}(i)    \qquad\,\qquad\qquad\qquad\qquad\qquad\qquad
  ~\\
  \nonumber\\
  \left.\qquad\qquad\qquad\right.
  \sum_{q\,=-\infty}^{\infty}  G_{lpq}(e)~  \left[\,G_{lp\,'q\,'}(e)\,\right]_{\textstyle{_{q\,'\,=\,q\,-\,2\,(p-p\,')}}}~\cos\left(\,2\,(p\,'\,-\,p)\,\omega_0\,\right)
  \,~\omega_{\textstyle{_{lmpq}}}\,
  \,k_l(\omega_{\textstyle{_{lmpq}}})~\sin\epsilon_l(\omega_{\textstyle{_{lmpq}}})~
  \,~,~~\qquad
  \nonumber
  \ea
  $\omega_0\,$ being the value of the pericentre. This formula is of a limited practical value, since $\,\omega_0\,$ seldom stays idle. For example, if we are computing tidal damping in a planet perturbed by the star, $\,\omega_0\,$ of the star as seen from the planet will be evolving due to the equinoctial precession of the planet equator.
  }

  Our formula (\ref{196}) differs from the appropriate expression in Kaula (1964, Eqn 28) that contains a redundant factor $\,(1+k_{\,l})/2\,$.

 \noindent
 In the special situation where
 \begin{itemize}
 \item[(a)~] $~l=2\,$,
 \item[(b)~] the body is incompressible, $\,$so $\,k_2\,=\,3\,h_2/5\,$, $\,$\footnote{~Static Love numbers of an incompressible spherical planet satisfy the relation $\;(2l+1)\,k_l\,=\,3\,h_l\;$. As explained in Appendix \ref{love}, an analogue of this equality for dynamical Love numbers is $\;(2l+1)\,k_l(\omega_{\textstyle{_{lmpq}}})\,=3\,h_l(\omega_{\textstyle{_{lmpq}}})\,$.}
 \item[(c)~] the spin is synchronous, with no libration,
 \end{itemize}
 the expression (\ref{196}) agrees with the formula (31) from Peale \& Cassen (1978). The comparison is carried out in Appendix \ref{appD}.$\,$\footnote{~In our expression, all terms are positive-definite, because the factors $\,\omega_{\textstyle{_{2mpq}}}\,k_2(\omega_{\textstyle{_{2mpq}}})\,\sin\epsilon_2(\omega_{\textstyle{_{2mpq}}})\,$ are even functions of the tidal mode $\,\omega_{\textstyle{_{2mpq}}}\,$. Peale \& Cassen (1978) have their terms proportional to the products
 $~\,\frac{\textstyle 3}{\textstyle 5}\,h_2\,\frac{\textstyle{2-2p+q-m}}{\textstyle{Q_{2mpq}}}\,$. These terms, too, are positive definite, because Peale \& Cassen (1978) are using a nonstandard convention $~1/{\textstyle{Q_{2mpq}}}\,\equiv\,\sin\epsilon_2(\omega_{\textstyle{_{2mpq}}})~$. Within this convention, the quality factors $\,{\textstyle{Q_{2mpq}}}\,$ are $\,${\it{not}}$\,$ positive definite.}

 \section{Conclusions}

 We have derived from the first principles a formula for the tidal dissipation rate in a homogeneous spherical body. $\,${\it{En$\,$ route}}$\,$ to that formula, we compared our intermediate results with those by Zschau (1978) and Platzman (1984). When restricted to the special case of an incompressible spherical planet spinning synchronously without libration, our final formula coincides with the commonly used expression from Peale \& Cassen (1978, Eqn. 31). 
 

 We propose to use our equation (\ref{196}) for rocky planets and moons, instead of the classic formula from Peale \& Cassen (1978, eqn 31), because 
 it correctly captures the frequency dependence of tidal dissipation for objects outside the 1:1 resonance.

 Several applications are provided in the work by Makarov and Efroimsky (2014).

 \section*{Acknowledgments}

 M.E. is indebted to Jeroen Tromp and Mikael Beuthe for pointing out the advantage of the Lagrange description, and to Gabriel Tobie for a very useful exchange on tidal heating.

 The authors are grateful to James G. Williams for meticulous reading of the manuscript and very judicious comments that were of great help.

 The authors' special thanks go to the referee, Patrick A. Taylor, whose thoughtful and comprehensive report enabled the authors to improve the quality of the paper significantly.

 This research has made use of NASA's Astrophysics Data System.\\

\pagebreak

 \appendix
 \begin{center}
  {\underline{\Large{\bf{Appendix}}}}
 \end{center}

 \section{The associated Legendre functions and their normalisation}\label{appB}

 The Legendre polynomials are usually defined by the Rodriguez formula:
 \ba
 \label{B2}
 P_l(x)
 \,=\,\frac{\textstyle 1}{\textstyle 2^{\,l}\,l!}~\frac{\textstyle d^{\,l}}{\textstyle dx^{\,l}}
 \left(\,x^2\,-\,1\,\right)^l~.~
 \ea
 The associated Legendre functions $\,P_{lm}(x)\,$ (termed associated Legendre $\,${\it{polynomials}}$\,$ when their argument is sine or cosine of some angle) were introduced by Ferrers (1877) as \footnote{~Sometimes in the literature they also use the functions
 \ba
 \nonumber
 P_{l}^{m}(x)=(-1)^{m}\,\left(1-x^2\right)^{m/2}\,\frac{d^m\,}{dx^m}\,P_l(x)=(-1)^{m}\,P_{lm}(x)\,~,~~~
 \ea
 as defined, e.g., in Abramowitz \& Stegun (1972, p. 332). There also exists a different convention wherein $\,P_l^m(x)\,$ lacks the $\,(-1)^{m}\,$ multiplier and thus coincides with $\,P_{lm}(x)\,$, as in Arfken \& Weber (1995, p. 623).}
 \ba
 \label{B3}
 P_{lm}(x)=\left(1-x^2\right)^{m/2}\,\frac{d^m\,}{dx^m}\,P_l(x)~\,~,\quad\mbox{where}\quad l\geq m\geq 0\,~.
 \ea
 The so-defined associated Legendre functions are sometimes called {\it{unnormalised}}, although a more accurate term would be: {\it{in Ferrers' normalisation}}. This
 normalisation reads as:
 \begin{subequations}
 \ba
 \int_{-1}^1  P_{lm}(x)~P_{\,l\,'\,m}(x) ~dx~=~ \frac{2}{2\,l\,+\,1}~\,\frac{(l+m)!}{(l-m)!}\,~\delta_{\,l\,l\,'}~
 \label{B4a}
 \ea
 or, equivalently:
 \ba
 \int_{-\pi/2}^{\pi/2}  P_{lm}(\sin\phi)~P_{\,l\,'\,m}(\sin\phi)~\cos\phi~d\phi~=~ \frac{2}{2\,l\,+\,1}\,~\frac{(l+m)!}{(l-m)!}~\,\delta_{\,l\,l\,'}~~,
 \label{B4b}
 \ea
 another equivalent form being
 \ba
 \int_{0}^{\pi}  P_{lm}(\cos\varphi)~P_{\,l\,'\,m}(\cos\varphi)~\sin\varphi~d\varphi~=~ \frac{2}{2\,l\,+\,1}\,~\frac{(l+m)!}{(l-m)!}\,~\delta_{\,l\,l\,'}~~.
 \label{B4c}
 \ea
 \label{B4}
 \end{subequations}
 The associated Legendre functions in Ferrers' normalisation should not be confused with the associated Legendre functions $\,\tilde{P}_{lm}(x)\,$ which are written in the {\it{Schmidt partial normalisation}}:
 \ba
 \int_{-1}^1  \tilde{P}_{lm}(x)~\tilde{P}_{\,l\,'\,m}(x) ~dx~=~ \frac{2}{2\,l\,+\,1}\,~(2\,-\,\delta_{\,0\,m})~\,\delta_{\,l\,l\,'} \,~.
 \label{B5}
 \ea
 For more on these normalisations, see Winch et al. (2005).

 \section{Keeping $\,\dot{\omega}\,$ implies either keeping $\,\dot{\cal{M}}_0\,~$  or defining $\,{\cal{M}}\,$ as $\,dn/dt\,$}\label{difficulty}

  Under disturbance, the mean anomaly is written as
 \ba
 {\cal{M}}\,=~{\cal{M}}_0\,+~\int^{\,t}_{t_0} n(t)~dt~~,~~~\mbox{where}\qquad
 n(t)\,\equiv\,\sqrt{G\,(M\,+\,M^*)~a^{-3}(t)\,}~,
 \label{}
 \ea
 so the expression (\ref{504a}) for the Fourier tidal modes acquires the form of
 \ba
 \omega_{\textstyle{_{lmpq}}}\;\equiv~\stackrel{\bf\centerdot~~~~}{v_{lmpq}}\,-~m\,\stackrel{\bf\centerdot\,}{\theta}
 ~=\;(l-2p)\;(\dot{\omega}\,+\,\dot{\cal{M}}_0)\,+\,q\,\dot{\cal{M}}_0
 \,+\,(l-2p+q)\;n\,+\,m\;(\dot{\Omega}\,-\,\dot{\theta})~~.\quad
 \label{kremer}
 \ea
 Kaula (1964, equation 40) makes an oversight by accepting the approximation
 \ba
 \stackrel{\bf\centerdot}{\cal{M}\,}\,=~\stackrel{\bf\centerdot}{{\cal{M}}_0}\,+~n~\approx~n~~.
 \label{105}
 \ea
 Indeed, as $\,\dot{\omega}\,$ and $\,\dot{\cal{M}}_0\;$ are often of the same order, it is incorrect to keep the former while neglecting the latter. We present two examples. In the first, $\,\dot{\omega}\,$ and $\,\dot{\cal{M}}_0\;$
 are of the same order but of opposite signs, so they largely compensate one another. This suggests a simultaneous neglect of $\,${\it{both}}$\,$ terms. In the second example, $\,\dot{\omega}\,$ and $\,\dot{\cal{M}}_0\;$ turn out to be of the same order and the same sign, so keeping one
 of these terms requires keeping the other.

 \subsection{Example 1. Tidal perturbation of a low-inclination, low-eccentricity orbit}

 Consider a low-inclined perturber. From the tides it creates, the perturber gets predominantly transversal orbital disturbance. We need two planetary equations in the Gauss form (Brouwer \& Clemence 1961, page 301, eqn 33): \footnote{~The system (33) in Brouwer \& Clemence (1961, page 301) contains an equation for the rate $\,d\epsilon/dt\,$, where (as explained on the preceding page in {\it{Ibid.}}) $\,\epsilon\,$ is understood as $\,\epsilon^{{{\,I}}}\,\equiv\,{\cal{M}}_0\,+\,\tilde\omega\,=\,{\cal{M}}_0\,+\,\omega\,+\,\Omega\,$.}
 \ba
 \stackrel{\bf\centerdot\,}{\omega}~=~\frac{\sqrt{1\,-\,e^2}}{n~a~e}~\left[\,-~R~\cos f~+~\left(\,1\,+\,\frac{r}{p}\,\right)~T~\sin f\,\right]~-~\frac{\,\sin(\omega\,+\,f)~\,\cot i\,}{n~a~\sqrt{1\,-\,e^2}}~\,\frac{\,r\,}{a}~W~~,
 \nonumber
 \ea
 \ba
 \stackrel{\bf\centerdot}{{\cal{M}}_0}~=~\frac{1\,-\,e^2}{n~a~e}~\left[\,\left(\,\cos f~-~2~\,\frac{\,r\,}{p}\,~e\,\right)\,R~-~\left(\,1\,+\,\frac{r}{p}\,\right)~T~\sin f\,\right]~~,
 \nonumber
 \ea
 where $\,f\,$ is the true anomaly, $\,p\,\equiv\,a\,(1\,-\,e^2)\,$ is the $\,${\it{semilatus rectum}}$\,$, while $\,R\,$, $\,T\,$, and $\,W\,$ are the radial, transversal, and normal-to-orbit forces, respectively. In a situation where the perturbation is predominantly transversal and the terms with $\,R\,$ and $\,W\,$ may be neglected, we obtain:
 \ba
 \stackrel{\bf\centerdot\,}{\omega}\,+~\stackrel{\bf\centerdot}{{\cal{M}}_0}~\approx~\frac{e}{1\,+\,\sqrt{1\,-\,e^2}}~\frac{\sqrt{1\,-\,e^2}}{n~a}~\left(\,1\,+\,\frac{r}{p}\,\right)~T~\sin f~~.
 \label{106}
 \ea
  A low-inclined moon gets predominantly transversal orbital disturbance from the tides it creates in the planet. Inserting the latter expression in the formula (\ref{kremer}) for the Fourier modes, we see that for the modes with a zero $\,q\,$ (like the semidiurnal tide parameterised with $\,lmpq\,=\,2200\,$) the input from the pericentre rate may be omitted, if the eccentricity is not too large. Indeed, for $\,q=0\,$, the term $\,q\dot{\cal{M}}_0\,$ vanishes, while the term $\,(l-2p)(\dot{\omega}+\dot{\cal{M}}_0)\,$ is now approximated
 with $\,(l-2p)\,$ multiplied by the expression (\ref{106}). Although $\,\dot{\omega}\,$ and $\,\dot{\cal{M}}_0\,$ can, separately, be substantial, their sum (\ref{106}) is smaller by the order of $\,e\,$. Being (in this particular case) of the same order but of opposite sign, $\,\dot{\omega}\,$ and $\,\dot{\cal{M}}_0\,$ largely compensate one another. Therefore, if we choose to drop $\,\dot{\cal{M}}_0\,$, we should also drop $\,\dot{\omega}\,$. In this special case, dropping of both will be legitimate.

 As a useful aside, we would remind that the mean longitude is defined through $\,L\,\equiv\,{\cal{M}}\,+\,\omega\,+\,\Omega\,$, its rate being $~\dot{L}\,\equiv\,
 \sqrt{G\,(M\,+\,M^*)~a^{-3}(t)\,}\,+\,\dot{\cal{M}}_0\,+\,\dot\omega\,+\,\dot\Omega~$. As we have just seen, the rates $\,\dot{\cal{M}}_0\,$ and $\,\dot\omega\,$
 largely compensate one another and may both be neglected in the considered case. If, above that, the rate of the node happens to be negligible, then the mean motion from
 the Kepler law will be close to the mean longitude rate.

 \subsection{Example 2. Orbital perturbation due to oblateness}

  The situation is different where the principal perturbation is due to the oblateness of the tidally perturbed primary.
  The mean rates (Vallado 2007, pp. 647 - 648)
 \begin{eqnarray}
 \nonumber
 \stackrel{\bf\centerdot}{{\cal{M}}_0}~=~\frac{\,3\,}{\,4\,}~n~J_2~\frac{R^2}{a^2}~\frac{\,2~-~3~\sin^2i\,}{(1~-~e^2)^{3/2}}
 ~~\qquad\mbox{and}\qquad~
 \stackrel{\bf\centerdot\,}{\omega}~=~\frac{\,3\,}{\,4\,}~n~J_2~\frac{R^2}{a^2}~\frac{\,4~-~5~\sin^2i\,}{(1~-~e^2)^{2}}~~.
 \end{eqnarray}
 are of the same order and sign. Therefore, when keeping $\,\stackrel{\bf\centerdot\,}{\omega}\,$, we must also include $\,\stackrel{\bf\centerdot}{{\cal{M}}_0}\,$.

 The easiest way to get rid of $\,\stackrel{\bf\centerdot}{{\cal{M}}_0}\,$ is to define the mean motion as $\,n\,\equiv\,{\bf{\dot{\cal{M}}}}\,$. This,  the so-called $\,${\it{anomalistic}}$\,$ mean motion will, however, differ from  $~\,\sqrt{G\,(M\,+\,M^*)~a^{-3}(t)\,}~$.

  \section{Universality of the Darwin-Kaula description}\label{universality}

 As we saw earlier, to obtain the decomposition (\ref{107b}) from the Fourier series (\ref{101b}), each term of the latter series must be endowed with a mitigating factor $\,k_{\textstyle{_l}}=k_{\textstyle{_l}}(\omega_{\textstyle{_{lmpq}}})\,$ of its own and, likewise, must acquire its own phase lag  $\,\epsilon_{\textstyle{_l}}=\epsilon_{\textstyle{_l}}(\omega_{\textstyle{_{lmpq}}})\,$. In the literature, some authors enquired whether this mitigate-and-lag method is general enough to describe tides. The answer to this question is affirmative, insofar as the tides are linear. Without going into details (to be found in Efroimsky 2012$\,$a,$\,$b), we would mention that an $\,l-$degree part of the operator (\ref{107a}) is a convolution called the {\it{Love operator}}:
 \ba
 U_{l}(\erbold,\,t)\;=\;\left(\frac{R}{r}
 \right)^{{\it l}+1}\int_{-\infty}^{t} {\bf\dot{\it{k}}}_{\textstyle{_l}}(t-t\,')~W_{\it{l}}
 (\eRbold\,,\;\erbold^{\;*},\;t\,')\,dt\,'~,
 \label{chuk}
 \ea
 Indeed, linearity of tides means that, at each time $\,t\,$, the overall magnitude of reaction depends linearly on the magnitudes of the disturbance at all preceding instants of time, $\,t\,'\leq t\,$. The emergence of inputs from earlier times stems from the inertia (``memory") of the material. A disturbance that took place at an instant $\,t\,'\,$ appears in the integral for $\,U_{l}(\erbold,\,t)\,$ with a weight $\,{\bf\dot{\it{k}}}_{\textstyle{_l}}(t-t\,')\,$ that depends on the elapsed time. Following Churkin (1998), who gave this formalism its present shape, we call these weights {\it{Love functions}}.

 In the frequency domain, the above convolution becomes:
 \ba
 \bar{U}_{\textstyle{_{l}}}(\omega)\;=\;\left(\,\frac{R}{r}\,\right)^{l+1}\bar{k}_{\textstyle{_{l}}}(\omega)\;\,\bar{W}_{\textstyle{_{l}}}(\omega)\;\;,
 \label{gek}
 \ea
 where $\,\omega=\omega_{\textstyle{_{lmpq}}}\,$ is the tidal mode (not the periapse); $\,\bar{U}_{\textstyle{_{l}}}(\omega)\,$ and $\,\bar{W}_{\textstyle{_{l}}}(\omega)\,$ are the Fourier or Laplace images of the potentials $\,{U}_{\textstyle{_{l}}}(\erbold,\,t)\,$ and $\,{W}_{\textstyle{_{l}}}(\Rbold,\,\erbold^*,\,t)\,$; while the complex Love numbers
 \ba
 \bar{k}_{\textstyle{_l}}(\omega)
 \;=\;|\bar{k}_{\textstyle{_l}}(\omega)|\;e^{\textstyle{^{-i\,\epsilon_{{_l}}(\omega)}}}\;=\;{k}_{\textstyle{_l}}(\omega)\;e^{\textstyle{^{-i\,\epsilon_{{_l}}(\omega)}}}~~
 \label{number}
 \ea
 are the Fourier or Laplace components of the Love functions $\,{\bf\dot{\it{k}}}_{\textstyle{_l}}(t-t\,')\,$. The actual dynamical Love numbers are the real parts of the complex Love numbers, $\,{k}_{\textstyle{_l}}(\omega)\,=\,|\bar{k}_{\textstyle{_l}}(\omega)|~$; $\,$while the tidal lags are the complex Love numbers' negative phases.

  The frequency dependencies $\,\bar{k}_{\textstyle{_{l}}}(\omega)\,$ and, consequently, $\,{k}_{\textstyle{_l}}(\omega)\,$ and $\,{\epsilon}_{\textstyle{_l}}(\omega)~$ can be derived from the expression for the complex compliance $\,\bar{J}(\chi)\,$ or the complex rigidity $\,\bar{\mu}(\chi)=1/\bar{J}(\chi)\,$ of the mantle (with $\,\chi=\chi_{\textstyle{_{lmpq}}}\,\equiv\,|\omega_{\textstyle{_{lmpq}}}|\,$ being the physical forcing frequency). The dependency $\,\bar{J}(\chi)\,$ follows from the rheological model.

 Evidently, the formula (\ref{gek}) is but a concise version of (\ref{107b}). Thus we see that the mitigate-and-lag method ensues directly from the linearity assumption.

 \section{The Eulerian and Lagrangian descriptions.\\
 Perturbative approach to a periodically deformed body}\label{EL}

 \subsection{Perturbative treatment}

 Under perturbation, two changes will happen in a point $\,\erbold\,$ at a time $\,t~$:
 \begin{itemize}
 \item[\bf 1.~] Physical fields will now acquire different values in this point at this time. \\
 ~\\
 For example, a moon fixed in a certain position relative to the planetary surface will render some distribution of its potential over the volume of the host planet. The same moon fixed in a different position will generate a different distribution of its potential. This, in its turn, will yield a different deformation of the planet and therefore a different spatial distribution of its tidal-response potential and of all other quantities.\\
 ~\\
 Thus, instead of the unperturbed Eulerian dependencies $\,q^0_{_E}(\rbold\,,~t)\,$, we now have
 \ba
 q_{_E}(\rbold\,,~t)~=~q^0_{_E}(\rbold\,,~t)~+~q\,'_{_E}(\rbold\,,~t)\,~.
 \label{}
 \ea

 \item[\bf 2.~] A different particle will now arrive in the point $\,\erbold\,$ at the time $\,t\,$. It will not be the same particle as the one expected there at the time $\,t\,$ in the absence of perturbation.\\
      ~\\
      On the other hand, a particle that starts in $\,\Xbold~$ at $~t=0\,$, will appear, at the time $\,t\,$, not in the point $\,\xbold=\fbold(\Xbold,\,t)\,$ but in some other place
 \ba
 \erbold~=~\xbold~+~\ubold
 ~=~\fbold(\Xbold,\,t)~+~\ubold(\Xbold,\,t)\,~.
 \label{}
 \ea
 \end{itemize}
 These two changes will influence the Lagrangian dependencies on the initial conditions. The dependency of each field will obtain a variation $~q\,'_{_L}(\Xbold,\,t)~$:
 \ba
 q_{_L}(\Xbold,\,t)~=~q^0_{_L}(\Xbold,\,t)~+~q\,'_{_L}(\Xbold,\,t)\,~.
 \label{phobos}
 \ea
 In the absence of perturbation, the particle $\,\Xbold\,$ was destined to arrive in $\,\xbold\,$, wherefore $\,q^0_{_L}(\Xbold,\,t)\,$ was defined through (\ref{110}). Under
 perturbation, the same particle $\,\Xbold\,$ is expected to end up in $\,\erbold\,$, so the Lagrangian dependency becomes
 \bs
 \ba
 q_{_L}(\Xbold,\,t)&\equiv&q_{_E}(\rbold\,,~t)
 \label{equivalence_a}
 \label{114a}\\
 \nonumber\\
 &=&q_{_E}(\,\fbold(\Xbold,\,t)\,+\,\ubold(\Xbold,\,t)\,,~t\,)
 \label{equivalence_b}
 \label{114b}\\
 \nonumber\\
 &=&q_{_E}(\,\fbold(\Xbold,\,t)\,,~t\,)~+~\ubold(\Xbold,\,t)~\nabla_{\textstyle{_x\,}} q_{_E}~+~O(\ubold^2)
 \label{equivalence_c}
 \label{114c}\\
 \nonumber\\
 &=&q^0_{_E}(\,\fbold(\Xbold,\,t)\,,~t\,)~+~q\,'_{_E}(\,\fbold(\Xbold,\,t)\,,~t\,)~+~\ubold(\Xbold,\,t)~\nabla_{\textstyle{_x\,}} q_{_E}~+~O(\ubold^2)~~.\qquad~\qquad
 \label{equivalence_d}
 \label{114d}
 \ea
 \label{equivalence}
 \label{114}
 \es
 Insertion of (\ref{phobos}) into the left-hand side of (\ref{equivalence}) will give us:
 \ba
 q^0_{_L}(\Xbold,\,t)~+~q\,'_{_L}(\Xbold,\,t)~=~q^0_{_E}(\,\fbold(\Xbold,\,t)\,,~t\,)~+~q\,'_{_E}(\,\fbold(\Xbold,\,t)\,,~t\,)~+~\ubold(\Xbold,\,t)~\nabla_{\textstyle{_x\,}} q_{_E}~+~O(\ubold^2)~~.
 \nonumber
 \ea
 Subtracting (\ref{equality_b}) from this formula, we arrive at a relation between the perturbations of the Lagrangian and Eulerian quantities:
 \ba
 q\,'_{_L}(\Xbold,\,t)~=~q\,'_{_E}(\,\fbold(\Xbold,\,t)\,,~t\,)~+~\ubold(\Xbold,\,t)~\nabla_{\textstyle{_x\,}} q_{_E}~+~O(\ubold^2)~~,
 \label{perturbations}
 \label{115}
 \ea
 where the first term on the right-hand side, $~q\,'_{_E}(\xbold,\,t)~$, expresses the change in the final spatial distribution of the field $\,q\,$. The other two terms show up because perturbation changes the mapping from $\,\Xbold\,$ to the current location.

 \subsection{An equivalent description}

 A slightly different, although equally valid viewpoint is possible. In a reference setting, at time $\,t\,$, an observer
{{located in}} $\,\xbold\,$ will see the arrival of a particle that started from $\,\Xbold\,$:
 \ba
 q^0_{_L}(\Xbold,\,t)~\equiv~q^0_{_E}(\xbold\,,~t)~~~~~~~~~~~
 \label{117}
 \ea
 In a perturbed situation, the same observer in $\,\xbold\,$ will register, at the time $\,t\,$, the arrival of a different particle, one that started from $\,\Xbold-\Ubold~$:
 \bs
 \ba
 q_{_L}(\Xbold-\Ubold,\,t)~\equiv~q_{_E}(\xbold\,,~t)~~,~~~~~~~~
 \label{118a}
 \ea
 which is:
 \ba
 q_{_L}(\Xbold,\,t)~-~\Ubold~\nabla_{_X}q_{_L}~+~O(U^2)~=~q_{_E}(\xbold\,,~t)~~.~~~~~~~~
 \label{118b}
 \ea
 \label{118}
 \es
 Subtraction of (\ref{117}) from (\ref{118b}) gives us the variations:
 \bs
 \ba
 q_{_L}(\Xbold,\,t)~-~q^0_{_L}(\Xbold,\,t)~-~\Ubold~\nabla_{_X}q_{_L}~+~O(U^2)~=~q_{_E}(\xbold\,,~t)~-~q^0_{_E}(\xbold\,,~t)
 \label{119a}
 \ea
 or, simply,
 \ba
 q\,'_{_L}(\Xbold,\,t)~=~q\,'_{_E}(\xbold\,,~t)~+~\Ubold~\nabla_{_X}q_{_L}~+~O(U^2)~~.
 \label{119b}
 \ea
 \label{119}
 \es
 Introducing the Jacobian $\;J\equiv\,\frac{\textstyle dV^t}{\textstyle dV^0}\,=\,\mbox{det}\,\frac{\textstyle\partial x_i}{\textstyle\partial X_j}\;\,$, ~we write:
 \ba
 \Ubold~\nabla_{_X}q_{_L}~=~\Ubold~J~\nabla_{\textstyle{_x\,}} q_{_E}~=~\ubold~\nabla_{\textstyle{_x\,}} q_{_E}~~,
 \label{120}
 \ea
 with $\,\ubold\equiv\Ubold J\,$.
  Thus (\ref{119b}) and (\ref{perturbations}) are equivalent insofar as $\,O(\Ubold^2)=O(\ubold^2)\,$.

 While the language of (\ref{perturbations}) is more conventional than that of (\ref{119}), the latter description is easier for physical interpretation. Suppose we are
 observing gradual cooling of a flow. In an unperturbed setting, a particle, that started in $\,\Xbold\,$ at the time $\,t=0\,$, will show up in $\,\xbold\,$ at the time $\,t\,$. Accordingly, a measurement of the temperature in $\,\xbold\,$ at the time $\,t\,$ will render, in the absence of perturbation, a value to which the particle $\,\Xbold\,$ has cooled down by this time --- see the equality (\ref{117}).

 Under perturbation, the rate of cooling of each particle will change. In addition, owing to the change of trajectories, a different particle
 will show up in $\,\xbold\,$ at the time $\,t\,$. Now this will be a particle that started its movement at $\,t=0\,$ from some point $\,\Xbold-\Ubold\,$.
 So a measurement of the temperature in the point $\,\xbold\,$ at the time $\,t\,$ will now give us a temperature value to which the particle $\,\Xbold-\Ubold\,$ has
 cooled down --- see the equation (\ref{118a}).

 The difference between the measurements performed in $\,\xbold\,$ in the perturbed and unperturbed cases will, according to (\ref{119b}), read as $\,q\,'_{_E}(\xbold\,,~t)\,=\,q\,'_{_L}(\Xbold,\,t)\,-\,\Ubold~\nabla_{_X}q_{_L}~+~O(U^2)\,$. The first term on the right renders the cooling down of the
 particle arriving in $\,\xbold\,$ at the time $\,t\,$, while the second and third terms reflect the fact that, under disturbance, we register a particle arriving from a point displaced by $\,\Ubold\,$, compared to the particle that would be brought to $\,\xbold\,$ by an unperturbed flow.

 With aid of (\ref{120}), the expression (\ref{119b}) can be equivalently rewritten as
 \ba
 D_t~=~\partial_t~+~\vbold\,\nabla_{\textstyle{_x}}~~,
 \label{121}
 \ea
 where $\,\vbold\equiv \partial\ubold/\partial t\,$, while $\,D\equiv d/dt\,$ is the $\,${\it{comoving}}$\,$ derivative.
 The physical interpretation of (\ref{121}) is obvious: the rate of cooling of a moving particle, $\,dq/dt\,$, can be measured by a quiescent observer. The observer, however, must amend his result, $\,\partial q/\partial t\,$, with a correction taking into account the fact that, being quiescent, he is measuring the difference between the temperature of different particles passing by, not of the same particle.

 \subsection{Periodically deformed solids. Linearisation}

 Hereafter, we shall restrict our consideration to the case of a periodically deformed solid. It is natural to associate the reference trajectory $\,\xbold~=~\fbold(\Xbold,\,t)\,$ with the equilibrium configuration. In this configuration, the particles stay idle, so $\,\xbold\,$ coincides with the initial value $\,\Xbold~$:
 \ba
 \xbold~=~\fbold(\Xbold,\,t)~~~,\qquad\mbox{where}\qquad\fbold(\Xbold,\,t)~=~\Xbold\qquad\mbox{for~~all~}~~t~~,
 \label{GER}
 \label{122}
 \ea
 while all properties keep in time their fiducial values:
 \ba
 q^0_{_L}(\Xbold,\,t)\,=\,q^0_{_E}(\xbold,\,t)\,=\,q^0\,~.
 \label{123}
 \ea
 Be mindful that the superscript $\,0\,$ did not originally mark a value fixed in time but a trajectory chosen to be reference. It is only now, that the role of a reference
 configuration is played by an equilibrium body, the superscript $\,0\,$ begins to denote an unchanging value.

 For a particle originally located in $\,\Xbold\,$, its perturbed trajectory $\,\erbold\,$ differs from its reference trajectory $\,\xbold\,$ by some $\,\ubold\,$:
 \ba
 \erbold~=~\xbold~+~\ubold
 ~=~\fbold(\Xbold,\,t)~+~\ubold(\Xbold,\,t)\,~.
 \label{124}
 \ea
 When the reference trajectory is the equilibrium, insertion of (\ref{GER}) into (\ref{124}) results in
 \bs
 \ba
 \erbold~=~\Xbold~+~\ubold(\Xbold,\,t)\,~,
 \label{125a}
 \ea
 which can also be written as
 \ba
 \erbold~=~\xbold~+~\ubold(\xbold,\,t)\,~,
 \label{125b}
 \ea
 \label{125}
 \es
 because, in this case, the unperturbed trajectory $\,\xbold\,$ always coincides with the initial value $\,\Xbold\,$.

 We work in a linearised approximation, neglecting the term $\,O(\ubold^2)\,$ in (\ref{perturbations}) and writing all expansions up to terms linear in the
 displacement $\,\ubold\,$ or velocity $\,\vbold\equiv d\ubold/dt\,$.

 For a short and simple explanation of the linearised Lagrangian and Eulerian descriptions of tides, see Wang (1997). A more comprehensive treatment is offered in
 the book by Dahlen \& Tromp (1998, Section 3.1.1).

 \subsection{Conservation of mass in the Lagrangian and Eulerian descriptions}

 Denote the Eulerian value of the mass density with $\,\rho_{_E}(\erbold,\,t)\,$. As mass cannot be destroyed or created, its amount in a comoving volume $\,V^t\,$ of a flow stays constant:
 \ba
 \frac{d}{dt}\,\int_{V^t}\rho_{_E}\,dV^t~=~0~~.
 \label{126}
 \ea
 For the reference history $\,\xbold=\fbold(\Xbold,\,t)\,$, this would imply:
 \bs
 \ba
 \int_{V^{t}_{ref}}\rho^{\,0}_{_E}(\xbold\,,~t)\,d^4\xbold~=~\int_{V^0}\rho^{\,0}_{_E}(\Xbold\,,~0)\,d^4\Xbold~~.
 \label{BELLa}
 \label{127a}
 \ea
 For a perturbed history $\,\erbold=\fbold(\Xbold,\,t)+\ubold(\Xbold,\,t)\,$, we have:
 \ba
 \int_{V^{t}_{pert}}\rho_{_E}(\erbold\,,~t)\,d^4\erbold~=~\int_{V^0}\rho_{_E}(\Xbold\,,~0)\,d^4\Xbold~~.
 \label{BELLb}
 \label{127b}
 \ea
 \label{BELL}
 \label{127}
 \es
 For each individual particle, its perturbed trajectory $\,\erbold\,$ stems from the same initial position $\,\Xbold\,$ as the appropriate reference trajectory $\,\xbold\,$, so the initial densities are the same:
 \ba
 \rho_{_E}(\Xbold\,,~0)\;=\;\rho^{\,0}_{_E}(\Xbold\,,~0)\,~.
 \label{128}
 \ea
 At later times, however, $\,\rho_{_E}(\erbold,\,t)\,$ and $\,\rho^{\,0}_{_E}(\xbold,\,t)\,$ have different functional forms.

  \subsubsection{The continuity law in the Eulerian description}\label{continuity}

 The right-hand sides of the formulae (\ref{BELLa}) and (\ref{BELLb}) coincide, as they render the mass of the same initial distribution $\,\rho_{_E}(\Xbold,\,0)=\rho^{\,0}_{_E}(\Xbold,\,0)\,$. Thus the left-hand sides of the two formulae also coincide:
 \ba
 0~=~\int_{V^{t}_{pert}}\rho_{_E}(\erbold\,,~t)\,d^4\erbold~-~\int_{V^t_{ref}}\rho^{\,0}_{_E}(\xbold\,,~t)\,d^4\xbold~~,
 \label{129}
 \ea
 as the mass stays unchanged, no matter whether the system follows the reference history or a perturbed one.
 Now switch from the perturbed coordinates, $\,\erbold\,$, to the reference ones, $\,\xbold~$:
 \bs
 \ba
 0&=&\int\,d^4\xbold~\left[\,J~\rho_{_E}(\erbold\,,~t)~-~\,\rho^{\,0}_{_E}(\xbold\,,~t)\right]
 \label{130a}\\
 \nonumber\\
  &=&\int\,d^4\xbold~\left[\,\left(\,1\,+\,\nabla_{\textstyle{_{{x}}}}\cdot\ubold\,\right)~\rho_{_E}(\erbold\,,~t)~-~\,\rho^{\,0}_{_E}(\xbold\,,~t)\right]~~,
 \label{130b}
 \ea
 \label{130}
 \label{DGG}
 \es
 where the Jacobian is:
 \bs
 \ba
 \label{131a}
 J~\equiv~\frac{dV^{t}_{pert}}{dV^{t}_{ref}}~=~\mbox{det}~\frac{\partial r_i}{\partial x_j}
  ~=~1~+~\nabla_{\textstyle{_{{x}}}}\cdot\ubold~+~O(\ubold^2)~=~1~+~J~\nabla_{\textstyle{_r}}\cdot\ubold~+~O(\ubold^2)
 ~~.~~~
 \ea
 From this we see that the Jacobian can also be written as  \footnote{~The expression (\ref{131b}) indicates that, within a linear approximation, we do not need to distinguish between $\,\nabla_{\textstyle{_r\,}}\,$ and  $\,\nabla_{\textstyle{_{{x}}}}\,=\,J\,\nabla_{\textstyle{_{{r}}}}\,=\,\nabla_{\textstyle{_{{r}}}}\,+\,(\nabla_{\textstyle{_{{r}}}}\cdot \ubold)\,\nabla_{\textstyle{_{{r}}}}\,+\,O(\ubold^2)\,$, when the operators are applied to a quantity that, by itself, is of the first order of smallness.}
 \ba
 J~=~\frac{\textstyle 1~+~O(\ubold^2)}{\textstyle 1~-~\nabla_{\textstyle{_r}}\cdot\ubold}~=~1~+~\nabla_{\textstyle{_r}}\cdot\ubold~+~O(\ubold^2)~~.
 \label{131b}
 \ea
 \label{131}
 \label{GID}
 \es
 From (\ref{130a}), we obtain the exact equality
 \ba
 \rho_{_E}(\erbold\,,~t)~J~=~\rho^{\,0}_{_E}(\xbold\,,~t)~~,
 \label{132}
 \ea
 a linearised version thereof being
 \bs
 \ba
 0&=&\rho_{_E}(\erbold\,,~t)~(1\,+\,\nabla_{\textstyle{_r}}\cdot\ubold)~-~\rho^{\,0}_{_E}(\erbold-\ubold\,,~t)~+~O(\ubold^2)
 \label{133a}
 \label{DFGa}\\
 \nonumber\\
 &=&~\rho_{_E}(\erbold\,,~t)~-~\rho^{\,0}_{_E}(\erbold\,,~t)~+~\rho_{_E}(\erbold\,,~t)~\nabla_{\textstyle{_r}}\cdot\ubold~+~\ubold~\nabla_{\textstyle{_r\,}}\rho^{\,0}_{_E}(\erbold\,,~t)~+~O(\ubold^2)~~.
 \label{133b}
 \label{DFGb}
 \ea
 \label{DFG}
 \label{133}
 \es
 For a small $\,t\,$, the deviation $\,\ubold\,$ between the two trajectories is linear in time, and so is the difference between the perturbed and reference density functions. Thus, we may change $\,\rho_{_E}(\erbold,\,t)\,\nabla_{\textstyle{_r}}\cdot\ubold~\,$ to $\,~\rho^{\,0}_{_E}(\erbold,\,t)\,\nabla_{\textstyle{_r}}\cdot\ubold~$, ~to obtain an expression correct to first order in $\,\ubold~$:
 \ba
 {\rho_{_E}}'~+~\nabla_{\textstyle{_r}}\cdot\left(\rho^{\,0}_{_E}\,\ubold\right)~=~0~~,
 \label{NBV}
 \label{134}
 \ea
 where the finite variation is
 \ba
 {\rho_{_E}}'(\erbold,\,t)~\equiv~\rho_{_E}(\erbold,\,t)~-~\rho^{\,0}_{_E}(\erbold,\,t)
 ~~.
 \label{135}
 \ea
 We would reiterate that the perturbative approach to Eulerian quantities implies comparison between their present spatial distributions. So the two histories are
 compared in the same point $\,\erbold\,$ and at the same time $\,t\,$.

 In (\ref{DFGb}), we could also have changed $\,\ubold\nabla_{\textstyle{_r\,}}\rho^{\,0}_{_E}(\erbold,\,t)~\,$ to $\,~\ubold\nabla_{\textstyle{_r\,}}\rho_{_E}(\erbold,\,t)\,$. Then, instead of (\ref{NBV}), we
 would have obtained $\,{\rho_{_E}}'~+~\nabla_{\textstyle{_r}}\cdot\left(\rho_{_E}\,\ubold\right)~=~0\,$, without the superscript $\,0\,$ in the second term. In the linear approximation, however,
 this would be no better than (\ref{NBV}). Traditionally, the form (\ref{NBV}) is preferred in the literature.

 However, when switching to a differential form of the conservation law, we no longer need to keep the superscript $\,0\,$, because the difference between
 $\,\rho_{_E}(\erbold,\,t)\,$ and $\,\rho^{\,0}_{_E}(\erbold,\,t)\,$ becomes infinitesimally small. So the differential law reads as:
 \ba
 \frac{\partial\rho_{_E}}{\partial t}~+~\nabla_{\textstyle{_r}}\cdot\left(\rho_{_E}\,\vbold\right)~=~0~~,
 \label{136}
 \ea
 where $\,\vbold\equiv\partial\ubold/\partial t\,$, the partial derivative giving the rate of change
 with coordinates fixed.

 Employment of the perturbative formula (\ref{NBV}) near a deformable free boundary requires some care. On the one hand, the
 reference density $\,\rho^{\,0}_{_E}(\erbold)\,$ makes an abrupt step there. On the other hand, due to deformation of the boundary, we may get a finite present density in a
 point where the reference density used to be zero, and vice versa.

 \subsubsection{The continuity law in the Lagrangian description}

 The Lagrangian density is introduced in the standard way (\ref{114a}):
 \ba
 \rho_{_L}(\Xbold,\,t)\,\equiv\,\rho_{_E}(\erbold,\,t)~~,
 \label{137}
 \label{formula}
 \ea
 so the formula (\ref{132}) becomes:
 \bs
 \ba
 \rho_{_L}(\Xbold\,,~t)~J~=~\rho^{\,0}_{_E}(\xbold\,,~t)~~.
 \label{138a}
 \ea
 For a periodically deformed solid, the reference density $\,\rho^{\,0}_{_E}(\xbold,\,t)\,$ is the density of the undeformed, stable configuration. So $\,\rho^{\,0}_{_E}(\xbold,\,t)\,=\,\rho^{\,0}_{_E}(\Xbold,\,0)\,=\,\rho^{\,0}(\Xbold)\,$ is time-independent, and the equality (\ref{138a}) becomes simply
 \ba
 \rho_{_L}(\Xbold\,,~t)~J~=~\rho^{\,0}_{_E}(\Xbold)~~.~~
 \label{138b}
 \ea
 \label{138}
 \es
 In accordance with the general formula (\ref{perturbations}), we interrelate the density variations as
 \bs
 \ba
 {\rho_{_L}}'(\Xbold\,,~t)~=~{\rho_{_E}}'(\xbold\,,~t)~+~\ubold~\nabla_{\textstyle{_x}} \rho^{\,0}_{_E}(\xbold\,,~t)~~.
 \label{139a}
 \ea
 In the considered case of small periodic variations, the reference trajectory is simply $\,\xbold=\Xbold\,$ at all times; so on the right-hand side of the above formula we have a gradient of a constant-in-time stationary distribution:  $~\,\nabla_{\textstyle{_x\,}} \rho^{\,0}(\xbold,\,t)\,=\,\nabla_{\textstyle{_X\,}} \rho^{\,0}(\Xbold)~$. Thence we obtain:
 \ba
 {\rho_{_L}}'(\Xbold\,,~t)~=~{\rho_{_E}}'(\Xbold\,,~t)~+~\ubold~\nabla_{\textstyle{_X\,}} \rho^{\,0}_{_E}(\Xbold) ~~.
 \label{139b}
 \ea
 \label{139}
 \es
 Combining this formula with (\ref{134}), we arrive at \footnote{~When combining (\ref{139b}) with (\ref{134}), we should not be confused by the fact that in (\ref{139b}) all
 quantities are functions of $\,X\,$, while in (\ref{134} - \ref{135}) these quantities show up as functions of $\,\erbold\,$. Nor should we be confused by $\,\nabla\,$
 denoting $\,\nabla_r\,$ in (\ref{134}) and $\,\nabla_X\,$ in (\ref{139b}). As our intention is simply to compare the functions, we are free to change the notations in
 (\ref{134} - \ref{135}) from $\,\erbold\,$ to $\,\Xbold\,$, whereafter (\ref{140}) will come out trivially. \label{foot}}
 \ba
 {\rho_{_L}}'~+~\rho^{\,0}_{_E}~\nabla_{\textstyle{_X}}\cdot\ubold~=~0~~,
 \label{140}
 \ea
 where $\,{\rho_{_L}}'={\rho_{_L}}'(\Xbold,\,t)\,$, while $\,\rho^{\,0}_{_E}=\rho^{\,0}_{_E}(\Xbold)\,$.

 \subsection{The Poisson equation}

 \subsubsection{In the Eulerian description}

 Perturbed or not, the density always obeys the Poisson equation; while the perturbing potential $\,W\,$ obeys the Laplace equation outside the perturber:
 \begin{subequations}
 \ba
 \nabla_{\textstyle{_r\,}}^{\,2}\,V_{_E}&=&-~4\,\pi\,G\,\rho_{_E}~~,
 \label{10a}\\
 \nonumber\\
 \nabla_{\textstyle{_r\,}}^{\,2}\,V^{\,0}_{_E}&=&-~4\,\pi\,G\,\rho^{\,0}_{_E}~~,
 \label{10b}\\
 \nonumber\\
 \nabla_{\textstyle{_r\,}}^{\,2}\,W_{_E}&=&0~~.~\qquad\qquad
 \label{10c}
 \ea
 \label{10}
 \end{subequations}
 Subtraction of (\ref{10b}) from (\ref{10a}) results in a Poisson equation for the density perturbation:
 \ba
 \left.~\quad~\right. \nabla_{\textstyle{_r\,}}^{\,2}\,{V_{_E}}'~=~-~4~\pi~G~{\rho_{_E}}'~~~
 \label{11a}
 \ea
 or, equivalently:
 \ba
 \left.~\quad~\right. \nabla_{\textstyle{_r\,}}^{\,2}\,{U_{_E}}~=~-~4~\pi~G~{\rho_{_E}}'~~,
 \label{11b}
 \label{154}
 \ea
 where we took into account the relations (\ref{9}) and (\ref{10c}).

 \subsubsection{In the Lagrangian description}

 Insertion of the formulae (\ref{145}) and (\ref{150}) into the Eulerian version of the Poisson equation, (\ref{11a}), results in the Lagrangian version of this equation:
 \bs
 \ba
 \left.~\quad~\right. \nabla_{\textstyle{_r\,}}^{\,2}\,\left(\,{V_{_L}}'\,-~\ubold\cdot\nabla_{\textstyle{_x\,}} V^{\,0}\,\right)~=~-~4~\pi~G~\left(\,{\rho_{_L}}'\,-~\ubold\cdot\nabla_{\textstyle{_x\,}} \rho^{\,0}    \,\right)~~.
 \label{155a}
 \ea
 A switch to differentiation over the initial position, $\,\nabla_{_{x}}\,$, would entail corrections of the order of
 $\,O(\ubold^2)\,$. In neglect of those, the equation may be written as
 \ba
 \left.~\quad~\right. \nabla_{\textstyle{_{x\,}}}^{\,2}\,\left(\,{V_{_L}}'\,-~\ubold\cdot\nabla_{\textstyle{_{x\,}}} V^{\,0}\,\right)~=~-~4~\pi~G~\left(\,{\rho_{_L}}'\,-~\ubold\cdot\nabla_{\textstyle{_{x\,}}} \rho^{\,0}    \,\right)~~.
 \label{155b}
 \ea
 \label{155}
 \es
 For an initially homogeneous body, the above formulae simplify to:
 \bs
 \ba
 \left.~\quad~\right. \nabla_{\textstyle{_r\,}}^{\,2}\,\left(\,{V_{_L}}'\,-~\ubold\cdot\nabla_{\textstyle{_x\,}} V^{\,0}\,\right)~=~-~4~\pi~G~{\rho_{_L}}'~~
 \label{156a}
 \ea
 and
 \ba
 \left.~\quad~\right. \nabla_{\textstyle{_x\,}}^{\,2}\,\left(\,{V_{_L}}'\,-~\ubold\cdot\nabla_{\textstyle{_{x}\,}} V^{\,0}\,\right)~=~-~4~\pi~G~{\rho_{_L}}'~~.
 \label{156b}
 \ea
 \label{156}
 \es

 \section{Boundary conditions}\label{appA}

 The boundary condition on the total Eulerian potential $\,V_{\textstyle{_{E}}}\,$ is trivial. To avoid infinite forces, the potential must be continuous:
 \ba
 {V}^{\,\textstyle{^{(exterior)}}}_{\textstyle{_{E}}}\,=~{V}^{\,\textstyle{^{(interior)}}}_{\textstyle{_{E}}}\,~.
 \label{A1}
 \ea
 The boundary condition on the potential's gradient emerges as a corollary of the Gauss theorem and therefore mimics a similar condition from electrostatics.$\,$\footnote{~Melchior (1972) attributes the derivation of the boundary condition to Michel Chasles.} Let a small area $\,\vec{\bf s}\,=\,s\,\hat{\bf{n}}\,$ of the free surface
 be sandwiched between the top and bottom of a cylinder of an infinitesimal height $\,u\,=\,\vec{\bf u}\cdot\vec{\bf s}/s\,$, with the vector $\,\vec{\bf u}\,$ being the
 tidal displacement. The top and bottom should each have the principal curvature radii coinciding with those of the free surface, but in the leading order this can be ignored, with the enclosed volume thus being $\,u\,s\,=\,\vec{\bf u}\cdot\vec{\bf s}\,$. In neglect of the contributions from the infinitesimally small side areas of the cylinder, employment of the Gauss theorem for the Eulerian potential gives:
 \ba
 -~4~\pi~G~\rho^0~\vec{\bf u}\,\cdot\,\vec{\bf s}~=~\int_s\,\nabla_{\textstyle{_x\,}} V_{\textstyle{_{E}}}\,\cdot\,d\vec{\bf s}~=~
 \nabla_{\textstyle{_x\,}} {V}^{\,\textstyle{^{(exterior)}}}_{\textstyle{_{E}}}\cdot\,\vec{\bf s}~-~\nabla_{\textstyle{_x\,}} {V}^{\,\textstyle{^{(interior)}}}_{\textstyle{_{E}}}\cdot\,\vec{\bf s}~~.~~~~
 \label{}
 \ea
 Over a surface between layers, the condition will read as
 \ba
 \left(\,4~\pi~G~\rho^{\,0}~\vec{\bf u}\,\cdot\,\vec{\bf s}\,\right)^{\,\textstyle{^{(exterior)}}}~-~\left(4~\pi~G~\rho^{\,0}~\vec{\bf u}\,\cdot\,\vec{\bf s}\,\right)^{\,\textstyle{^{(interior)}}}~=~
 \nabla_{\textstyle{_x\,}} {V}^{\,\textstyle{^{(exterior)}}}_{\textstyle{_{E}}}\cdot\,\vec{\bf s}~-~\nabla_{\textstyle{_x\,}} {V}^{\,\textstyle{^{(interior)}}}_{\textstyle{_{E}}}\cdot\,\vec{\bf s}~~~~
 \label{}
 \ea
 or, equivalently,
 \ba
 \left[~-~4~\pi~G~\rho^{\,0}~{\bf u}~+~\frac{\partial~}{\partial\hat{\bf{n}}}\, {V}_{\textstyle{_{E}}}\,\right]^{\,\textstyle{^{(exterior)}}}\,=~
 \left[~-~4~\pi~G~\rho^{\,0}~{\bf u}~+~\frac{\partial~}{\partial\hat{\bf{n}}}\, {V}_{\textstyle{_{E}}}\,\right]^{\,\textstyle{^{(interior)}}}\,.~~~
 \label{A2}
 \ea
 In application to tides, it can be interpreted like this: the discontinuity in attraction is equal to the attraction of the deformation bulge (Legros et al. 2006). Since $\,V_0\,$, $\,W\,$ and their normal gradients are continuous on the boundary, the conditions on $\,U\,$ and $\,V\,'\,$ look exactly like (\ref{A1} - \ref{A2}). Specifically, in Section \ref{zp} we need the conditions on the total variation $\,V\,'~$:
 \ba
 {{V_{{_{E}}}}'}^{\,\textstyle{^{(exterior)}}}\,=~{{V_{{_{E}}}}'}^{\,\textstyle{^{(interior)}}}
 \label{A3}
 \ea
 \ba
 \left[~-~4~\pi~G~\rho^0~\vec{\bf u}~+~\frac{\partial~}{\partial\hat{\bf{n}}}\, {V_{_E}}'\,\right]^{\,\textstyle{^{(exterior)}}}\,=~
 \left[~-~4~\pi~G~\rho^0~\vec{\bf u}~+~\frac{\partial~}{\partial\hat{\bf{n}}}\, {V_{_E}}'\,\right]^{\,\textstyle{^{(interior)}}}\,~.
 \label{A4}
 \ea
 The Eulerian and Lagrangian potentials are interrelated through
 \ba
 {V_{{_{E}}}}'\,=~{V_{_L}}'\,-\,\ubold\cdot\nabla_{\textstyle{_x\,}} V^0\,~.
 \label{A5}
 \ea
 Thence, in the Lagrangian description, the conditions will acquire the form of
 \ba
 \left[\,{V_{_L}}'\,-~\ubold\cdot\nabla_{\textstyle{_x\,}} V^{\,0}\,\right]^{\,\textstyle{^{(exterior)}}}\,=~\left[\,{V_{_L}}'\,-~\ubold\cdot\nabla_{\textstyle{_x\,}} V^{\,0}\,\right]^{\,\textstyle{^{(interior)}}}
 \label{A6}
 \ea
 and
  \ba
 \left[\,\frac{\partial~}{\partial\hat{\bf{n}}}\,\left({V_{_L}}'-\,\ubold\cdot\nabla_{\textstyle{_x\,}} V^{\,0}\,\right)\,-\,4\,\pi\,G\,\rho^0\,{\bf u}\,\right]^{\,\textstyle{^{(exterior)}}}=\,
 \left[\,\frac{\partial~}{\partial\hat{\bf{n}}}\,\left({V_{_L}}'-\,\ubold\cdot\nabla_{\textstyle{_x\,}} V^{\,0}\,\right)\,-\,4\,\pi\,G\,\rho^0\,{\bf u}\,\right]^{\,\textstyle{^{(interior)}}}.~\,~
 \label{A7}
 \ea
 When the boundary is welded or its normal is parallel to $\,\nabla_{\textstyle{_x\,}} V^{\,0}\,$, the term $~-\ubold\cdot\nabla_{\textstyle{_x\,}} V^{\,0}~$ becomes
 continuous (Wang 1997). It, thus, can be removed from (\ref{A6}), rendering the incremental Lagrangian potential continuous. This term, however, cannot be omitted in (\ref{A7}).

 \section{Interrelation between dynamical Love numbers,\\
 for an incompressible homogeneous sphere}\label{love}\label{appC}

 For an incompressible homogeneous spherical body, the $\,${\it{static}}$\,$ Love numbers read as
 \ba
 k_l=\frac{3}{2\,(l\,-\,1)}\;\,\frac{1}{1\,+\,{\cal{A}}_l}~\qquad~\mbox{and}~\qquad~h_l=\frac{2\,l\,+\,1}{2\,(l\,-\,1)}\;\,\frac{1}{1\,+\,{\cal{A}}_l}~\,~,
 \label{C1}
 \ea
 where
 \ba
 {\cal{A}}_l\,\equiv\,\frac{\textstyle{(2\,l^{\,2}+\,4\,l\,+\,3)\,\mu}}{\textstyle{l}\,\mbox{g}\,\rho\,R}
 \,=\,\frac{\textstyle{3\,(2\,{\it{l}}^{\,2}+\,4\,{\it{l}}\,+\,3)\,\mu}}{\textstyle{4\,l\,\pi\,G\,\rho^2\,R^2}}
 \,=\,\frac{\textstyle{3\,(2\,{\it{l}}^{\,2}+\,4\,{\it{l}}\,+\,3)}}{\textstyle{4\,l\,\pi\,G\,\rho^2\,R^2\,J}}
 ~~.~\quad
 \label{C2}
 \ea
 $\mu\,$ and $\,J\,=\,1/\mu\,$ being the relaxed
 rigidity and compliance, and $\,G
 \,$ being Newton's gravity constant. The formulae (\ref{C1}) yield a well-known relation connecting the {{static}} Love numbers:
 \ba
 (2\,l\,+\,1)~k_l~=~3~h_l\,~.
 \label{C3}
 \ea
 Expressions (\ref{C1}) are obtained by solving a system comprising the static version of the Second Law of Newton and the constitutive equation
 interconnecting the stress and strain through the rigidity $\,\mu\,$. A wonderful theorem, called $\,${\it{the correspondence principle}}$\,$ or $\,${\it{the
 elastic-viscoelastic analogy}}$\,$, tells us that in many situations the dynamical versions of the Second Law of Newton and constitutive equation, when written in
 the frequency domain as algebraic equations for operational moduli, mimic the static versions of these equations. In order for this correspondence to take place, the accelerations and inertial forces should be negligibly small (see, e.g., Appendix B to Efroimsky 2012$\,$a). In that case, the complex Love numbers $\,\bar{k}_l(\omega)\,$ and $\,\bar{h}_l(\omega)\,$ will be expressed through the complex operational moduli $\,\bar{\mu}\,$ or $\,\bar{J}\,$ in the same algebraic manner as the static $\,{k}_l\,$ and $\,{h}_l\,$ are expressed via the static $\,{\mu}\,$ or $\,{J}\,$. Also recall that the static expressions (\ref{C1}) were derived under an extra assumption of
 incompressibility. If this assumption is also valid in the dynamical case, then the complex $\,\bar{k}_l(\omega)\,$ and $\,\bar{h}_l(\omega)\,$ are expressed through the
 complex $\,\bar{\mu}\,$ or $\,\bar{J}\,$ by formulae mimicking (\ref{C1}), whence an expression like (\ref{C3}) ensues for $\,\bar{k}_l(\omega)\,$ and $\,\bar{h}_l(\omega)\,$. Its imaginary part will read as:
 \ba
 (2\,l\,+\,1)~k_l(\omega)~\sin\epsilon(\omega)~=~3~h_l(\omega)~\sin\epsilon(\omega)\,~,
 \label{C4}
 \ea
 where $\,k_l(\omega)\,\equiv\,|\bar{k}_l(\omega)|\,$, $~h_l(\omega)\,\equiv\,|\bar{h}_l(\omega)|\,$ and $\,\omega\,=\,\omega_{\textstyle{_{lmpq}}}\,$.

 To draw to a close, we would emphasise that in the static expression (\ref{C2}) the letters $\,\mu\,$ and $\,J\equiv 1/\mu\,$ stand for the static (relaxed) values of the rigidity and compliance. In a dynamical analogue of this expression, the same letters will denote the unrelaxed values.

 \section{Heat production over tidal modes. A sketchy derivation of the formula (\ref{196}), in neglect of the ``degeneracy"}\label{sketch}

 To compute the dissipation rate at separate tidal modes, it is necessary to insert the expansions (\ref{1b}) and (\ref{505b}) into the formula (\ref{25b}) for the heating rate. This will render a comprehensive version of the somewhat symbolic sum (\ref{27}) and will enable us to understand what the modified sum $\,\sum^{\textstyle{^{\,\sharp}}}\,$ actually means. A full calculation is presented in Appendix \ref{appD} below. Here we present a simplified sketch of that derivation.

 Recall that several different Fourier modes $\,\omega_{\textstyle{_{lmpq}}}\,$ can share the same value $\,\omega\,$. Borrowing a term from quantum mechanics, we call this $\,${\it{degeneracy of modes}}. As a prelusory exercise, we calculate dissipation at different modes, neglecting the degeneracy. In other words, suppose that all Fourier modes $\,\omega\equiv\omega_{\textstyle{_{lmpq}}}\,$ have different values. Under this simplifying assumption, the expression under the integral in  (\ref{25b}) becomes:
 \bs
 \ba
 \nonumber
 \sum_{l=2}^{\infty}(2l+1)\,\left\langle W_l(t)\,\dot{U}_l(t)\right\rangle     \qquad\qquad\qquad\qquad\qquad\qquad\qquad\qquad\qquad\qquad\qquad\qquad\qquad~~~
 \ea
 \ba
 \nonumber
 &=&\sum_{\omega,\,\omega\,'}(2l+1)~\left\langle W_l(\omega\,')\,\cos\left[\omega\,' t+\varphi_{\textstyle{_{W_l}}}(\omega\,')\right]\,(\,-\,\omega)\,U_l(\omega)\,
 \sin\left[\omega t+\varphi_{\textstyle{_{U_l}}}(\omega)\right]\,\right\rangle
 ~\\
 \nonumber\\
 \nonumber\\
 \nonumber
 &=&-\sum_{\omega,\,\omega\,'}(2l+1)~\frac{\omega}{2}\,W_l(\omega\,')\,U_l(\omega)~
 \left\langle
 \,\sin\left[\,
 (\omega\,-\,\omega\,')~t~+~
 \varphi_{\textstyle{_{U_l}}}(\omega)-\varphi_{\textstyle{_{W_l}}}(\omega\,')\,\right]~~~~\,  \right.\\
 \nonumber\\
 && ~~~~~~~~~~~~~~~~~~~~~~~~~~~~~~~~~~~~~~\,
 \left.
 +\,\sin\left[\,(\omega\,+\,\omega\,')~t~+~\varphi_{\textstyle{_{U_l}}}(\omega)+\varphi_{\textstyle{_{W_l}}}(\omega\,')\,\right]\,\right\rangle~~,\qquad
 \label{36a}
 \ea
 where $~\langle\,.\,.\,.\,\rangle~$ denotes time averaging. Of the two sine functions on the right-hand side, we would have kept only the first one, had the Fourier tidal modes been positive-definite. In the tidal theory, however, the Fourier modes $\,\omega\,=\,\omega_{\textstyle{_{lmpq}}}\,$ can assume either sign, so both sine functions must be taken into account:
 \ba
 \nonumber
 \sum_{l=2}^{\infty}(2l+1)\,\left\langle W_l(t)\,\dot{U}_l(t)\right\rangle~=~        \qquad\qquad\qquad\qquad\qquad\qquad\qquad\qquad\qquad\qquad\qquad\qquad\qquad~~~
 \ea
 \ba
 -\sum_{\omega}\,(2l+1)\,\frac{\omega}{2}\,\left\{\,W_l(\omega)\,U_l(\omega)~
 \sin\left[\varphi_{\textstyle{_{U_l}}}(\omega)-\varphi_{\textstyle{_{W_l}}}(\omega)\right]
 +\,W_l(-\omega)\,U_l(\omega)~
 \sin\left[\varphi_{\textstyle{_{U_l}}}(\omega)+\varphi_{\textstyle{_{W_l}}}(-\omega)\right]
 \,\right\}~~\,~~
 \label{36b}
 \ea
 \ba
 = \sum_{\omega}\,(2l+1)\,\frac{\omega}{2}\,W^{\,2}_l(\omega)\,k_l(\omega)~\sin\epsilon_l(\omega)~+~\sum_{\omega}\,(2l+1)\,\frac{\omega}{2}\,W_l(\omega)\,W_l(-\omega)\,k_l(\omega)
 ~\sin\epsilon\,'_l(\omega)\,~,\qquad
 \label{36c}
 \ea
 \label{36}
 \es
 where we recalled that the dynamical Love number is an even function of the Fourier mode.

 On the right-hand side of (\ref{36c}), the first sum is a much expected input coinciding with the expression obtained by other authors -- see, e.g., the first line of formula (10) in Platzman (1984).$\,$\footnote{~The second line in Platzman's formula renders oceanic and atmospheric inputs.} This input is proportional to $\,k_l(\omega)~\sin\epsilon_l(\omega)\,$, where
 \ba
 \epsilon_l(\omega)\,\equiv\,\varphi_{\textstyle{_{W_l}}}(\omega)~-~\varphi_{\textstyle{_{U_l}}}(\omega)
 \label{37}
 \ea
 is the tidal phase lag at the frequency $\,\omega\,=\,\omega_{\textstyle{_{lmpq}}}\,$.

 The second sum in (\ref{36c}) comes into being due to the fact that the Fourier modes are not positive-definite. This input contains a factor of $\,k_l(\omega)~\sin\epsilon\,'_l(\omega)\,$, where the angle $\,\epsilon\,'_l(\omega)\,$ is, generally, different from the phase lag
 (\ref{37}) appropriate to the mode $\,\omega\,=\,\omega_{\textstyle{_{lmpq}}}\,$. Indeed,
 \ba
 \nonumber
 \epsilon\,'_l(\omega)\,\equiv\,-~\left[\,\varphi_{\textstyle{_{U_l}}}(\omega)~+~\varphi_{\textstyle{_{W_l}}}(-\omega)\,\right]\,=~
 \varphi_{\textstyle{_{W_l}}}(\omega)~-~\varphi_{\textstyle{_{U_l}}}(\omega)~-~\varphi_{\textstyle{_{W_l}}}(\omega)
 ~-~\varphi_{\textstyle{_{W_l}}}(-\omega)
 ~\\ \nonumber\\
 =~\epsilon_l(\omega)~-~\left[\,\varphi_{\textstyle{_{W_l}}}(\omega)~+~\varphi_{\textstyle{_{W_l}}}(-\omega)\,\right]\,~.
 \label{38}
 \ea
 At first glance, this result is most unphysical. Usually, to calculate dissipation rate, we have to sum, over physical frequencies or over Fourier modes, terms proportional to the sines of phase lags at those modes. The addition of a finite phase to those lags looks bizarre.
 However, an accurate calculation carried out in the Appendix \ref{appD} shows that the phase consists of two parts. One is equal to $\,\left[\,(\,-\,1)^{\,l}\,-\,1\,\right]
 ~{\pi}/{2}\,$, so its presence renders an overall factor of $\,(-1)^{\,l}\,$. The other part of the phase is $\,(m\,'\,+\,m)\,\lambda\,$, so after integration over the
 surface, it results in a $\,\delta(m\,'\,+\,m)\,$ factor,  \footnote{~The finite phase assumes the value of $\,\left[(\,-\,1)^{\,l}-1\right]\, {\pi}/{2}\,+\,(m\,'+m)\lambda\,$, with the integer $\,m\,'\,$ being the order of $\,\omega\,'=\omega_{\textstyle{_{lm'p'q'}}}\,$, and $\,m\,$ being that of
 $\,\omega=\omega_{\textstyle{_{lmpq}}}\,$. The presence of $\,\left[\,(\,-\,1)^{\,l}\,-\,1\,\right]\,{\pi}/{2}\,$ in the phase is equivalent to multiplying
 the sum by $\,(-1)^{\,l}\,$. So, for even $\,l\,$, this part of the phase can be ignored. The presence of the term $\,(m\,'\,+\,m)\,\lambda\,$ in the phase tells us that, after integration over the surface, only the terms with $\,m\,'\,=\,m\,=\,0\,$ stay. Hence, after integration, we are effectively left with
 $\,\epsilon'_l(\omega)\,=\,(-1)^l\,\epsilon_l(\omega)\,\delta(m\,'\,+\,m)\,$.}  where $\,m\,$ is the second index of $\,\omega_{\textstyle{_{lmpq}}}\,=\,\omega\,$, while $\,m\,'\,$ is the second index of $\,\omega_{\textstyle{_{lm'p'q'}}}\,=\,-\,\omega~$:
 \ba
 \nonumber
 \langle P\rangle~=~\frac{1}{4\,\pi\,G\,R}~\sum_{l=2}^{\infty}(2\,l\,+\,1)\,\int dS~\langle\,W(t)~\dot{U}(t)\,\rangle~=\qquad\qquad\qquad\qquad\qquad\qquad\qquad\qquad\qquad\qquad\qquad
 \ea
\ba
 \frac{1}{4\,\pi\,G\,R}\,\sum_\omega\,(2\,l\,+\,1)~\frac{\omega}{2}\,\int dS~W_l(\omega)\,\left[\,W_l(\omega)\,+\,(-1)^l\,\delta(m\,'\,+\,m)\,W_l(-\omega)\right]~k_l(\omega)~\sin\epsilon_l(\omega)\,~.~\qquad
 \label{39}
 \label{192}
 \ea
 The indices $\,m\,$ and $\,m\,'\,$ being nonnegative (see the equation \ref{1}), the emergence of $\,\delta(m\,'\,+\,m)\,$ indicates that the summation
 in the second part must be reduced to $\,m\,=\,m\,'\,=\,0~$:
 \ba
 \nonumber
 \langle P\rangle~=~\frac{1}{4\,\pi\,G\,R}~\int dS~
 \left[\,
 {\sum_{\omega\,=\,\omega_{\textstyle{_{_{lmpq}}}}}}~\frac{\omega}{2}~(2l+1)~k_l(\omega)~\sin\epsilon_l(\omega)~W^2_l(\omega)\right.~~~~~~~~~~~~~~~~~~~~~~~~~\\
 \nonumber\\
 \left.
 +~{\sum_{\omega\,=\,\omega_{\textstyle{_{_{l0pq}}}}}}~\frac{\omega}{2}~(2l+1)~k_l(\omega)~\sin\epsilon_l(\omega)~
 ~(-1)^l\,W_l(\omega)\,W_l(-\omega)
  \,\right]\,~.\quad
 \label{40}
 \label{193}
 \ea
 We then see what the superscript $\,\sharp\,$ introduced in (\ref{27} - \ref{30}) actually implies:
 \ba
 {\sum_\omega}^{\textstyle{^{\textstyle\,\sharp~}}}.\,.\,.~W_l^2(\omega)~\equiv~{\sum_{\omega\,=\,\omega_{\textstyle{_{_{lmpq}}}}}}~.\,.\,.~W^2_l(\omega)~+~
 {\sum_{\omega\,=\,\omega_{\textstyle{_{_{l0pq}}}}}}~.\,.\,.~(-1)^l~
 W_l(\omega)~W_l(-\omega)\,~,
 \label{41}
 \ea
 where the first sum on the right-hand side is complete (i.e., goes over all modes), while the second sum is only over the modes with a vanishing second index.

 Now, what is $\,W_l(\omega)\,$? Na{\"i}vely, $\,W_l(\omega)\,\equiv\,W_l(\omega_{\textstyle{_{lmpq}}})\,$ should be the real magnitude of the term $\,W_{\textstyle{_{lmpq}}}\,$ of Kaula's series (\ref{1b}). In reality, we have degeneracy of modes, so in each of the two Fourier series (for $\,W\,$ and for $\,U\,$) we first must group together the terms corresponding to each actual value of mode, and only afterward should we multiply the series by one another and perform time averaging.
 This calculation is presented in Appendix \ref{appD}.

 \section{Heat production over tidal modes. An accurate derivation of the formula (\ref{196})}\label{appD}

  As was explained above, the actual $\,W_l(\omega)\,$ will be a sum of several $\,W_l(\omega_{\textstyle{_{lmpq}}})\,$, over all the sets $\,lmpq\,$ furnishing the same value of $\,\omega\,$. Thence, in (\ref{41}) and elsewhere, summation over $\,\omega\,$ will be not a sum over $\,lmpq\,$, but a sum over all $\,${\it{distinct}}$\,$ values of $\,\omega_{\textstyle{_{lmpq}}}\,$.

 These details would be irrelevant, were we summing terms linear in $\,W_l(\omega_{\textstyle{_{lmpq}}})\,$. In that case, to group terms and then to sum the groups would be the same as to sum all the terms at once. We however are dealing with the expression (\ref{41}) quadratic in $\,W_l\,$, wherefore the said details matter a lot. Incorporation of those complicates the calculation technically, though the main idea remains the same as in (\ref{40}).

 \subsection{Prefatory algebra}

 First, it would be convenient to rewrite the formula (\ref{1b}) for the perturbing potential as
 \ba
 W(\eRbold\,,\;\erbold^{\;*})\,=\sum_{l=2}^{\infty} ~W_l~~, ~\qquad~ ~\qquad~ ~\qquad~ ~\qquad~ ~\qquad~ ~\qquad~ ~\qquad~ ~\qquad~
 \label{A10}
 \ea
 where
 \ba
 W_l\,=\sum_{m'p'q'}\,W_{lm'p'q'}\,
  =~\sum_{m'=0}^{l}\;\sum_{p'=0}^{l}\;\sum_{j=\,-\,\infty}^{\infty}
  A_{lm'p'q'}~\cos\left(v_{lm'p'q'}\,-~m'\,(\lambda\,+\,\theta)~+~\psi_{lm'}\right)\,~,
  \label{A11}
 \ea
 \ba
 A_{lm'p'q'}~=~-~\frac{G\,M^*}{a}\;\left(\,\frac{R}{a}\,\right)^{\textstyle{^l}}\;\frac{(l - m')!}{(l + m')!}\;
  \left(\,2\,-\,\delta_{0m'}\,\right)\;P_{lm'}(\sin\phi)\;F_{lm'p'}(i)\;G_{lp'q'}(e)~~,
 \label{A12}
 \ea
 \ba
 \psi_{lm'}~=~\left[\,(\,-\,1)^{\,l-m'}\,-\,1\,\right]~\frac{\pi}{4}\;~.~\qquad~ ~\qquad~ ~\qquad~ ~\qquad~ ~\qquad~ ~\qquad~ ~\qquad~
 \label{A13}
 \ea
 Similarly, the formula (\ref{505b}) for the additional tidal potential at $\,\erbold=\Rbold\,$ should be cast as
 \ba
 U(\eRbold\,,\;\erbold^{\;*})\,=\sum_{l=2}^{\infty}\;U_l~~, ~\qquad~ ~\qquad~ ~\qquad~ ~\qquad~ ~\qquad~ ~\qquad~ ~\qquad~ ~\qquad~
 \label{A14}
 \ea
 where
 \ba
 U_l\,=\sum_{mpq}\,U_{lmpq}\,
 =\sum_{m=0}^{l}\;\sum_{p=0}^{l}\;\sum_{q=\,-\,\infty}^{\infty}
 B_{lmpq}~\cos\left(v_{lmpq}\,-~m\,(\lambda\,+\,\theta)~+~\psi_{lm}~-~\epsilon_{lmpq}\right)~,~\quad
 \label{A15}
 \ea
 \ba
 B_{lmpq}~=~k_l(\omega_{lmpq})~A_{lmpq}\;~,
 \label{A16}
 \ea
 $\psi_{lm}\,$ is given by (\ref{A13}), while $\,\epsilon_{lmpq}\,$ is the phase lag (\ref{506b}).
 In (\ref{A16}) we omitted the multiplier $\,\left(\,{R}/{r}\,\right)^{\textstyle{^{l+1}}}\,$, as we are interested in the values of $\,U\,$ over the surface where $\,r=R\,$.

 The product of the quantities $\,W_l\,$ and $\,\dot{U}_l\,$ reads as
 \ba
 W_l\,\,\dot{U}_l\,= ~\qquad~ ~\qquad~ ~\qquad~ ~\qquad~ ~\qquad~ ~\qquad~ ~\qquad~ ~\qquad~ ~\qquad~ ~\qquad~ ~\qquad~ ~\qquad~ ~\qquad~ ~\qquad~ \,~\qquad~
 \label{A17}
 \ea
 \ba
 \nonumber
 -\sum_{m'p'q'}\sum_{mpq}\,A_{lm'p'q'}\,\cos\left(v_{lm'p'q'}-\,m'(\lambda+\theta)+\psi_{lm'}\right)
 \omega_{lmpq}B_{lmpq}~\sin\left(v_{lmpq}-m(\lambda+\theta)+\psi_{lm}-\epsilon_{lmpq}\right)
 \ea
 \ba
 \nonumber
 =~-~\frac{1}{2}\,\sum_{m'p'q'}\sum_{mpq}\,A_{lm'p'q'}\,\omega_{lmpq}\,B_{lmpq}~
 \sin\left(\,v_{lm'p'q'}+v_{lmpq}-(m'+m)\,(\lambda+\theta)+\,\psi_{lm'}+\,\psi_{lm}\,-\,\epsilon_{lmpq}\,\right)
 \ea
 \ba
 +~\frac{1}{2}~\sum_{m'p'q'}\sum_{mpq}\,A_{lm'p'q'}~\omega_{lmpq}~B_{lmpq}~\sin\left(~v_{lm'p'q'}-v_{lmpq}-\,(m'-m)\,(\lambda+\theta)\,+\,\psi_{lm'}\,-\,\psi_{lm}\,+\,\epsilon_{lmpq}~\right)~~.
 \nonumber
 \ea
 The formula (\ref{25b}) for the heating rate contains a time-average of the product $\,W_l\,\dot{U}_l\,$, integrated over the surface. In the first sum of (\ref{A17}), integration over the longitude $\,\lambda\,$ will leave only the terms with $\,m'\,=\,m\,=\,0\,$. (Recall that $\,m'\,$ and $\,m\,$ are nonnegative.) In the second sum, integration over $\,\lambda\,$ will eliminate all terms except the ones with $\,m'\,=\,m\,$. Therefore,
 \begin{subequations}
 \ba
 \nonumber
 W_l\,\,\dot{U}_l\,=
 &-&\frac{1}{2}~\sum_{p'q'}\sum_{pq}\,A_{l0p'q'}~\omega_{l0pq}~B_{l0pq}~
 \sin\left(~v_{l0p'q'}\,+\,v_{l0pq}\,+\,2\,\psi_{l0}\,-\,\epsilon_{l0pq}~\right)\\
 \nonumber\\
 &+&\frac{1}{2}~\sum_m\sum_{p'q'}\sum_{pq}\,A_{lmp'q'}~\omega_{lmpq}~B_{lmpq}~\sin\left(~v_{lmp'q'}\,-\,v_{lmpq}\,+\,\epsilon_{lmpq}~\right)~+~~.~.~.~~~,~\qquad\qquad~
 \label{A18a}
 \ea
 the ellipsis denoting the
 terms which are to vanish after integration over the longitude $\,\lambda\,$. The presence of the phase $\,2\psi_{l0}=
 \left[(\,-\,1)^{\,l}-1\right]\,{\pi}/{2}\,$ in the first sum is equivalent to multiplying the sum by $\,(-1)^{\,l}\,$. Combined with the formula
 (\ref{2}) for $\,v_{lmpq}\,$, this observation gives us:
 \ba
 \nonumber
 W_l\,\dot{U}_l\,= ~\qquad~\qquad~\qquad~\qquad~\qquad~\qquad~\qquad~\qquad~\qquad~\qquad~\qquad~\qquad~\qquad~\qquad~\qquad~\qquad~\qquad~~\\
     \nonumber\\
     \nonumber
 -~\frac{(-1)^{\,l}}{2}~\sum_{p'q'}\sum_{pq}\,A_{l0p'q'}~\omega_{l0pq}~B_{l0pq}~
 \sin\left(\,2\,(l-p\,'-p)~\omega+\,(2\,(l-p\,'-p)\,+\,q\,'\,+\,q)\,{\cal{M}}-\,\epsilon_{l0pq}\right) ~\qquad~\\
 \nonumber\\
 +~\frac{1}{2}\,\sum_m\sum_{p'q'}\sum_{pq}\,A_{lmp'q'}~\omega_{lmpq}~B_{lmpq}~\sin\left(\,2\,(p-p\,')\,\omega+\,(2p-q-2p\,'+q\,')\,{\cal{M}}+\,
 \epsilon_{lmpq}\,\right)\,+~.~.~.~~,~\qquad
 \label{A18b}
 \ea
 ${\cal{M}}\,$ and $\,\omega\,$ being the mean anomaly and the argument of the pericentre of the perturber. In Section \ref{diff}, we {\it{defined}} the mean motion $\,n\,$ as the mean anomaly rate $\,\stackrel{\bf\centerdot}{\cal{M}\,}\,$. Thence \footnote{~The epoch $\,t_0=0\,$ is set at the instant of periapse crossing whence $\,{\cal{M}}\,$ is reckoned. So we write the mean anomaly simply as $\,{\cal{M}}\,=\,n\,t\,$. Calculating the present rate of dissipation, we average over one or several
 cycles of tidal flexure, and not over the entire time span since the epoch (which may be distant). For a Keplerian orbit, $\,{\cal{M}}\,$ changes uniformly with time and always assumes a value of $\,2\,\pi\,N\,$ at a periapse crossing, with $\,N\,$ being integer. So we can always set $\,{\cal{M}}=0\,$ at a recent crossing and can reckon time from there (i.e, set $\,t_0=0\,$ at that moment).\\
 $\,\left.\quad\right.\,$ In realistic situations, $\,{\cal{M}}\,=\,\int_{t_0}^t n(t\,')\,dt\,'~$. ~(No $\,{\cal{M}}_0\,$ term here, as we agreed to define $\,n\,$ as $\,\stackrel{\bf\centerdot}{\cal{M}\,}$.) Despite the apsidal precession, $\,{\cal{M}}\,$ will still be changing by $\,2\pi\,$ between two subsequent periapse crossings. It
 however will no longer be right to substitute the integral with a product of $\,n\,$ by a time interval. Despite this, we still can set $\,t_0=0\,$ and $\,{\cal{M}}=0\,$ at a recent crossing, and can approximate the integral with $\,n\,t\,$ over several rotations sufficient for our averaging.}
 \ba
 \nonumber
 W_l\,\,\dot{U}_l\,=~\qquad~\qquad~\qquad~\qquad~\qquad~\qquad~\qquad~\qquad~\qquad~\qquad~\qquad~\qquad~\qquad~\qquad~\qquad~\qquad~\qquad~\\
     \nonumber\\
     \nonumber
 \,\frac{(-1)^{\,l}}{2}\,\sum_{p'q'}\sum_{pq}A_{l0p'q'}~\omega_{l0pq}~B_{l0pq}~
 \sin\left(\,-\,2\,(l-p\,'-p)~\omega-\,(2\,(l-p\,'-p)\,+\,q\,'\,+\,q)\,n\,t\,+\,\epsilon_{l0pq}\,\right)\qquad\quad\\
 \nonumber\\
 \left.\,\right.+\,\frac{1}{2}~\sum_m\sum_{p'q'}\sum_{pq}\,A_{lmp'q'}~\omega_{lmpq}~B_{lmpq}~\sin\left(~2\,(p-p\,')\,\omega+\,(2p-q-2p\,'+q\,')\,n\,t\,+\,\epsilon_{lmpq}
 ~\right)\,+~.~.~.~~,\qquad
 \label{A18c}
 \ea
 where we have moved the ``minus" sign inside the sine function in the first sum.

 Selecting the secular terms in the above sum, we notice that some of those contain not only the phase lag but also finite phases proportional to the
 initial value of the pericentre. Naturally, such ugly terms come in pairs containing equal but opposite initial phase. Each such pair renders a mutual input proportional to the sine of the phase lag $\,\epsilon_{lmpq}\,$.

 To demonstrate this, single out the initial phases in each term of (\ref{A18c}). The second sum in (\ref{A18c}) will be split into two subsums,
 one corresponding to $\,p\,'\neq p\,$, another to $\,p\,'=p~$:
 \ba
 \nonumber
 W_l\,\,\dot{U}_l&=&
 \frac{(-1)^{\,l}}{2}\,\sum_{p'q'}\sum_{pq}A_{l0p'q'}~\omega_{l0pq}~B_{l0pq}~
 \sin(\,-\,2\,(l-p\,'-p)~\omega_0\,-\,(\omega_{\textstyle{_{l0p'q'}}}\,+\,\omega_{\textstyle{_{l0pq}}})\,t
 \,+\,\epsilon_{l0pq}\,)\\
 \nonumber\\
 \nonumber
 &&+~\frac{1}{2}~\sum_m\sum_{p\,q}\sum_{\stackrel{\textstyle{^{~p\,'\,q\,'}}}{p\,'\neq p~}}\,A_{lmp'q'}~\omega_{lmpq}~B_{lmpq}~\sin(~2\,(p-p\,')\,\omega_0\,+\,(\omega_{\textstyle{_{lmp'q'}}}\,-
 \,\omega_{\textstyle{_{lmpq}}})\,t\,+\,\epsilon_{lmpq}~)\\
 \nonumber\\
 &&\left.\,\right.+\,\frac{1}{2}~\sum_m\sum_{pqq'}\,A_{lmpq'}~\omega_{lmpq}~B_{lmpq}~\sin\left(\,(q\,'\,-\,q)\,n\,t\,+\,\epsilon_{lmpq}~\right)\,+~.~.~.\,~\qquad
 \label{A18d}
 \label{ksa}
 \ea
  \label{A18}
 \end{subequations}

 \subsection{Selection of secular terms in (\ref{ksa}). The role of rheology}

 The purpose of the below-presented algebraic development of (\ref{ksa}) is to single out secular terms, i.e., the ones whose time average vanishes. Involvement of time
 averages immediately brings in rheological properties of the tidally perturbed body. Indeed, the averaging remains legitimate over so long a time scale as keeps the
 rheological response linear, not plastic. Very roughly, we may link the borderline to the Maxwell time of the mantle, $\,\tau_{_M}\,$. If the period of the apsidal motion
 is shorter than $\,\tau_{_M}\,$ or, at least, does not exceed it by more than an order or two, we have to keep the apsidal rate $\,\dot{\omega}\,$ in the expression
 (\ref{504}) for the Fourier mode and, accordingly, have to average the product $\,W_l\,\,\dot{U}_l\,$ over the apsidal period. If however the apsidal timescale exceeds
 $\,\tau_{_M}\,$ considerably, then averaging of the product $\,W_l\,\,\dot{U}_l\,$ over such a time scale will lack physical meaning, and the apsidal rate in the expression
 (\ref{504}) should be set zero. As we shall see below, this will make difference in our further selection of the secular parts of the first two subsums in (\ref{ksa}).

 \subsubsection{Symmetrisation}

 Processing the third subsum in (\ref{ksa}) is easy: we retain the secular terms (those with $\,q=q\,'\,$) and, in anticipation of time averaging, ignore the rest of the subsum.

 The second subsum in (\ref{ksa}) can be split into two equal halves, with $\,pq\,$ and $\,p\,'q\,'\,$ swapped in the second half:
 \bs
 \ba
 \frac{1}{2}~\sum_m\,\sum_{\stackrel{\textstyle{^{\,p\,,\,p\,'}}}{p\,'\neq p~}}\,\sum_{q\,,\,q\,'}\,~A_{lmp\,'q\,'}~\omega_{lmpq}~B_{lmpq}~\sin(~2\,(p-p\,')\,\omega_0\,+\,(\omega_{\textstyle{_{lmp\,'q\,'}}}\,-\,\omega_{\textstyle{_{lmpq}}})\,t\,+\,
 \epsilon_{lmpq}~)~=
 \nonumber
 \ea
 \ba
 \nonumber
 \frac{1}{4}~\sum~\left[\,A_{lmp\,'q\,'}~\omega_{lmpq}~B_{lmpq}~\sin(~2\,(p-p\,')\,\omega_0\,+\,(\omega_{\textstyle{_{lmp\,'q\,'}}}\,-
 \,\omega_{\textstyle{_{lmpq}}})\,t\,+\,\epsilon_{lmpq}~)~+ \right.~\quad~\qquad~\qquad\\
 \nonumber\\
 \left. A_{lmpq}~\omega_{lmp\,'q\,'}~B_{lmp\,'q\,'}
 ~\sin(~2\,(p\,'-p)\,\omega_0\,+\,(\omega_{\textstyle{_{lmpq}}}\,-
 \,\omega_{\textstyle{_{lmp\,'q\,'}}})\,t\,+\,\epsilon_{lmp\,'q\,'}~)\,\right]~=~\qquad\qquad
 \label{}
 \ea
 \ba
 \nonumber
 \frac{1}{4}~\sum~\left[\,A_{lmp\,'q\,'}~\omega_{lmpq}~A_{lmpq}~k_l(\omega_{lmpq})~\sin(~2\,(p-p\,')\,\omega_0\,+\,(\omega_{\textstyle{_{lmp\,'q\,'}}}\,-
 \,\omega_{\textstyle{_{lmpq}}})\,t\,+\,\epsilon_{lmpq}~)~+ \right.\quad\qquad\\
 \nonumber\\
 \left. A_{lmpq}~\omega_{lmp\,'q\,'}~A_{lmp\,'q\,'}~k_l(\omega_{lmp\,'q\,'})
 ~\sin(~2\,(p\,'-p)\,\omega_0\,+\,(\omega_{\textstyle{_{lmpq}}}\,-
 \,\omega_{\textstyle{_{lmp\,'q\,'}}})\,t\,+\,\epsilon_{lmp\,'q\,'}~)\,\right]~\,.\qquad
 \label{}
 \ea
  \label{above}
 \es
 Prior to time averaging of (\ref{above}), we, generally, have $~\omega_{lmpq}~k_l(\omega_{lmpq})\,\neq\,\omega_{lmp\,'q\,'}~k_l(\omega_{lmp\,'q\,'})~$ and $~\epsilon_{lmpq}
 \,\neq\,\epsilon_{lmp\,'q\,'}~$. The averaging, however, leaves us only the terms with $\,\omega_{lmpq}\,=\,\omega_{lmp\,'q\,'}\,$. Hence the secular part of the above
 expression is:
 \ba
 \nonumber
 \frac{1}{4}\sum_m\sum_{\stackrel{\textstyle{^{\,p\,,\,p\,'}}}{p\,'\neq p}}\sum_{q}\,\left[\,A_{lmp\,'q\,'}\,\right]_{\textstyle{_{
 \omega_{\textstyle{_{_{lmpq}}}}=\,\omega_{\textstyle{_{_{lmp\,'q\,'}}}}
 }}}\omega_{lmpq}\,B_{lmpq}\,\left[\,
 \sin (2\,(p-p\,')\,\omega_0\,+\,\epsilon_{lmpq})\,+\,\sin (\,-2\,(p-p\,')\,\omega_0+\epsilon_{lmpq})\,\right]~~\,~\qquad\\
 \nonumber\\
 =~\frac{1}{2}~\sum_m\,\sum_{\stackrel{\textstyle{^{\,p\,,\,p\,'}}}{p\,'\neq p~}}\,\sum_{q}\,\left[\,A_{lmp\,'q\,'}\,\right]_{\textstyle{_{
 \omega_{lmpq}\,=\,\omega_{lmp\,'q\,'}
 }}}~\omega_{lmpq}~B_{lmpq}~\cos(\,2\,(p\,'\,-\,p)\,\omega_0\,)~\sin\epsilon_{lmpq}~~.~\qquad~\,~\qquad\qquad\qquad
 \label{mug}
 \ea
 where the value of $\,q\,'\,$ is defined by the fact that the secular terms obey
 \ba
 0\,=\,\omega_{lmpq}\,-\,\omega_{lmp\,'q\,'}\,=\,-\,2\,(p-p\,')\,\dot{\omega}\,+~\left[\,-\,2\,(p\,-\,p\,')\,+\,q\,-\,q\,'\,\right]\,n~~.
 \label{gum}
 \ea

 Similarly to (\ref{above}), the first subsum in (\ref{ksa}) can be rewritten as
 \bs
 \ba
 \nonumber
 \frac{(-1)^{\,l}}{2}\,\sum_{\,p\,',\,q\,'\,}\sum_{\,p,\,q}A_{l0p\,'q\,'}~\omega_{l0pq}~B_{l0pq}~
 \sin(\,-\,2\,(l-p\,'-p)~\omega_0\,-\,(\omega_{\textstyle{_{l0p\,'q\,'}}}\,+\,\omega_{\textstyle{_{l0pq}}})\,t
 \,+\,\epsilon_{l0pq}\,)~=~\qquad\\
 \nonumber\\
 \nonumber\\
 \nonumber
 \frac{(-1)^{\,l}}{4}\sum_{\stackrel{\textstyle{^{\,p\,,\,p\,'\,}}}{
 }}\sum_{\,q\,,\,q\,'}~\left[\,A_{l0p\,'q\,'}~\omega_{l0pq}~B_{l0pq}~
 \sin (\,-\,2\,(l-p\,'-p)\,\omega_0\,-\,(\omega_{\textstyle{_{l0p'q'}}}+\omega_{\textstyle{_{l0pq}}})\,t\,
 +\,\epsilon_{l0pq}\,) \right.~\qquad~\qquad~\\
 \nonumber\\
 \left.~+~\,A_{l0pq}~\omega_{l0p\,'q\,'}~B_{l0p\,'q\,'}~\sin (\,-\,2\,(l-p-p\,')\,\omega_0\,-\,(\omega_{\textstyle{_{l0pq}}}\,+\,\omega_{\textstyle{_{l0p\,'q\,'}}})~t\,+\,\epsilon_{l0p\,'q\,'})\,\right]~=~\qquad~
 \label{}\\
 \nonumber\\
 \nonumber\\
 \nonumber
\frac{(-1)^{\,l}}{4}\sum_{\stackrel{\textstyle{^{\,p\,,\,p\,'\,}}}{
 }}\sum_{\,q\,,\,q\,'}~\left[\,A_{l0p\,'q\,'}~\omega_{l0pq}~A_{l0pq}~k_l(\omega_{l0pq})\;
 \sin (\,-\,2\,(l-p\,'-p)\,\omega_0\,-\,(\omega_{\textstyle{_{l0p'q'}}}+\omega_{\textstyle{_{l0pq}}})\,t\,
 +\,\epsilon_{l0pq}\,) \right.\quad\quad \\
 \nonumber\\
 \left.~+~\,A_{l0pq}~\omega_{l0p\,'q\,'}~A_{l0p\,'q\,'}~k_l(\omega_{l0p\,'q\,'})~\sin (\,-\,2\,(l-p-p\,')\,\omega_0\,-\,(\omega_{\textstyle{_{l0pq}}}\,+\,\omega_{\textstyle{_{l0p\,'q\,'}}})~t\,+\,\epsilon_{l0p\,'q\,'})\,\right]~\,.\qquad
 \label{belladonna}
 \ea
 \label{below}
 \es
 After time-averaging of (\ref{belladonna}), the factor $\,\omega_{l0p\,'q\,'}\,$ accompanying the second sine becomes $~-\,\omega_{l0pq}\,$, while $\,\epsilon_{l0p\,'q\,'}\,$ becomes $~-\,\epsilon_{l0pq}\,$. Hence the secular part of the above expression is:
 \ba
 \frac{(-1)^{\,l}}{2}\,\sum_{\stackrel{\textstyle{^{~p\,,\,p\,'}}}{p\,'\neq p~}}\,\sum_{q}\,\left[A_{l0p\,'q\,'}\right]_{\textstyle{_{\omega_{l0pq}\,=
 \,-\,\omega_{l0p\,'q\,'}}}}~\omega_{l0pq}~B_{l0pq}~\cos(\,2\,(l\,-\,p\,'\,-\,p)\,\omega_0\,)~\sin\epsilon_{l0pq}~\,~,~\qquad~\,~
 \label{cub}
 \ea
 where the value of $\,q\,'\,$ is determined from the condition of the terms being secular:
 \ba
 0\,=\,\omega_{lmpq}\,+\,\omega_{lmp\,'q\,'}\,=\,2\,(l\,-\,p\,-\,p\,')\,\dot{\omega}\,+~\left[\,2\,(l\,-\,p\,-\,p\,')\,+\,q\,+\,q\,'\,\right]\,n~~.
 \label{pug}
 \ea

 \subsubsection{The case of a quiescent pericentre}

 When the period of the apsidal motion is much (by orders of magnitude) longer than the Maxwell time $\,\tau_{_M}\,$, we cannot count on a linear tidal response at such long
 time scales. Instead, we should treat the pericentre as fixed and should nullify $\,\dot{\omega}\,$ in the expression (\ref{504}) for the Fourier mode, as well as in
 (\ref{gum}) and (\ref{pug}). From (\ref{mug}) and (\ref{gum}), we see that the time average of the second subsum in (\ref{A18d}) acquires the form of
 \ba
 \frac{1}{2}~\sum_m\,\sum_{\stackrel{\textstyle{^{\,p\,,\,p\,'}}}{p\,'\neq p~}}\,\sum_{q}\,\left[\,A_{lmp\,'q\,'}\,\right]_{\textstyle{_{
  q\,'\,=\,q-2(p-p\,')
 }}}~\omega_{lmpq}~B_{lmpq}~\cos(\,2\,(p\,'\,-\,p)\,\omega_0\,)~\sin\epsilon_{lmpq}~~.
 \label{guido}
 \ea
 From (\ref{cub}) and (\ref{pug}), we infer that the average of the first subsum in (\ref{A18d}) is
 \ba
 \frac{(-1)^{\,l}}{2}\,\sum_{\stackrel{\textstyle{^{~p\,,\,p\,'}}}{
 }}\,\sum_{q}\,\left[A_{l0p\,'q\,'}\right]_{\textstyle{_{
 q\,'\,=\,-q-2(l-p-p\,')}}}~\omega_{l0pq}~B_{l0pq}~\cos(\,2\,(l\,-\,p\,'\,-\,p)\,\omega_0\,)~\sin\epsilon_{l0pq}~\,~.~\qquad~\,~
 \label{vero}
 \ea
 Altogether, the secular part of the expression (\ref{A18d}) acquires the form of
 \bs
 \ba
 \nonumber
 ^{\textstyle{^{(\dot{\omega}=0)}}}\langle\,W_l\,\,\dot{U}_l\,\rangle &=&\frac{(-1)^{\,l}}{2}~\sum_{p\,q}\sum_{\stackrel{\textstyle{^{~p\,'}}}{
 }}\,\left[A_{l0p\,'q\,'}\right]_{\textstyle{_{
 q\,'\,=\,-q-2(l-p-p\,')}}}~~\omega_{l0pq}~B_{l0pq}~\cos(\,2\,(l\,-\,p\,'\,-\,p)\,\omega_0\,)~\sin\epsilon_{l0pq}
 \\
 \nonumber\\
 \nonumber
 &+&\frac{1}{2}~\sum_m\sum_{p\,q}\sum_{~p\,'\neq p}\,\left[\,A_{lmp\,'q\,'}\,\right]_{\textstyle{_{
  q\,'\,=q-2(p-p\,')
 }}}~\omega_{lmpq}~B_{lmpq}~\cos(\,2\,(p\,'\,-\,p)\,\omega_0\,)~\sin\epsilon_{lmpq}~\\
 \nonumber\\
 &+&\frac{1}{2}~\sum_m\sum_{pq}\,A_{lmpq}~\omega_{lmpq}~B_{lmpq}~\sin\epsilon_{lmpq}~+\,~.~.~.~\,~,\qquad
 \label{quiescenta}
 \ea
 where the last two sums may now be reunited into one:
 \ba
 \nonumber
 ^{\textstyle{^{(\dot{\omega}=0)}}}\langle\,W_l\,\,\dot{U}_l\,\rangle &=&\frac{(-1)^{\,l}}{2}\sum_{p\,p\,'\,q} \left[A_{l0p\,'q\,'}\right]_{\textstyle{_{
 q\,'=-q-2(l-p-p\,')}}}~\omega_{l0pq}~B_{l0pq}~\cos(\,2\,(l-p\,'-p)\,\omega_0\,)~\sin\epsilon_{l0pq}
 \\
 \nonumber\\
 &+&\frac{1}{2}\,\sum_m\sum_{p\,p\,'\,q}\,\left[A_{lmp\,'q\,'}\right]_{\textstyle{_{
  q\,'\,=q-2(p-p\,')
 }}}\,\omega_{lmpq}~B_{lmpq}\,\cos(\,2\,(p\,'-p)\,\omega_0\,)\,\sin\epsilon_{lmpq}\,+\,.\,.\,.\,~,~\qquad~~~
 \label{quiescentb}
 \ea
 \label{quiescent}
 \es
 the ellipsis denoting the
 terms which will vanish after integration over the longitude $\,\lambda\,$.

  \subsubsection{The case of a moving pericentre}

 Consider a situation where the evolution rate of the pericentre is relatively fast, i.e., not much slower than the Maxwell timescale of the mantle material. In this
 situation, the time derivative $\,\dot{\omega}\,$ should be taken into account in the expression for the Fourier tidal modes.

 Just as in the case of an idle pericentre, here we begin with the second subsum in (\ref{A18d}). As we saw above, its secular part can be transformed into the expression (\ref{mug}), where the index $\,q\,'\,$ is no longer independent but is constrained by the condition (\ref{gum}). For a vanishing $\,\dot{\omega}\,$, that condition became $\,q\,'\,=\,q\,-\,2\,(p\,-\,p\,')\,$, thus making the secular part look as (\ref{guido}).

 Now the situation is different, and the equality $\,q\,'\,=\,q\,-\,2\,(p\,-\,p\,')\,$ no longer ensues from (\ref{gum}).
There, though, is an alternative method of satisfying (\ref{gum}). For a chosen set of integers $\,\{p,\,p\,',\,q,\,q\,'\}\,$, it is possible to pick up commensurate values of $\,\dot{\omega}\,$ and $\,n\,$, obeying
 \ba
 \frac{\,\dot{\omega}\,}{n\,}\,=~-~\,\frac{\,2\,(p\,-\,p\,')\,-\,(q\,-\,q\,')\,}{2\,(p\,-\,p\,')}~=~-~1~+~\frac{q\,-\,q\,'}{\,2\,(p\,-\,p\,')\,}
 \,~.
 \label{f}
 \ea
 This equality will then be true also for any set $\,\{Kp,\,Kp\,',\,Kq,\,Kq\,'\}\,$ with a nonzero integer $\,K\,$. We should not, however, accept the formulae at their face value, but should always keep in mind the physical meaning of the orbital elements of our concern. While $\,${\it{formally}}$\,$ solution of a perturbed Keplerian problem
 does not prohibit the pericentre rate from being comparable to $\,n\,$, such a proximity is unlikely to be attained in real life -- except in the situations where an orbit
 is undegoing a rapid and considerable change. Usually, $\,\dot\omega\,$ is $\,${\it{much}}$\,$ smaller in absolute value than $\,n\,$. Therefore, $\,(q-q\,')/(2(p-p\,'))\,$ should be very close to, but not exactly equal to $\,1\,$. There are no combinations of integers to make this happen, because $\,2(p-p\,')\,$ takes values between $\,-2\,l\,$ and $\,2\,l\,$, so the smallest absolute value of the ratio $\,(q-q\,')/(2(p-p\,'))\,$ is $\,1/(2\,l)\,$. For $\,l=2\,$, this would make (\ref{f}) look $\,{\,\dot{\omega}\,}/{n\,}\,=~-~\,3/4\,$, while for a larger $\,l\,$ the ratio $\,{\,\dot{\omega}\,}/{n\,}\,$ would be even closer to unity, a situation irrelevant to actual settings. Therefore, the entire second subsum in (\ref{A18d}) must drop out.

 Consider the first subsum in (\ref{A18d}). As we saw, its secular part can be transformed into the expression (\ref{cub}), where the index $\,q\,'\,$ is constrained by the condition (\ref{pug}). For a vanishing $\,\dot{\omega}\,$, the condition used to entail $\,q\,'=\,-\,q-2(l-p-p\,')\,$, so the secular part used to assume the form (\ref{vero}). This will no longer be so in the presence of $\,\dot{\omega}\,$. We then have two ways of satisfying the condition (\ref{pug}). One is to
 impose, simultaneously, the conditions
 \ba
 l\,=\,p\,+\,p\,'\,~~,\qquad q\,=\,-\,q\,'\,~,
 \label{c}
 \ea
 in which case the secular part of the discussed subsum will be simplified to
 \ba
 \frac{(-1)^{\,l}}{2}\,\sum_{p\,q}\,\left[A_{l0p\,'q\,'}\right]_{\textstyle{_{\,q\,'\,=\,-q}}}^{\textstyle{^{\,l=p\,'+p}}}~\,\omega_{l0pq}~B_{l0pq}~\sin\epsilon_{l0pq}~\,~.~\qquad~\,~
 \label{subsum}
 \ea
 Another way is to pick up, for some set $\,\{l,\,p,\,p\,',\,q,\,q\,'\}\,$, commensurate $\,\dot{\omega}\,$ and $\,n\,$ obeying
 \ba
 \frac{\,\dot{\omega}\,}{n\,}~=~-~1~-~\,\frac{q\,+\,q\,'}{2\,(l\,-\,p\,-\,p\,')}\,~.
 \label{}
 \ea
 This will work also for any set $\{Kl,\,Kp,\,Kp\,',\,Kq,\,Kq\,'\}$ with a nonzero integer $K$.
 Just as in the formula (\ref{f}), this option is not viable, as no set $\{l,\,p,\,p\,',\,q,\,q\,'\}$ makes the
 pericentre rate much lower than $\,n\,$. So the only realistic possibility is implemented by (\ref{c} - \ref{subsum}).

 All in all, the secular part of the expression (\ref{A18d}) acquires the form
 \ba
 \nonumber
 ^{\textstyle{^{(\dot{\omega}\neq0)}}}\langle W_l\,\,\dot{U}_l\rangle\,= ~\qquad~\qquad~\qquad~\qquad~\qquad~\qquad~\qquad~\qquad~\qquad~\qquad~\qquad~\qquad~\qquad~\qquad~\qquad~\qquad\\
 \nonumber\\
 \frac{(-1)^{\,l}}{2}\,\sum_{\,p\,q\,}\,\left[A_{l0p\,'q\,'}\right]_{\textstyle{_{q\,'\,=\,-q}}}^{\textstyle{^{\,l=p\,'+p}}}~\omega_{l0pq}~B_{l0pq}~\sin\epsilon_{l0pq}
 ~+~\frac{1}{2}~\sum_m\sum_{\,p\,q\,}\,A_{lmpq}~\omega_{lmpq}~B_{lmpq}~\sin\epsilon_{lmpq}\,+~.~.~.~~,\quad
 \label{moving}
 \ea
 the ellipsis signifying the part vanishing after integration over the longitude $\,\lambda\,$.

 \subsubsection{Comparison of the two cases}

 Comparing formulae (\ref{quiescentb}) and (\ref{moving}), we see that a whole group of terms, absent in the moving-pericentre case, shows up when the pericentre stays idle.
 These are the terms obtained by summation over $\,p\,'\,$. When averaging over the apsidal period is legitimate, most of them vanish and the value of $\,p\,'\,$ gets fixed.
 Otherwise these terms stay.

 As we explained above, averaging is physically meaningful over timescales which do not exceed the Maxwell time by too many orders of magnitude. Over longer periods, linear
 viscoelastic reaction is not guaranteed. Specifically, we cannot be sure that full rebound is possible. At such long times, it would be physical to
 regard the periapse idle -- which will change the averaging procedure and will leave unaveraged the terms numbered with all $\,p\,'\,$.

 To understand whether the periapse should be regarded quiescent or not, its period must be compared with the Maxwell time. The latter is shorter (and may thus
 fall short of the apsidal period) at higher temperatures. So for heated-up planets the periapse may have to be regarded as idle, thus giving birth to the  $\,p\,'\,\neq\,p\,$ terms and making the heat production more intensive.


 \subsection{The low-inclination approximation}\label{low}

 A considerable simplification becomes available if we agree to omit the $\,O(i^2)\,$ terms. To appreciate this, examine the expressions for the inclination functions $\,F_{lmp}(i)\,$.

 For $\,l=2\,$, we have:
 $$
 F_{201}(i)~=~-~\frac{1}{2}~+~O(i^2)\quad,\qquad F_{220}(i)~=~3~+~O(i^2)~~,
 $$
 the other $\,F_{2mp}(i)\,$ being of the order of $\,i\,$ or higher. Moreover, as the functions $\,F_{200}(i)\,$ and $\,F_{202}(i)\,$ are of the order of $\,i^2\,$, the
 only relevant product with $\,m=0\,$ is the diagonal one: $\,F_{201}^{\,2}(i)\,=\,{1}/{4}\,+\,O(i^2)~$, the cross products $\,F_{20p\,'}(i)\,F_{20p}(i)\,$ being $\,O(i^2)\,$. Similarly, since $\,F_{221}(i)\,$ and $\,F_{222}(i)\,$ are of the order of $\,i^2\,$ or higher, the only relevant product with $\,m=2\,$ is diagonal: $\,F_{220}^{\,2}(i)~=~9~+~O(i^2)~$,

 Performing the same check for $\,l=3\,$, we write down:
 $$
 F_{311}(i)~=~-~\frac{3}{2}~+~O(i^2)\quad,\qquad F_{330}(i)~=~15~+~O(i^2)~~,
 $$
 and notice that the other $\,F_{3mp}(i)\,$ are of the order of $\,i\,$ or higher. Above that, as the functions $\,F_{310}(i)\,$, $\,F_{312}(i)\,$, and  $\,F_{313}(i)\,$ are of the order of $\,i^2\,$ or higher, the only surviving product with $\,m=1\,$ is the diagonal one, $\,F^{\,2}_{311}(i)\,=\,{9}/{4}\,+\,O(i^2)\,$, the cross products being $\,O(i^2)\,$. By the same token, since $\,F_{331}(i)\,$, $\,F_{332}(i)\,$, and  $\,F_{333}(i)\,$ are of the order of $\,i^2\,$ or higher, the only remaining term with $\,m=3\,$ is diagonal:
 $\,F^{\,2}_{330}(i)\,=\,225\,+\,O(i^2)\,$.

 A similar examination of the functions with $\,l=4\,$ demonstrates that only the diagonal terms $\,F_{402}^2(i)\,$, $\,F_{421}^2(i)\,$, and
 $\,F_{440}^2(i)\,$ survive, the cross terms being $\,O(i^2)\,$ or higher-order.

 We have performed this check for all $\,l\leq 7\,$, and made sure that only the $\,p\,'=p\,$ terms exceed $\,O(i^2)\,$. As the terms with $\,l>7\,$ are unlikely to play any role in practical calculations ever,$\,$\footnote{~Each product $\,A_{lmp\,'q\,'}\,B_{lmpq}\,$ is proportional to $\,(R/a)^{2l}\,$, and an extra factor of $\,R/a\,$ will show up later, when we plug everything into
 the expression for heat production rate. Indeed, integration over the surface will give us a factor of $\,R^2\,$ in (\ref{BBB}). Then the overall multiplier $\,1/(4\pi GR)\,$ standing in (\ref{25b}) must be taken into account. All together, this leaves us with an extra $\,R\,$, thus making the final expression for dissipation rate proportional to $\,(R/a)^{2l+1}\,$.}$\,$ we stop our check here and reduce our sum to the diagonal ($\,p\,'=p\,$) terms only.

 Then, for a quiescent pericentre, the secular part of (\ref{A18d}) will be:
 \ba
 \nonumber
 ^{\textstyle{^{(\dot{\omega}=0)}}}\langle\,W_l\,\,\dot{U}_l\,\rangle &=&\frac{(-1)^{\,l}}{2}~\sum_{p\,q}\,\left[A_{l0pq\,'}\right]_{\textstyle{_{
 q\,'\,=\,-q-2(l-2p)}}}~~\omega_{l0pq}~B_{l0pq}~\cos(\,2\,(l\,-\,2\,p)\,\omega_0\,)~\sin\epsilon_{l0pq}
 ~\\
 \nonumber\\
 &+&\frac{1}{2}~\sum_m\sum_{pq}\,A_{lmpq}~\omega_{lmpq}~B_{lmpq}~\sin\epsilon_{lmpq}~+\,~.~.~.~\,~,\qquad\qquad\qquad\qquad\qquad\qquad\qquad
 \label{qui}
 \ea
 For a moving pericentre, the secular part of (\ref{A18d}) will assume the form of
 \ba
 \nonumber
 ^{\textstyle{^{(\dot{\omega}\neq0)}}}\langle W_l\,\,\dot{U}_l\rangle\,= ~\qquad~\qquad~\qquad~\qquad~\qquad~\qquad~\qquad~\qquad~\qquad~\qquad~\qquad~\qquad~\qquad~\qquad~\qquad~\qquad\\
 \nonumber\\
 \frac{(-1)^{\,l}}{2}\,\sum_{\,p\,q\,}\,\left[A_{l0pq\,'}\right]_{\textstyle{_{q\,'\,=\,-q}}}^{\textstyle{^{\,p=l/2}}}~\omega_{l0pq}~B_{l0pq}~\sin\epsilon_{l0pq}
 ~+~\frac{1}{2}~\sum_m\sum_{\,p\,q\,}\,A_{lmpq}~\omega_{lmpq}~B_{lmpq}~\sin\epsilon_{lmpq}\,+~.~.~.~~.\quad
 \label{mov}
 \ea
 Recall that the ellipsis denotes the part vanishing after integration over the longitude. In the first sum, the condition $\,p=l/2\,$ means that the first sum is nil for
 odd values of $l$.

 The formulae (\ref{qui}) and (\ref{mov}) have been derived here for reference purposes solely. While they enable one to write down much shorter expressions for the
 tidal dissipation rate, below we shall write the expressions which are general and can be used for an arbitrary $\,i\,$.

 \subsection{The dissipation rate}

 \subsubsection{The case of a quiescent pericentre}

 Insertion of the formulae
 (\ref{A12}), (\ref{A16}) and (\ref{quiescentb}) into the equation (\ref{180b}) gives us the damping rate in the case when the pericentre is fixed and no averaging over
 the apsidal period is required:
  \ba
  \nonumber
  \langle\,P\,\rangle~=~\frac{1}{4\pi G R}~\sum_l(2l+1)~\int dS~\left\langle W_l\,\dot{U}_l \right\rangle~=~
  \ea
  \ba
  \nonumber
  \frac{1}{4\pi G R}\,\left(\frac{G\,M^*}{a}\right)^2\,
  \sum_{l=2}^{\infty}\,(2l+1)\,\left(\frac{R}{a}\right)^{\textstyle{^{2\,l}}}\left\{~\frac{(-1)^{\,l}}{2}~
       \sum_{p=0}^{l} \sum_{p\,'=0}^{l} F_{l0p}(i)~F_{l0p\,'}(i)\,\sum_{q\,=-\infty}^{\infty}\,G_{lpq}
  (e)~
  \right.
  \ea
  \ba
  \nonumber
   \left[\,G_{lp\,'q\,'}(e)\,\right]_{\textstyle{_{q\,'\,=\,-\,q\,-\,2\,(l-p-p\,')}}}\,~\omega_{\textstyle{_{l0pq}}}\,
  \,k_l(\omega_{\textstyle{_{l0pq}}})~\sin\epsilon_l(\omega_{\textstyle{_{l0pq}}})~
 \cos\left(\,2\,(l\,-\,p\,'\,-\,p)\,\omega_0\,\right)
 \int\,dS\;\left[\,P_{l0}(\sin\phi)\,\right]^2
 \label{}
 \ea
  \ba
  \nonumber
  +~\frac{1}{2}~\sum_{m=0}^{l}\,
  \left[~
  \frac{(l - m)!}{({\it l} + m)!}\;
  \left(\,2~-~\delta_{0m}\,\right)
  ~\right]^{\textstyle{^{2}}}~\sum_{p=0}^{l}F_{lmp}(i)\;\sum_{p\,'=0}^{l}F_{lmp\,'}(i)~\sum_{q\,=-\infty}^{\infty}  G_{lpq}(e)~\,\qquad\qquad\qquad
       ~\\   \nonumber\\ \nonumber\\
  \left.
   \left[\,G_{lp\,'q\,'}(e)\,\right]_{\textstyle{_{q\,'\,=\,q\,-\,2\,(p-p\,')}}}~\cos\left(\,2\,(p\,'\,-\,p)\,\omega_0\,\right)
   \,~\omega_{\textstyle{_{lmpq}}}\,
  \,k_l(\omega_{\textstyle{_{lmpq}}})~\sin\epsilon_l(\omega_{\textstyle{_{lmpq}}})\,
 \int\,dS\,\left[\,P_{{\it{l}}m}(\sin\phi)\,\right]^2
 \right\}~~.~~~~
 \label{}
 \ea
  Using the normalisation (\ref{B4b}) and recalling that $dS=R^2\,\cos\phi\,d\phi\,d\lambda\,$, write the integral as
 \ba
 \int\,dS\;\left[\,P_{{\it{l}}m}(\sin\phi)\,\right]^2\,=~R^2\,2~\pi~\frac{2}{2l\,+\,1}\,~\frac{(l+m)!}{(l-m)!}~\,,
 \label{BBB}
 \ea
 whence
   \ba
  \nonumber
  \langle\,P\,\rangle~=~
  \ea
  \ba
  \nonumber
  \frac{G\,{M^*}^{\,2}}{2~a}\,
  \sum_{l=2}^{\infty}\,\left(\frac{R}{a}\right)^{\textstyle{^{2\,l\,+\,1}}}\left\{~(-1)^{\,l}~~\delta_{0m}~
       \sum_{p=0}^{l} \sum_{p\,'=0}^{l} F_{lmp}(i)~F_{lmp\,'}(i)~
              \quad
  \right.
  \ea
  \ba
  \nonumber
  \,\sum_{q\,=-\infty}^{\infty}\,G_{lpq}(e)~\left[\,G_{lp\,'q\,'}(e)\,\right]_{\textstyle{_{q\,'\,=\,-\,q\,-\,2\,(l-p-p\,')}}}\,~  \cos\left(\,2\,(l\,-\,p\,'\,-\,p)\,\omega_0\,\right)
 \label{}
 \ea
  \ba
  \nonumber
  +~\sum_{m=0}^{l}~
    \frac{(l - m)!}{({\it l} + m)!}\;
  \left(\,2~-~\delta_{0m}\,\right)^2
  ~\sum_{p=0}^{l}F_{lmp}(i)\;\sum_{p\,'=0}^{l}F_{lmp\,'}(i)~\,\qquad\qquad~\,\qquad\qquad
       ~\\   \nonumber\\ \nonumber\\
  \left.
 ~\sum_{q\,=-\infty}^{\infty}  G_{lpq}(e)~  \left[\,G_{lp\,'q\,'}(e)\,\right]_{\textstyle{_{q\,'\,=\,q\,-\,2\,(p-p\,')}}}~\cos\left(\,2\,(p\,'\,-\,p)\,\omega_0\,\right)
   ~\right\}\,~\omega_{\textstyle{_{lmpq}}}\,
  \,k_l(\omega_{\textstyle{_{lmpq}}})~\sin\epsilon_l(\omega_{\textstyle{_{lmpq}}})~
  ~ ~.~\qquad
 \label{A139}
 \ea
 The expression will be simplified considerably, if we notice that the first sum in the curly brackets can be absorbed by the second sum. To appreciate this, introduce auxiliary integers
 \ba
 \tilde{p}~\equiv~l~-~p\,'\,~\quad\mbox{and}~\qquad\tilde{q}~\equiv~-~q\,'\,~.
 \label{}
 \ea
 Just as $\,p\,'\,$, the integer $\,\tilde{p}\,$ assumes the values from $\,0\,$ through $\,l\,$. Like $\,q\,'\,$, the integer $\,\tilde{q}\,$ runs from
 $\,-\,\infty\,$ to $\,\infty\,$. In terms of $\,\tilde{p}\,$ and $\,\tilde{q}\,$, the first sum becomes
 \bs
  \ba
   (-1)^{\,l}\,\delta_{0m}\sum_{p=0}^{l} \sum_{\tilde{p}=0}^{l} F_{lmp}(i)\,F_{lm(l-\tilde{p})}(i)
   \sum_{q\,=-\infty}^{\infty}G_{lpq}(e)\,\left[\,G_{l(l-\tilde{p})(-\tilde{q})}(e)\,\right]_{\textstyle{_{\tilde{q}\,=\,q\,-\,2\,(p\,-\,\tilde{p})}}}~  \cos\left(2(\tilde{p}-p)\,\omega_0\,\right) ~\qquad
 \label{}
 \ea
  \ba
  =~\delta_{0m}\sum_{p=0}^{l} \sum_{\tilde{p}=0}^{l} F_{lmp}(i)\,F_{lm\tilde{p}}(i)
   \sum_{q\,=-\infty}^{\infty}G_{lpq}(e)\,\left[\,G_{l\tilde{p}\tilde{q}}(e)\,\right]_{\textstyle{_{\tilde{q}\,=\,q\,-\,2\,(p\,-\,\tilde{p})}}}~  \cos\left(2(\tilde{p}-p)\,\omega_0\,\right)\,~,~~ \qquad
 \label{A141b}
 \ea
 \label{A141}
 \es
 where we made use of the relations \footnote{~The equality (\ref{middle}) is borrowed from Giacaglia (1976a, eqn. 23), while (\ref{earlier}) comes from Gooding \& Wagner (2008, eqn 10). The formula (\ref{further}) stems from the integral expression (e.g., Giacaglia 1976b, eqn 4)
  \ba
  F_{lmp}(i)~=~\frac{1}{2\,\pi}~\int_{-\pi}^{\pi}P^{m}_l(\sin\phi)~e^{\textstyle{^{i\,[m\lambda\,-\,(l-2p)u]}}}\,du~~.
  \label{}
  \ea
  }
  \ba
 G_{lpq}(e)~=~G_{l(l-p)(-q)}(e)~~,
 \label{middle}
 \ea
 \ba
 F_{lmp}(i)~=~(-1)^{l-m}~F_{lm(l-p)}(\pi-i)
 \label{earlier}
 \ea
 and
 \ba
 F_{l0p}(i)~=~F_{l0p}(\pi-i)~~.
 \label{further}
 \ea
 Evidently, (\ref{A141b}) is equal to the $\,m=0\,$ part of the second sum in the curly brackets in (\ref{A139}). Denoting $\,\tilde{p}\,$ with $\,p\,'\,$
 (which is legitimate, as these are just dummy indices), we can absorb the expression (\ref{A141b}) into the second sum. The outcome is:
  \ba
  \nonumber
  \langle\,P\,\rangle~=~
  \frac{G\,{M^*}^{\,2}}{a}\,
  \sum_{l=2}^{\infty}\,\left(\frac{R}{a}\right)^{\textstyle{^{2\,l\,+\,1}}}
  \sum_{m=0}^{l}~
  \frac{(l - m)!}{({\it l} + m)!}\;
  \left(\,2 -\delta_{0m}\,\right)
  \,\sum_{p=0}^{l}F_{lmp}(i)\;\sum_{p\,'=0}^{l}F_{lmp\,'}(i)
  \ea\ba
  \sum_{q\,=-\infty}^{\infty}  G_{lpq}(e)~  \left[\,G_{lp\,'q\,'}(e)\,\right]_{\textstyle{_{q\,'\,=\,q\,-\,2\,(p-p\,')}}}~\cos\left(\,2\,(p\,'\,-\,p)\,\omega_0\,\right)
  \,~\omega_{\textstyle{_{lmpq}}}\,
  \,k_l(\omega_{\textstyle{_{lmpq}}})~\sin\epsilon_l(\omega_{\textstyle{_{lmpq}}})~
  \,~.~~\qquad
  \label{A146}
  \ea

 \subsubsection{The case of a uniformly moving pericentre}\label{comparison}

 Similarly to the above, insertion of the equations (\ref{moving}), (\ref{A12}) and (\ref{A16}) into the formula (\ref{180b}) will give us the heating rate in the case when the pericentre is moving and our time averaging includes that over the apsidal period:
  \ba
  \nonumber
  \langle\,P\,\rangle~=~\frac{1}{4\pi G R}~\sum_l(2l+1)~\int dS~\left\langle W_l\,\dot{U}_l \right\rangle~=~
  \ea
  \ba
  \nonumber
  \frac{1}{4\pi G R}\,\left(\frac{G\,M^*}{a}\right)^2\,
  \sum_{l=2}^{\infty}\,(2l+1)\,\left(\frac{R}{a}\right)^{\textstyle{^{2\,l}}}\left\{~\frac{(-1)^{\,l}}{2}~
       \sum_{p=0}^{l}F^{\,2}_{l0p}(i)~(-1)^{l}
  \right.
  \ea
  \ba
  \nonumber
   \,\sum_{q\,=-\infty}^{\infty}\,G^{\,2}_{lpq}(e) ~\omega_{\textstyle{_{l0pq}}}\,
  \,k_l(\omega_{\textstyle{_{l0pq}}})~\sin\epsilon_l(\omega_{\textstyle{_{l0pq}}})~
 \int\,dS\;\left[\,P_{l0}(\sin\phi)\,\right]^2\,+
 \label{}
 \ea
  \ba
  \nonumber
  \frac{1}{2}~\sum_{m=0}^{l}\,
  \left[~
  \frac{(l - m)!}{({\it l} + m)!}\;
  \left(\,2~-~\delta_{0m}\,\right)
  ~\right]^{\textstyle{^{2}}}~\sum_{p=0}^{l}F^{\,2}_{lmp}(i)\,\qquad\qquad\qquad
       ~\\   \nonumber\\ \nonumber\\
  \left.
   \sum_{q\,=-\infty}^{\infty}  G^{\,2}_{lpq}
  (e)~\omega_{\textstyle{_{lmpq}}}\,
  \,k_l(\omega_{\textstyle{_{lmpq}}})~\sin\epsilon_l(\omega_{\textstyle{_{lmpq}}})~
 \int\,dS\;\left[\,P_{{\it{l}}m}(\sin\phi)\,\right]^2
 ~\right\}\,~ ~,~~
 \label{}
 \ea
 where we also employed the equalities (\ref{middle} - \ref{further}). Further insertion of (\ref{BBB}) leads us to:
  \bs
  \ba
  \nonumber
  \langle\,P\,\rangle~=~
    \frac{G\,{M^*}^{\,2}}{2~a  }\sum_{l=2}^{\infty}\left(\,\frac{R}{a}\,\right)^{\textstyle{^{2\,l\,+\,1}}}\,\sum_{m=0}^{l}
  \sum_{p=0}^{l}F^{\,2}_{lmp}(i)
  \ea
  \ba
  \sum_{q\,=-\infty}^{\infty}G^{\,2}_{lpq}(e)
  \left[
 \delta_{0m}
    ~+~\frac{(l - m)!}{({\it l} + m)!}
  \left(2-\delta_{0m}\right)^{2}\right] \,\omega_{\textstyle{_{lmpq}}}
  k_l(\omega_{\textstyle{_{lmpq}}})~\sin\epsilon_l(\omega_{\textstyle{_{lmpq}}})~=~\quad
 \label{}
 \ea
 \ba
  \frac{G\,{M^*}^{\,2}}{a  }~\sum_{l=2}^{\infty}\left(\frac{R\,}{a}\right)^{\textstyle{^{2l+1}}}\,\sum_{m=0}^{l}\,
  \sum_{p=0}^{l}F^{\,2}_{lmp}(i)
  \sum_{q\,=-\infty}^{\infty}G^{\,2}_{lpq}(e)
  \frac{(l - m)!}{({\it l} + m)!}
  \left(2-\delta_{0m}\right)\,\omega_{\textstyle{_{lmpq}}}
  k_l(\omega_{\textstyle{_{lmpq}}})~\sin\epsilon_l(\omega_{\textstyle{_{lmpq}}})\,~.\qquad
 \label{our}
 \ea
 \es
It would be instructive to compare this expression with the corresponding result from the classical paper by Peale and Cassen (1978). To this end, three items must be kept in mind.
 \begin{itemize}
 \item[\bf 1.~] The authors of {\it{Ibid.}} tacitly assumed that averaging should be carried out not only over the tidal period but also over the apsidal period --- this can be understood from how their formulae (21) transformed into (22). This is why their resulting formula (31) is appropriate to compare with our expression (\ref{our}).
 \item[\bf 2.~] As the derivation in {\it{Ibid.}} was intended for the incompressible case and for $\,l=2\,$ solely, we should use, for the purpose of comparison, the
 equality $\,k_2(\omega_{\textstyle{_{2mpq}}})=3h_2(\omega_{\textstyle{_{2mpq}}})/5\,$ derived in Appendix \ref{appC}.
 \item[\bf 3.~] In {\it{Ibid.}}, only the case of synchronous rotation was addressed, with $\,
     \omega_{2mpq}=\,(2-2p)\dot{\omega}\,+\,(2-2p+q)\,n\,+m\,(\dot{\Omega}-\dot{\theta})\,\approx\,(2-2p+q-m)\,n\,$, where $\,n\,$ is the apparent mean
 motion of the perturber.  Librations were ignored.
 \end{itemize}
 Taking all this into account, we write, for the purpose of comparison, an appropriately simplified version of our expression (\ref{our}):
  \ba
  \nonumber
  ^{\textstyle{^{(synchronous)}}}\langle\,P\,\rangle^{\textstyle{^{(incompress)}}}_{\textstyle{_{~l=2}}}~=~
  \ea
  \ba
  \frac{G\,{M^*}^{\,2}}{a  }\left(\frac{R\,}{a}\right)^{\textstyle{^{2l+1}}}\sum_{m=0}^{2}
  \sum_{p=0}^{2}F^{\,2}_{2mp}(i)
  \sum_{q\,=-\infty}^{\infty}G^{\,2}_{2pq}(e)
  \frac{(2 - m)!}{(2 + m)!}
  \left(2-\delta_{0m}\right)\,\omega_{\textstyle{_{2mpq}}}
  ~\frac{3}{5}~h_2(\omega_{\textstyle{_{2mpq}}})~\sin\epsilon_2(\omega_{\textstyle{_{2mpq}}})\,~.~~~
  \label{}
  \ea
 As the time lag in our formula (\ref{506a}) is always positive-definite, the sign of the phase lag $\,\epsilon_l(\omega_{\textstyle{_{lmpq}}})\,$ coincides with that of the tidal mode $\,\omega_{\textstyle{_{lmpq}}}\,$, wherefore the product $\,\omega_{\textstyle{_{lmpq}}}~\sin\epsilon_l(\omega_{\textstyle{_{lmpq}}})\,$ can always be written down as a product of absolute values:
 \ba
 \omega_{\textstyle{_{lmpq}}}~\sin\epsilon_l(\omega_{\textstyle{_{lmpq}}})~=~|\,\omega_{\textstyle{_{lmpq}}}\,|\,\cdot\,|\,\sin\epsilon_l(\omega_{\textstyle{_{lmpq}}})\,|~=~
 \frac{\chi_{\textstyle{_{lmpq}}}}{Q_{\textstyle{_{lmpq}}}}~~,
 \label{}
 \ea
 where $\,\frac{\textstyle 1~}{\textstyle Q_{\textstyle{_{lmpq}}}}\,\equiv\,|\,\sin\epsilon_l(\omega_{\textstyle{_{lmpq}}})\,|\,$ is the inverse quality factor, while $\,\chi_{\textstyle{_{lmpq}}}\,\equiv\, |\,\omega_{\textstyle{_{lmpq}}}\,|\,$ is the positive-definite physical forcing frequency. For synchronous spin and $\,l=2\,$, the forcing frequency is $\,\chi_{\textstyle{_{2mpq}}}\approx\, |\,2\,-\,2\,p\,+q\,-m\,|\,n\,$, whence the quadrupole input into the power is:
  \ba
  \nonumber
  ^{\textstyle{^{(synchronous)}}}\langle\,P\,\rangle^{\textstyle{^{(incompress)}}}_{\textstyle{_{~l=2}}}~=~
  \ea
  \ba
  \frac{G\,{M^*}^{\,2}}{a  }\left(\frac{R\,}{a}\right)^{\textstyle{^{2l+1}}}\sum_{m=0}^{2}
  \sum_{p=0}^{2}F^{\,2}_{2mp}(i)
  \sum_{q\,=-\infty}^{\infty}G^{\,2}_{2pq}(e)
  \frac{(2 - m)!}{(2 + m)!}
  \left(2-\delta_{0m}\right)
  ~\frac{3}{5}~h_2~\frac{|\,2-2p+q-m\,|}{Q_{\textstyle{_{2mpq}}}}~n~~,\qquad
  \label{dc}
  \ea
  with an absolute value in the numerator.

 Peale \& Cassen (1978) had in their formula (31) simply $\,(\,2\,-\,2\,p\,+q\,-m\,)\,$ instead of $\,|\,2\,-\,2\,p\,+q\,-m\,|\,$. Nonetheless, their result was correct, because
 they were using a nonstandard convention $\,1/Q_{lmpq}\,=\,\sin\epsilon_l(\omega_{lmpq})\,$ wherein the inverse quality factors incorporated the signs of the lags and, thus, were not positive definite.
 

 \begin{deluxetable}{lr}
 \tablecaption{Symbol key \label{nota.tab}}
 \tablewidth{0pt}
 \tablehead{
 \multicolumn{1}{c}{Notation}  &
 \multicolumn{1}{c}{Description}\\
 }
 \startdata
 $\rbold^{\,*}$ & \dotfill the position of the star relative to the centre of the planet \\
 $r^{\,*}$ & \dotfill the star-planet distance\\
 $\phi^{\,*}$ & \dotfill the declination of the star {\underline{relative to the equator of the planet}} \\
 $\lambda^{\,*}$ & \dotfill the right ascension of the star {\underline{relative to a fixed meridian on the planet}} \\
 $\Rbold$ & \dotfill a point on the surface of the planet \\
 $\rbold$ & \dotfill a point outside the planet, located above the surface point $\Rbold$ \\
 $R$ & \dotfill the radius of the planet \\
 $r$ & \dotfill distance from the centre of the planet to an exterior point $\,\rbold$\vspace{3mm}\\
 $\phi$ & \dotfill the latitude of the point $\,\Rbold\,$ on the surface of the planet \\
 $~$ &  (also the declination of an exterior point $\,\rbold\,$ located above the surface point $\Rbold$) \vspace{3mm}\\
 $\lambda$ & \dotfill the longitude of a point $\,\Rbold\,$ on the surface of the planet  \\
 $~$ &  (also the right ascension of an exterior point $\,\rbold\,$ located above the surface point $\Rbold$)  \vspace{3mm}\\
 $W(\Rbold,\,\erbold^{\,*})$ & \dotfill tide-raising potential at a surface point $\,\Rbold\,$ of the planet\\
 $W_l(\Rbold,\,\erbold^{\,*})$ & \dotfill ~the $l$-degree part of the tide-raising potential at a surface point $\Rbold$ of the planet\\
 $U(\rbold,\,\erbold^{\,*})$ & \dotfill additional tidal potential in a point $\,\rbold\,$ outside the planet\\
 $U_l(\rbold,\,\erbold^{\,*})$ & \dotfill ~$.\,.\,.$~the $l$-degree part of the additional tidal potential in a point $\rbold$ outside the planet\\
 $V^{\,0}$ & \dotfill the constant-in-time spherically symmetrical potential of an undeformed planet\\
 $V\,'$ & \dotfill the total perturbation of the potential of the planet ($\,V\,'\,=\,W\,+\,U\,$)\\
 $V$ & \dotfill the overall potential of the planet ($\,V\,=\,V^{\,0}\,+\,V\,'\,=\,V^{\,0}\,+\,W\,+\,U\,$)\\
 $\omega_{\textstyle{_{lmpq}}}$ & \dotfill the Fourier modes of the tide\\
 $\chi_{\textstyle{_{lmpq}}}$ & \dotfill the physical frequencies of the tidal stresses and strains ~($\,\chi_{\textstyle{_{lmpq}}}\,=\,|\,\omega_{\textstyle{_{lmpq}}}\,|\,$)\\
 $\chi$ & \dotfill a shorter notation for $\,\chi_{\textstyle{_{lmpq}}}\,$\\
 $M$ & \dotfill the mass of the planet \\
 $M^{\,*}$ & \dotfill the mass of the star \\
 $a$ & \dotfill the semimajor axis \\
 $e$ & \dotfill the eccentricity \\
 $i$ & \dotfill the obliquity (the inclination of the star as seen from the planet) \\
 $\omega$ & \dotfill the argument of the pericentre of the star as seen from the planet\\
 $\omega$ & \dotfill when there is no risk of confusion, $\omega$ is also used as a short notation for $\omega_{\textstyle{_{lmpq}}}$\\
 $\Omega$ & \dotfill the longitude of the node of the star as seen from the planet\\
 ${\cal{M}}$ & \dotfill the mean anomaly of the star as seen from the planet\\
 $n$ & \dotfill the mean motion ~\\
 $G$ & \dotfill Newton's gravitational constant \\
 $\theta$ & \dotfill the rotation angle of the planet\\
 $\stackrel{\bf\centerdot}{\theta\,}$ & \dotfill the spin rate of the planet\\
 $k_l,\,h_l\,$ & \dotfill the degree-$l\,$ static Love numbers of the planet\\
 $k_l(\omega_{\textstyle{_{lmpq}}}),\,h_l(\omega_{\textstyle{_{lmpq}}})\,$ & \dotfill the degree-$l\,$ dynamical Love numbers of the planet\\
 $\epsilon_l(\omega_{\textstyle{_{lmpq}}})$ & \dotfill the degree-$l\,$ tidal phase lag\\
 $\Delta t_l(\omega_{\textstyle{_{lmpq}}})$ & \dotfill the degree-$l\,$ tidal time lag\\
 $Q_l(\omega_{\textstyle{_{lmpq}}})$ & \dotfill the degree-$l\,$ tidal quality factor defined as $\,Q_l(\omega_{\textstyle{_{lmpq}}})\,\equiv\,1/\sin\epsilon_l(\omega_{\textstyle{_{lmpq}}}\,$)\\
 $Q_{\textstyle{_{lmpq}}}$ & \dotfill the $\,Q\,$ factor, in the notation of Kaula (1964) and Peale \& Cassen (1978)\\
 $F_{lmp}(i)$ & \dotfill the inclination functions\\
 $G_{lpq}(e)$ & \dotfill the eccentricity polynomials\\
 $\mu$ & \dotfill the mean rigidity of the mantle\\
 $J$ & \dotfill the mean compliance of the mantle ($\,J\,=\,1/\mu\,$)\\
 $J$ & \dotfill when there is no risk of confusion, $J$ also denotes the Jacobian\\
 $\rho$ & \dotfill the mean density of the planet\\
 g & \dotfill the surface gravity on the planet\\
 $\ubold$ & \dotfill tidal displacement in the planet\\
 $\vbold$ & \dotfill the velocity of tidal displacement in the planet\\
 $P$ & \dotfill the power exerted on the planet by the tidal stresses\\
 $\langle P\rangle$ & \dotfill the time-averaged power (averaged over one or several cycles of flexure)\\
 \enddata
 \end{deluxetable}

 \end{document}